%% file: p-full.tex
\documentclass[10pt]{llncs}

\usepackage[pdftex]{graphicx,color}	

\usepackage{amssymb}
\usepackage{url}

\usepackage{myMac-llncs}	

\RequirePackage{wrapfig}	

\input{lmac}

\usepackage{subfig}	

	\setboolean{DRAFT_FLAG}{true} 	
	\setboolean{DRAFT_FLAG}{false}  

	\renewcommand{\myPara}[1]{
		\addtocounter{myParagraph}{1}
		{\bf {\cocyan{\P \themyParagraph. #1}}}
		}

	\newcommand{\ifFullPaper}[2]{ #2 }
	\renewcommand{\ifFullPaper}[2]{ #1 }	
	\renewcommand{\ifFullPaper}[2]{ #2 }	

\begin{document}

\mainmatter	

\title{
	Resolution-Exact Planner for Thick Non-Crossing 2-Link Robots\footnote{
	This work is supported by NSF Grants CCF-1423228 and CCF-1564132.
}}
\titlerunning{Thick Non-Crossing 2-Link Robots}

\author{
Chee K. Yap \and  Zhongdi Luo \and Ching-Hsiang Hsu}
\institute{
	Department of Computer Science\\
	Courant Institute, NYU\\ New York, NY 10012, USA\\
        \email{\{yap,zl562,chhsu\}@cs.nyu.edu}
	}

\authorrunning{Yap, Luo, Hsu}


\toctitle{XXX: Lecture Notes in Computer Science}
\tocauthor{YYY: Authors' Instructions}
\maketitle

\begin{abstract}
	We consider the path planning problem for a 2-link
	robot amidst polygonal obstacles.
	Our robot is parametrizable by the lengths $\ell_1, \ell_2>0$ of its
	two links,
	the thickness $\tau \ge 0$ of the links, and an angle $\kappa$ that
	constrains the angle between the 2 links to be strictly greater than
	$\kappa$.  The case $\tau>0$ and $\kappa \ge 0$
	corresponds to ``thick non-crossing'' robots.
	This results in a novel 4DOF configuration space
	$\RR^2\times (\TT\setminus\Delta(\kappa))$ where $\TT$ is the torus
	and $\Delta(\kappa)$ the diagonal band of width $\kappa$.
	
	We design a resolution-exact planner for this robot using
	the framework of Soft Subdivision Search (SSS). 
	First, we provide an analysis of the space of forbidden
	angles, leading to a soft predicate for classifying configuration
	boxes.  We further exploit the T/R splitting technique 
	which was previously introduced for self-crossing thin 2-link robots. 
	
	Our open-source implementation in Core Library
	achieves real-time performance for a suite
	of combinatorially non-trivial obstacle sets.
	Experimentally, our algorithm is significantly better
	than any of the state-of-art sampling algorithms we looked at,
	in timing and in success rate.
\end{abstract}

\setcounter{page}{1}
\section{Introduction}
	
	Motion planning is one of the key topics of robotics
	\cite{lavalle:planning:bk,choset-etal:bk}.
	The dominant approach to motion planning for the last two decades 
	has been based on sampling, as represented by
	PRM \cite{kslo:prm} or RRT \cite{kuffner2000rrt} and their many
	variants.  An alternative (older) approach is based on subdivision 
	\cite{brooks-perez:subdivision:83,zhu-latombe:hierarchical:91,barbehenn-hutchinson:incremental-planner:95}.
	Recently, we introduced the notion of \dt{resolution-exactness}
	which might be regarded\footnote{
	  In the theory of computation, a computability
	  concept that has no such converse (e.g., recursive enumerability)
	  is said to be ``partially complete''.
	}
	as the well-known idea of ``resolution completeness'' but
	with a suitable converse
	\cite{wang-chiang-yap:motion-planning:13,yap:sss:13}.
	Briefly, a planner is resolution-exact if
	in addition to the usual inputs of path planning,
	there is an input parameter $\vareps>0$, and
	there exists a $K>1$ such that the planner
	will output a path if there exists one with clearance $K\vareps$;
	it will output NO-PATH if there does not exist one
	with clearance $K/\vareps$.  Note that its output is
	indeterminate if the optimal
	clearance lies between $K/\vareps$ and $K\vareps$.
	This provides the theoretical basis for exploiting
	the concept of \dt{soft predicates}, which is roughly speaking
	the numerical approximation of exact predicates.
	Such predicates avoid the hard problem of deciding zero,
	leading to much more practical algorithms than exact algorithms.
	To support such algorithms,
	we introduce an algorithmic framework
	\cite{yap:sss:13,sss2}
	based on subdivision called \dt{Soft Subdivision Search} (SSS).
	The present paper studies an SSS algorithm for a 2-link robot.
	\NOignore{
	    \refFig{100-noncrossing} shows a path
	    found by this robot in a nontrivial environment.
	}%

	\NOignore{
	\begin{figure}[ht]
		\vspace*{-0.8cm}
		\centering
		\subfloat[Trace of robot origin]{
	\includegraphics[width=0.28\textwidth]{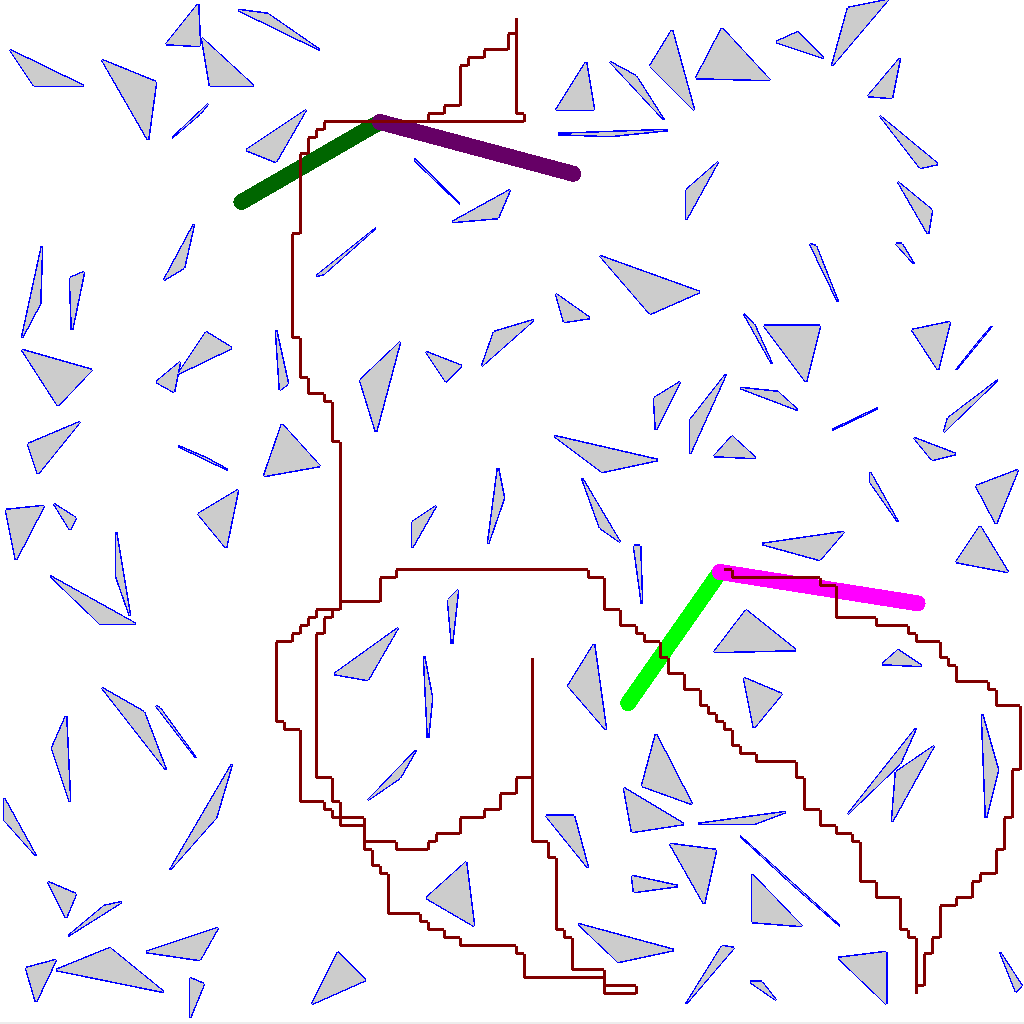}}
		\subfloat[Sub-sampled path]{
	\includegraphics[width=0.28\textwidth]{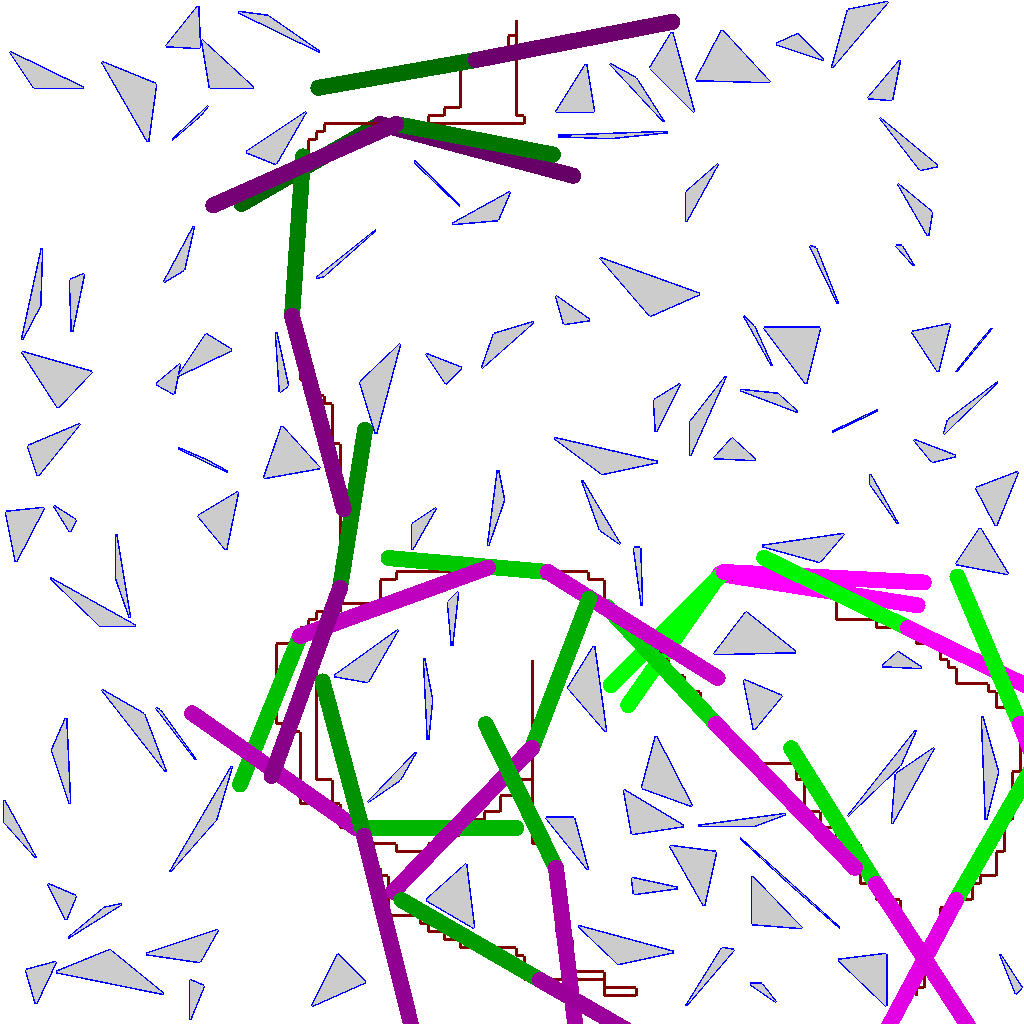}}
		\subfloat[Subdivision boxes]{
	 \includegraphics[width=0.28\textwidth]{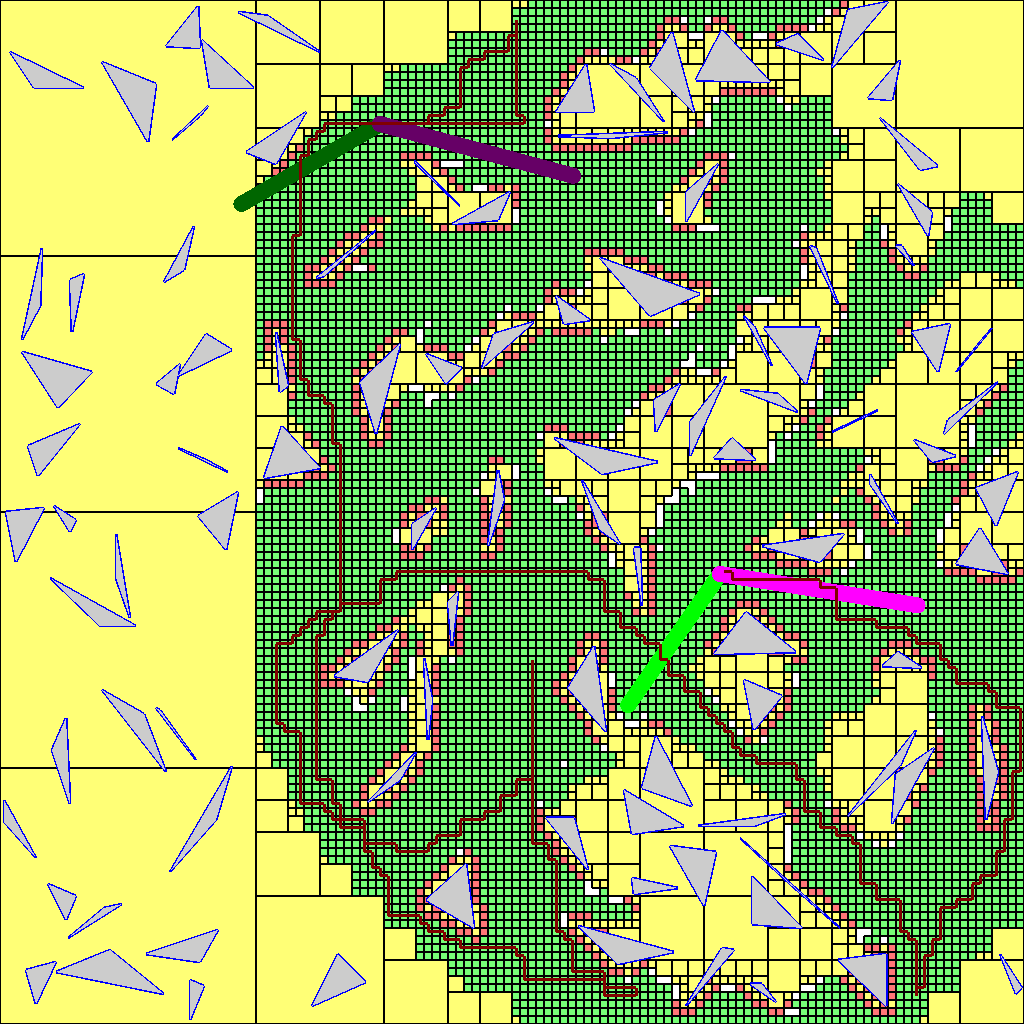}}
	 \caption{100 Random Triangles Environment:
	   non-crossing path found $(\kappa=115\degree)$ }
		\label{fig:100-noncrossing}
		\vspace*{-0.8cm}
	\end{figure}
	}%
	
\ifFullPaper{
	    
	    	\begin{figure}[htb]
	    	  \begin{center}
		   \scalebox{0.25}{
	    	     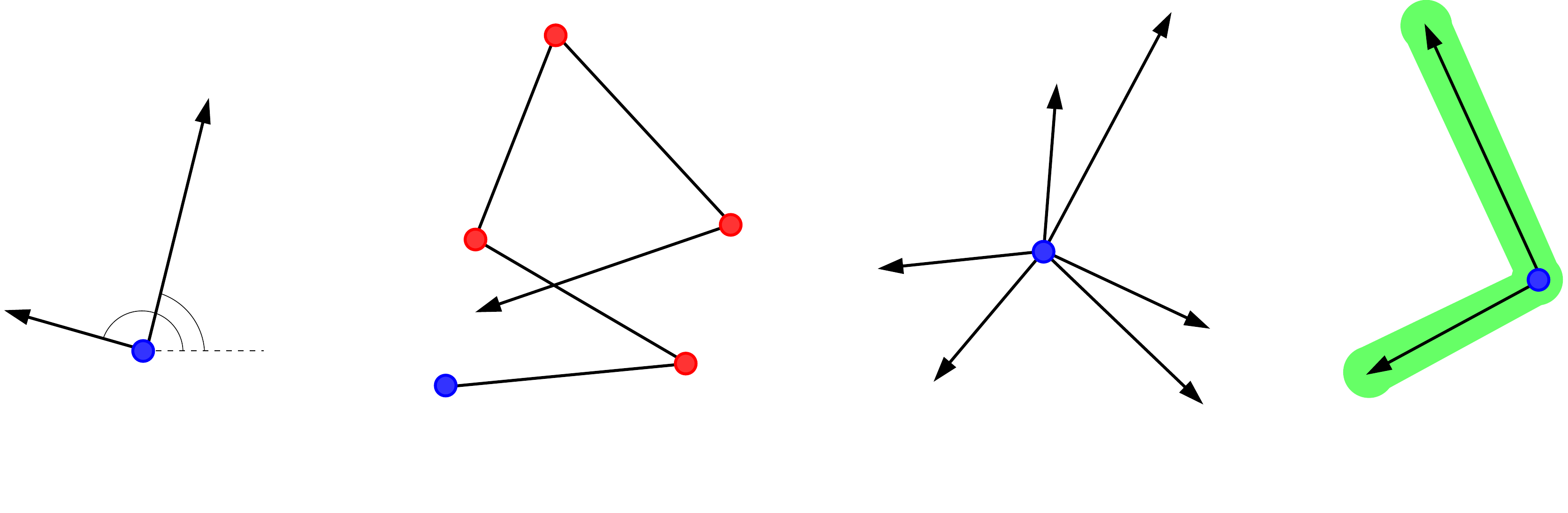}
	    	   \caption{Link Robots}
	    	   \label{fig:linkrobot}
	    	  \end{center}
	    	\end{figure} 	
	    }{}
	
	Link robots offer a compelling class of non-trivial robots
	for exploring path planning
	(see \cite[chap.~7]{devadoss-orourke:bk}).
	In the plane, the simplest example of a non-rigid robot
	is the \dt{2-link robot}, $R_2=R_2(\ell_1,\ell_2)$,
	with links of lengths $\ell_1, \ell_2>0$.
\ifFullPaper{
	  The two links are connected through a rotational joint $A_0$ 
	  called the \dt{robot origin} as illustrated in \refFig{linkrobot}(a).
	  }{
	  The two links are connected through a rotational joint $A_0$ 
	  called the \dt{robot origin}.
	  }
	The 2-link robot is in the intersection of two well-known
	families of link robots,
	\dt{chain robots} and \dt{spider robots}
\ifFullPaper{ as illustrated in \refFig{linkrobot}(b,c) }{}
	(see \cite{luo-chiang-lien-yap:link:14}).
	One limitation of link robots is that links are
	unrealistically modeled by line segments.  On the other hand,
	a model of mechanical links involving complex details
	may require algorithms that currently do not exist or have
	high computational complexity.  As a compromise, we introduce
	\dt{thick links} by forming the Minkowski sum of each link with
	a ball of radius $\tau>0$ (\dt{thin links} correspond to $\tau=0$).
\ifFullPaper{ See \refFig{linkrobot}(d). }{}
	To our knowledge, no exact algorithm for thick $R_2$ is known;
	for a single link $R_1$, an exact algorithm based on
	retraction follows from \cite{odun-sharir-yap:vorII:87}.
	In this paper, we further parametrize $R_2$
	by a ``bandwidth'' $\kappa$ which
	constrains the angle between the 2 links to
	be greater than or equal to $\kappa$ (``self-crossing'' links
	is recovered by setting $\kappa<0$).
	Thus, our full robot model
		$$R_2(\ell_1,\ell_2,\tau,\kappa)$$
	has four parameters; our algorithms
	are uniform in these parameters.

\ifFullPaper{	
	\begin{figure}[ht]
	  	\centering
		\subfloat[{\small $\alpha$ (above), $\beta$ (below)}]{
	    \includegraphics[width=0.30\textwidth]{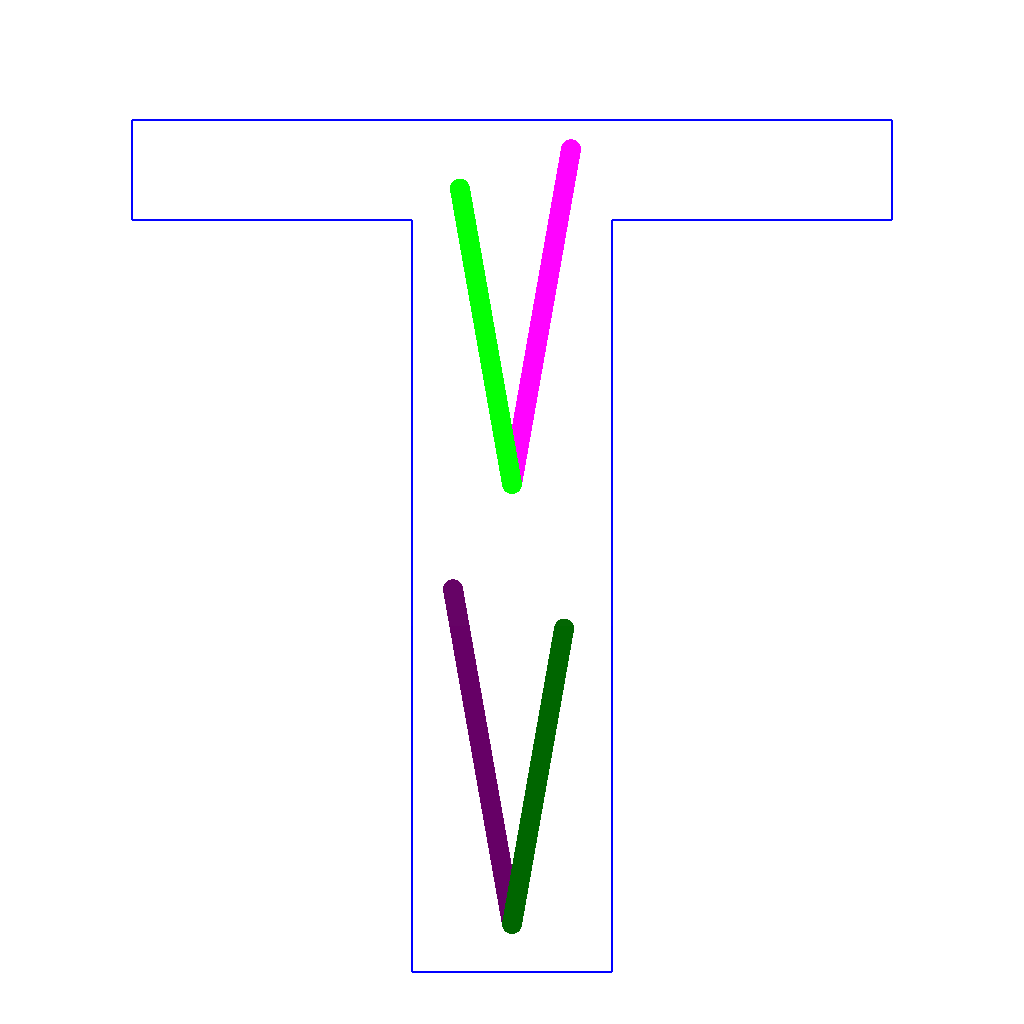}}
	    \subfloat[{\small Self-crossing path }]{
	    \includegraphics[width=0.30\textwidth]{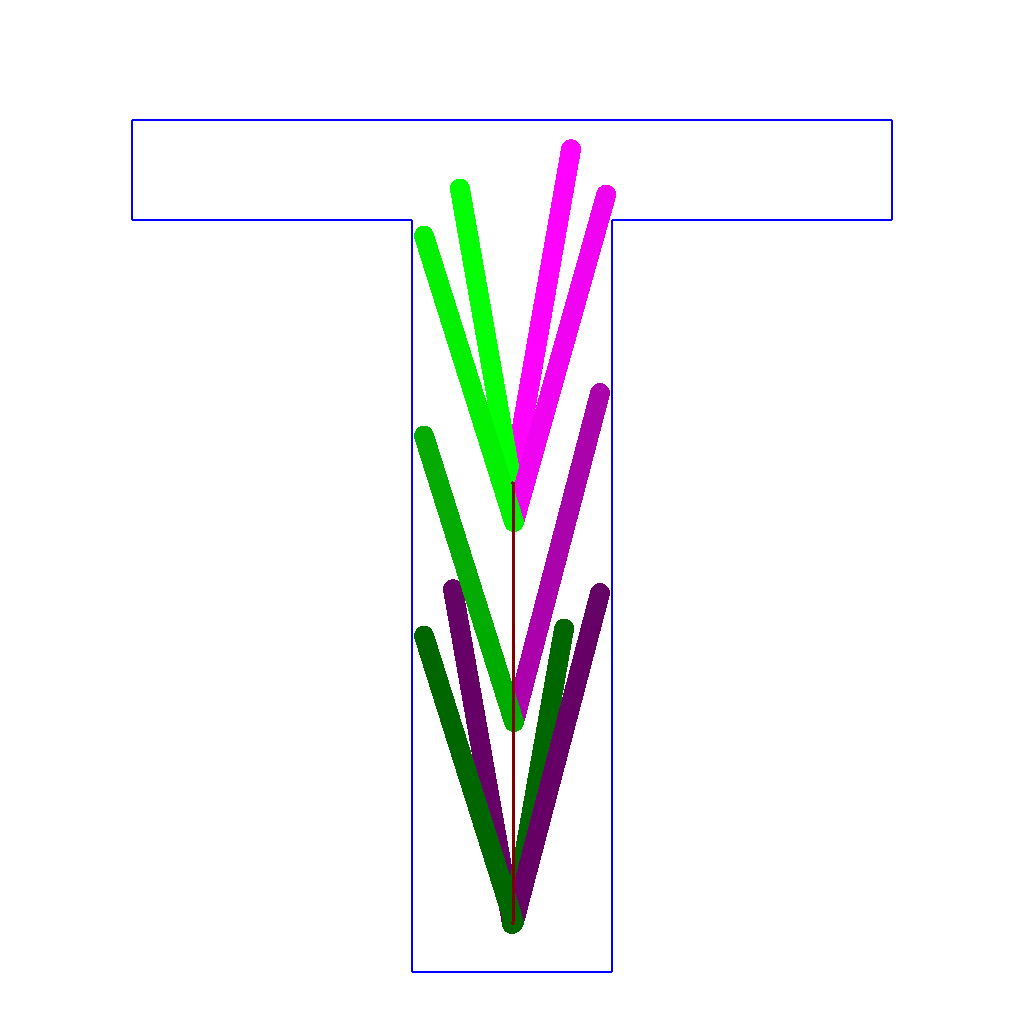}}
	    \subfloat[{\small Non-crossing path }]{
	    \includegraphics[width=0.30\textwidth]{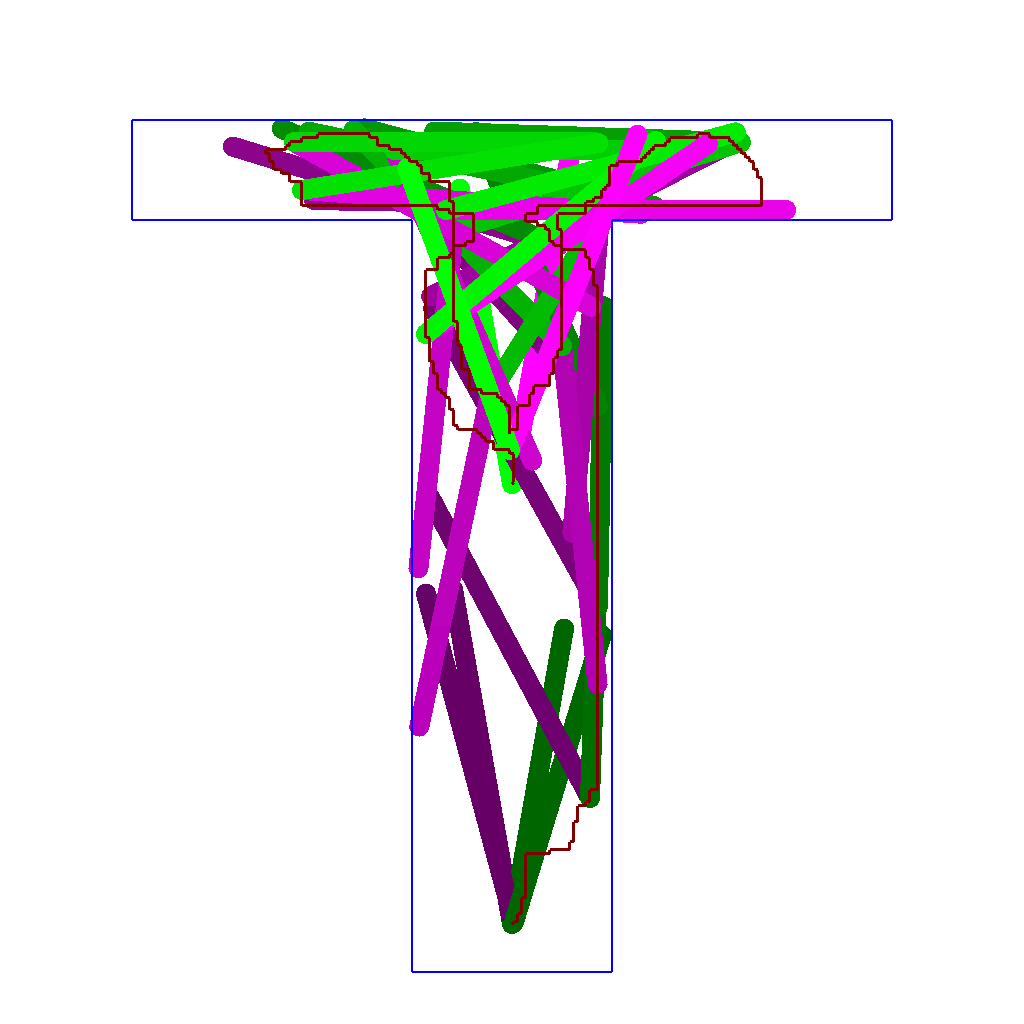}}
		\caption{Path from configurations $\alpha$ to $\beta$ in T-Room Environment}
		\label{fig:Troom}
	\end{figure}
}
{
	\begin{figure}[ht]
		\vspace*{-0.8cm}
	  	\centering
		\subfloat[{\small $\alpha$ (above), $\beta$ (below)}]{
	    \includegraphics[width=0.27\textwidth]{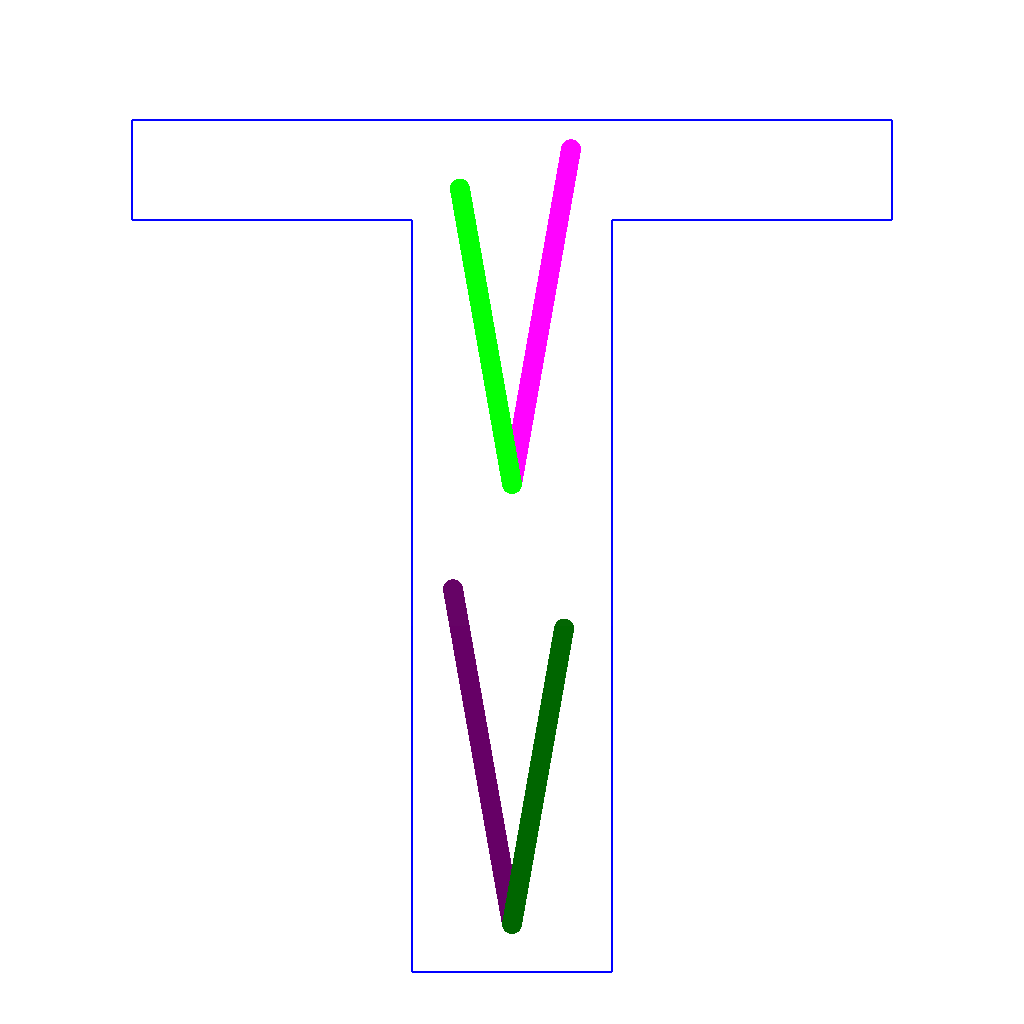}}
	    \subfloat[{\small Self-crossing path }]{
	    \includegraphics[width=0.27\textwidth]{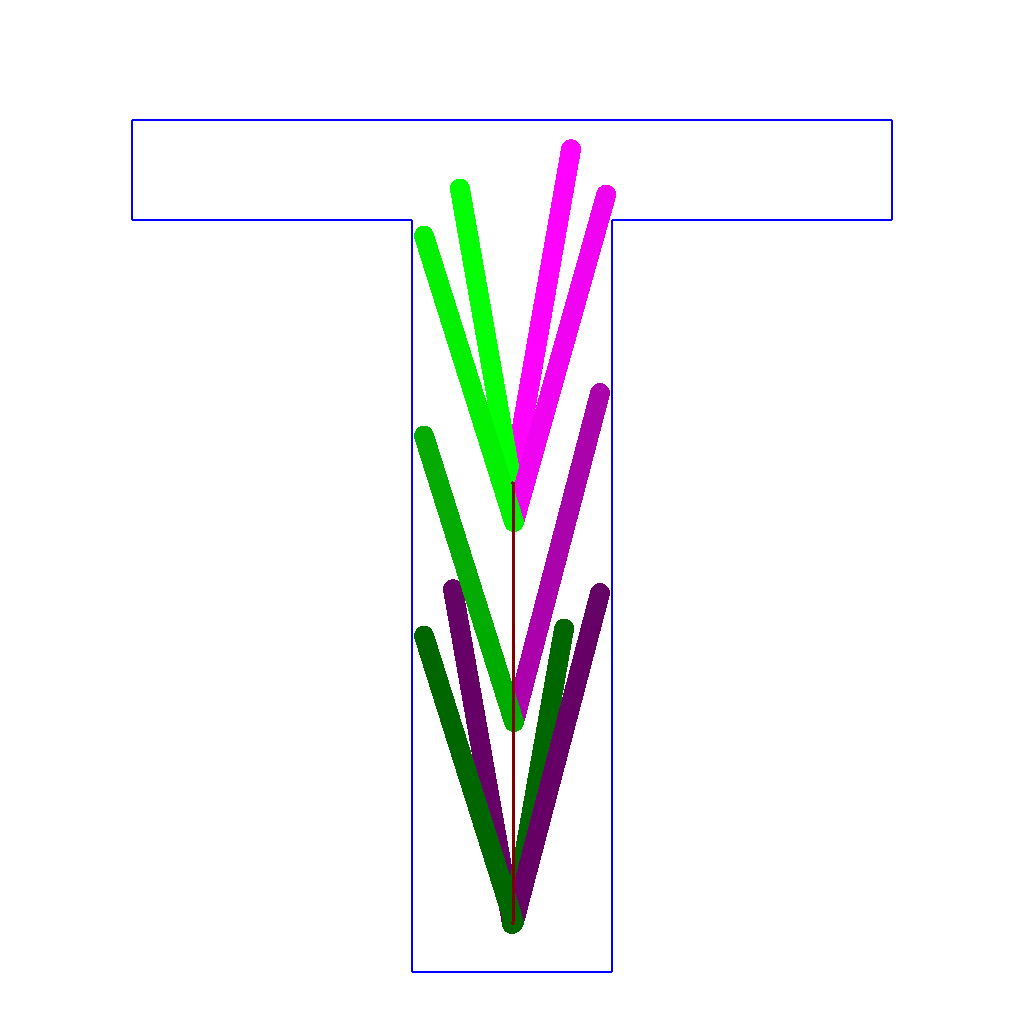}}
	    \subfloat[{\small Non-crossing path }]{
	    \includegraphics[width=0.27\textwidth]{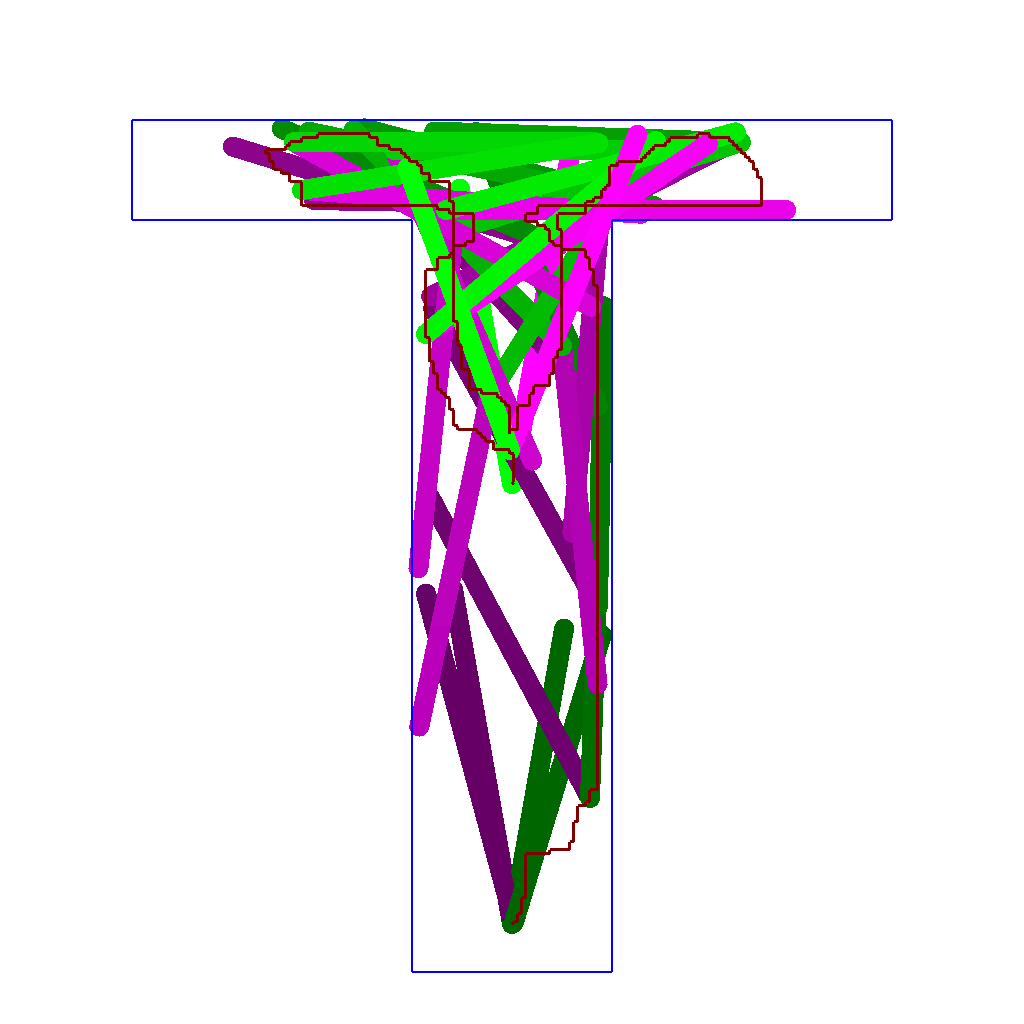}}
		\caption{Path from configurations $\alpha$ to $\beta$ in T-Room Environment}
		\label{fig:Troom}
		\vspace*{-0.8cm}
	\end{figure}
}
	
	To illustrate the non-crossing constraint, we use
	the simple ``T-room'' environment in \refFig{Troom}.
	Suppose the robot has to move from the start configuration $\alpha$
	to the goal configuration $\beta$ as indicated in
	\refFig{Troom}(a).
	There is an obvious path from $\alpha$ to $\beta$ as illustrated in
	\refFig{Troom}(b):
	the robot origin moves directly from its start to goal positions,
	while the link angles simultaneously adjust to their goal angles.
	However, such paths require the two links to cross each other.
	To achieve a ``non-crossing'' solution from $\alpha$ to $\beta$, 
	we need a less obvious path such as found by our algorithm in
	\refFig{Troom}(c):
	the robot origin must first move {\em away} from the goal, towards
	the T-junction, in order to maneuver the 2 links into an appropriate
	relative order before moving toward the goal configuration.

	We had chosen $\vareps=2$ in 
	\refFig{Troom}(b,c); also, $\kappa$ is $7$ for the
	non-crossing instance.
	But if we increase either $\vareps$ to $3$ or $\kappa$ to $8$,
	then the non-crossing instance would report NO-PATH.
	%
	It is important to know that the NO-PATH output from resolution-exact
	algorithms is not never due to exhaustion (``time-out'').
	It is a principled answer, guaranteeing the non-existence
	of paths with clearance $>K\cdot \vareps$ (for some $K>1$
	depending on the algorithm). 
	In our view, the narrow passage problem is, in the limit,
	just the \dt{halting problem} for path planning:
	algorithms with narrow passage problems will also have
	non-halting issues when there is no path.
	Our experiments suggests that the
	``narrow passage problem'' is barely an issue for our
	particular 4DOF robot, but it could be a severe issue
	for sampling approaches.
	But no amount of experimental data can truly express the conceptual
	gap between the various sampling heuristics and
	the {\em a priori} guaranteed methods such as ours.

	\ignore{
	Other than $\vareps$, our algorithm use no other input arguments
	than the logically necessary ones.  In contrast,
	some sampling algorithms require pre-processing arguments
	such as ``initial sampling size'' that could be
	tuned to optimize performance for each environment.
        }%
	
\ifFullPaper{The T-Room Environment has trivial combinatorial complexity,
	designed to illustrate the non-crossing phenomenon.
	But our algorithm scales well with the combinatorial
	complexity of the environment; all our solutions are ``realtime''.
}{}
\ifFullPaper{An interesting environment is \refFig{100-noncrossing} with
	100 randomly generated triangles.

	\begin{figure}[ht]
		\centering
		\subfloat[Trace of the robot origin]{
	\includegraphics[width=0.32\textwidth]{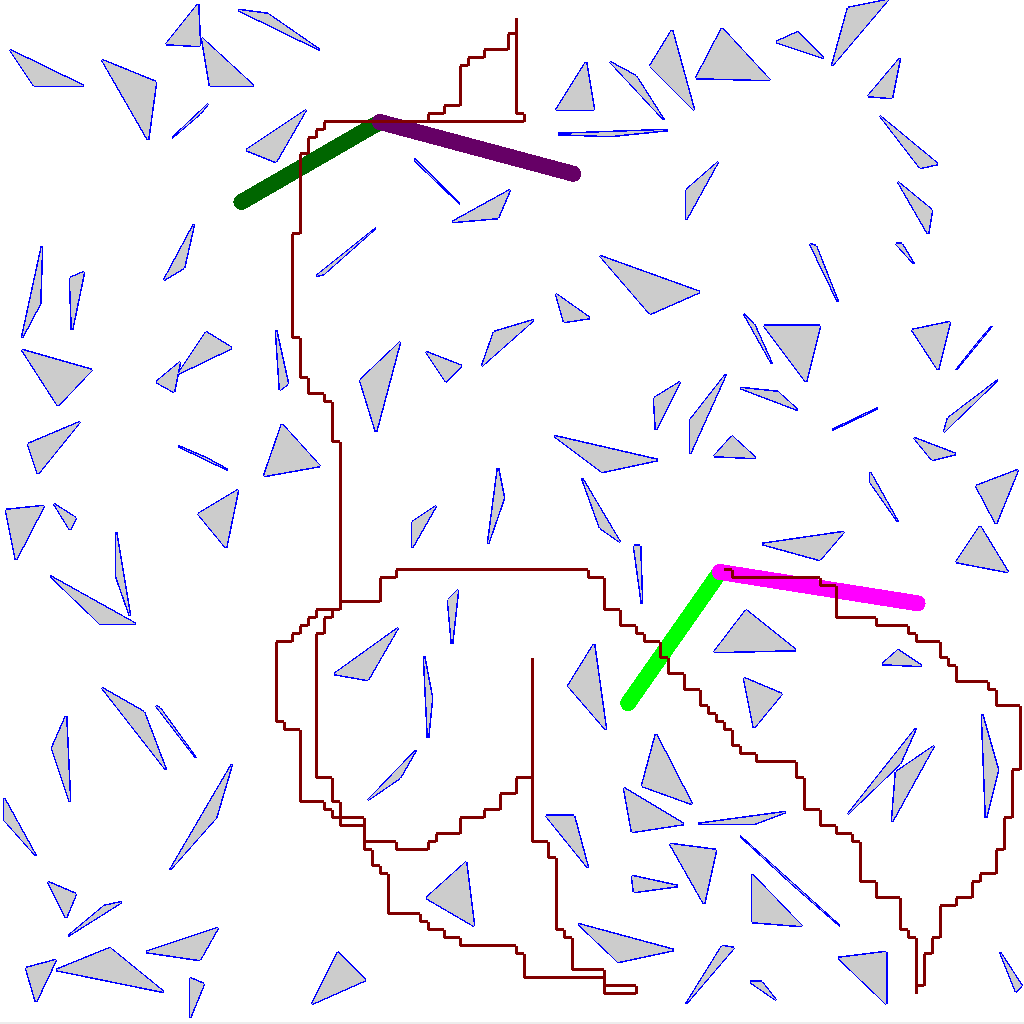}}
		\subfloat[Sub-sampled path]{
	\includegraphics[width=0.32\textwidth]{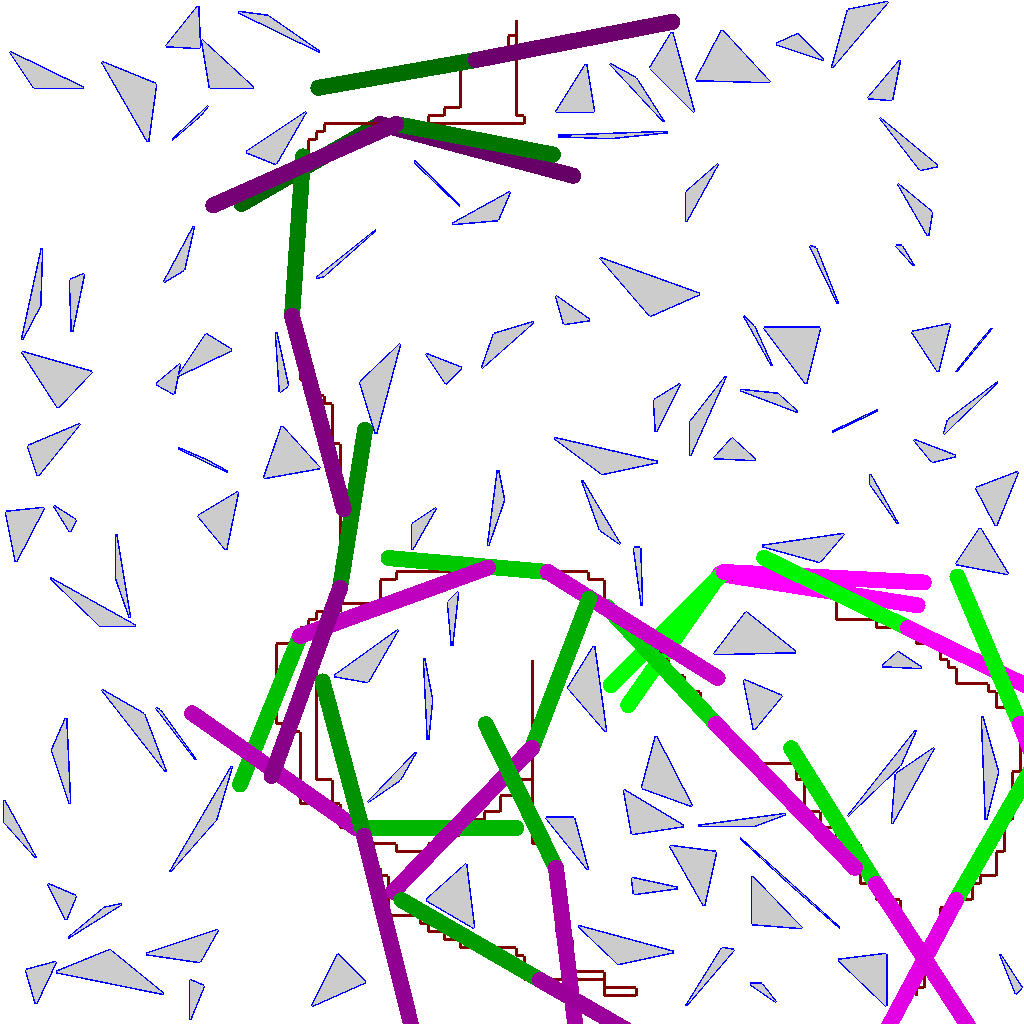}}
		\subfloat[Subdivision boxes]{
	 \includegraphics[width=0.32\textwidth]{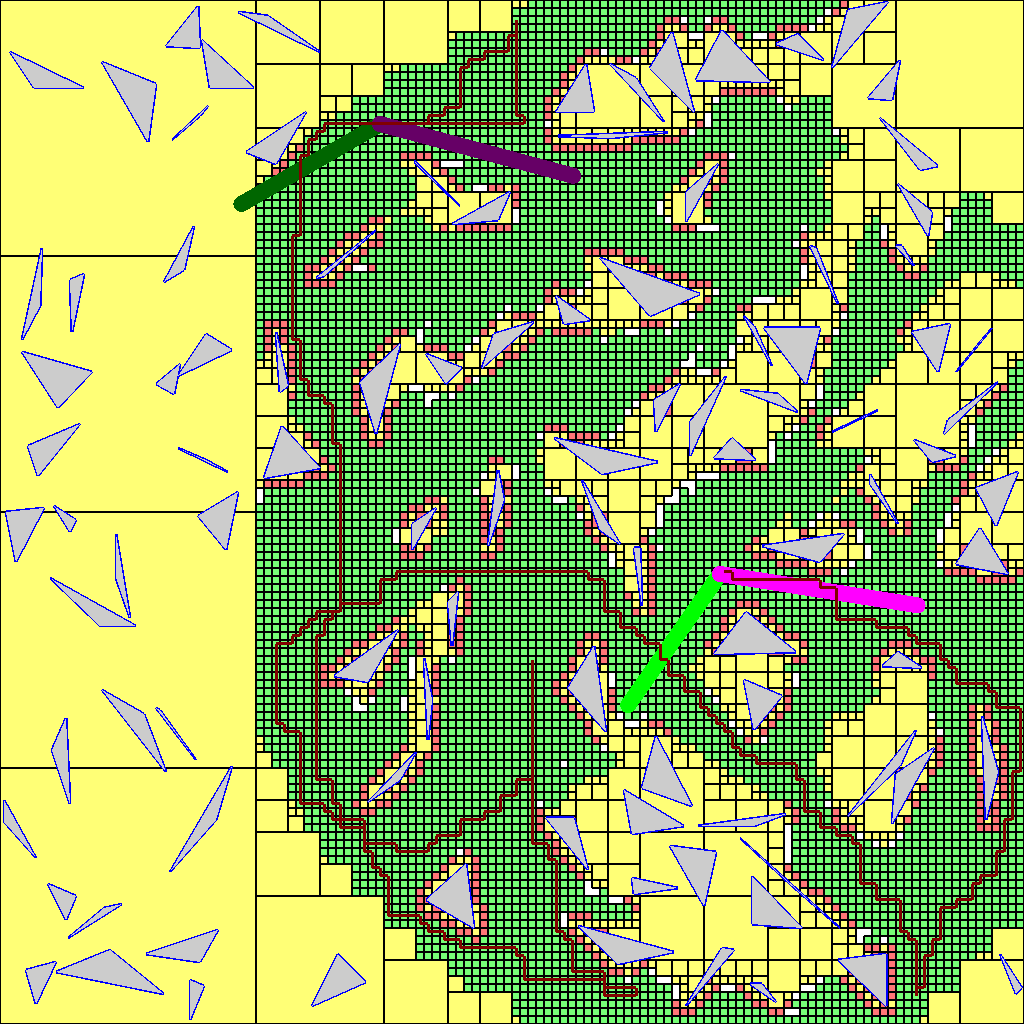}}
	 \caption{100 Random Triangles Environment:
	   non-crossing path found $(\kappa=115\degree)$ }
		\label{fig:100-noncrossing}
	\end{figure}
	}{}
	
	\ignore{
	\begin{figure}[ht]
		\centering
		\subfloat[Trace of the robot origin]{
	\includegraphics[width=0.36\textwidth]{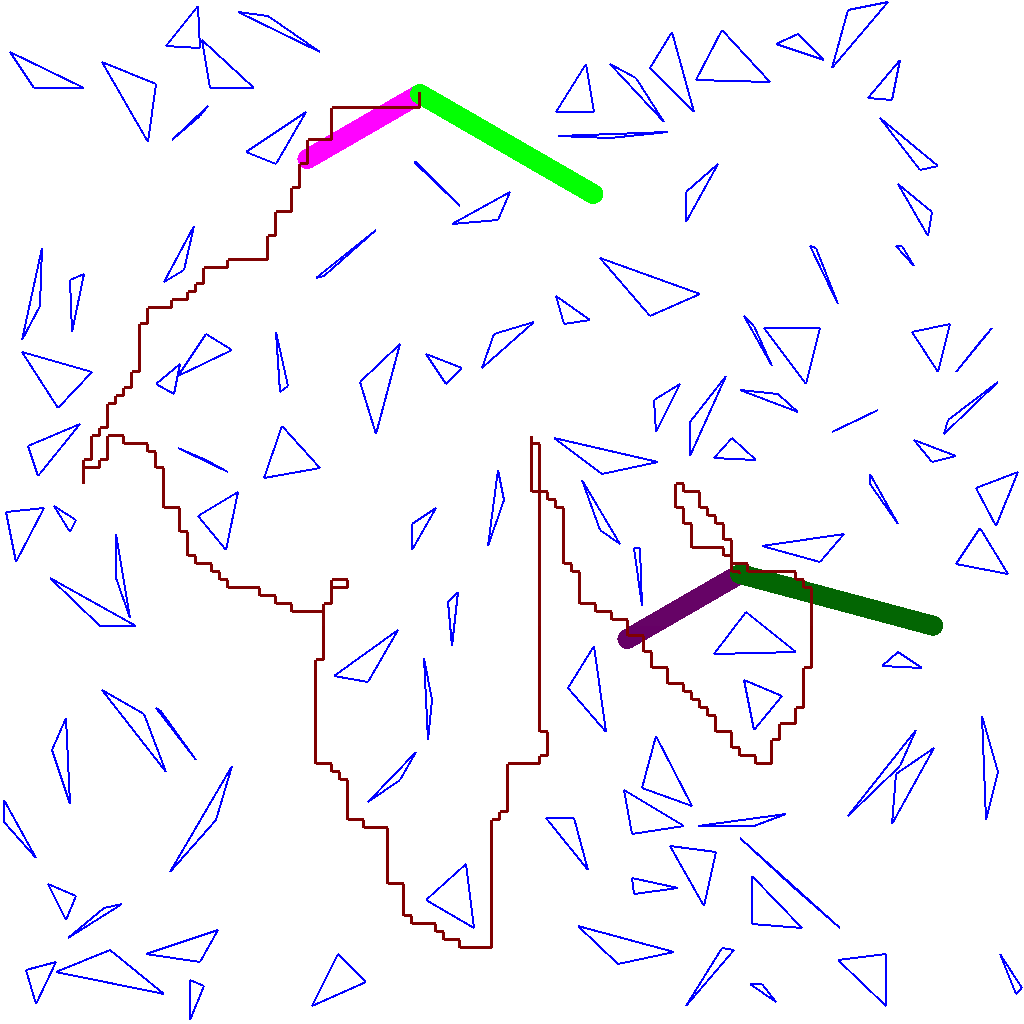}}
		\subfloat[Sub-sampled path]{
	\includegraphics[width=0.36\textwidth]{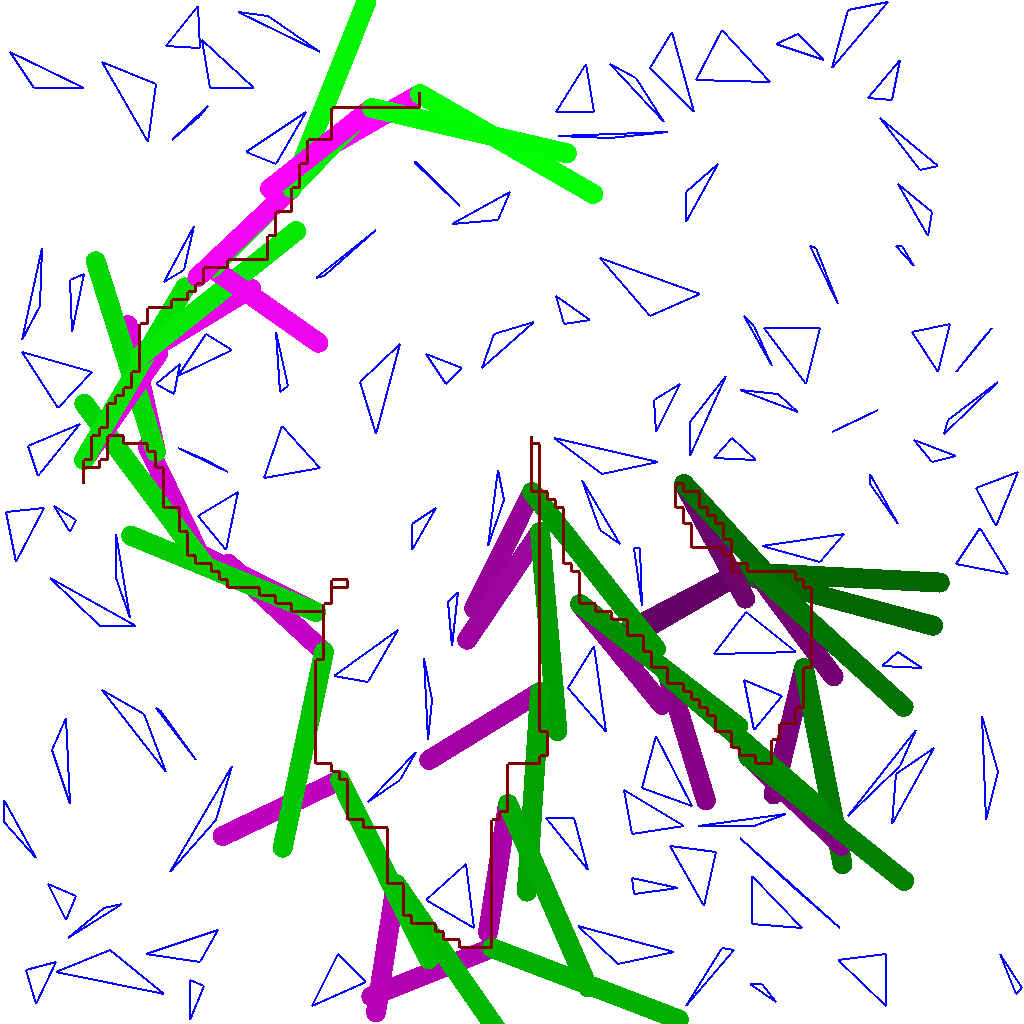}}
		\subfloat[Subdivision boxes]{
	\includegraphics[width=0.36\textwidth]{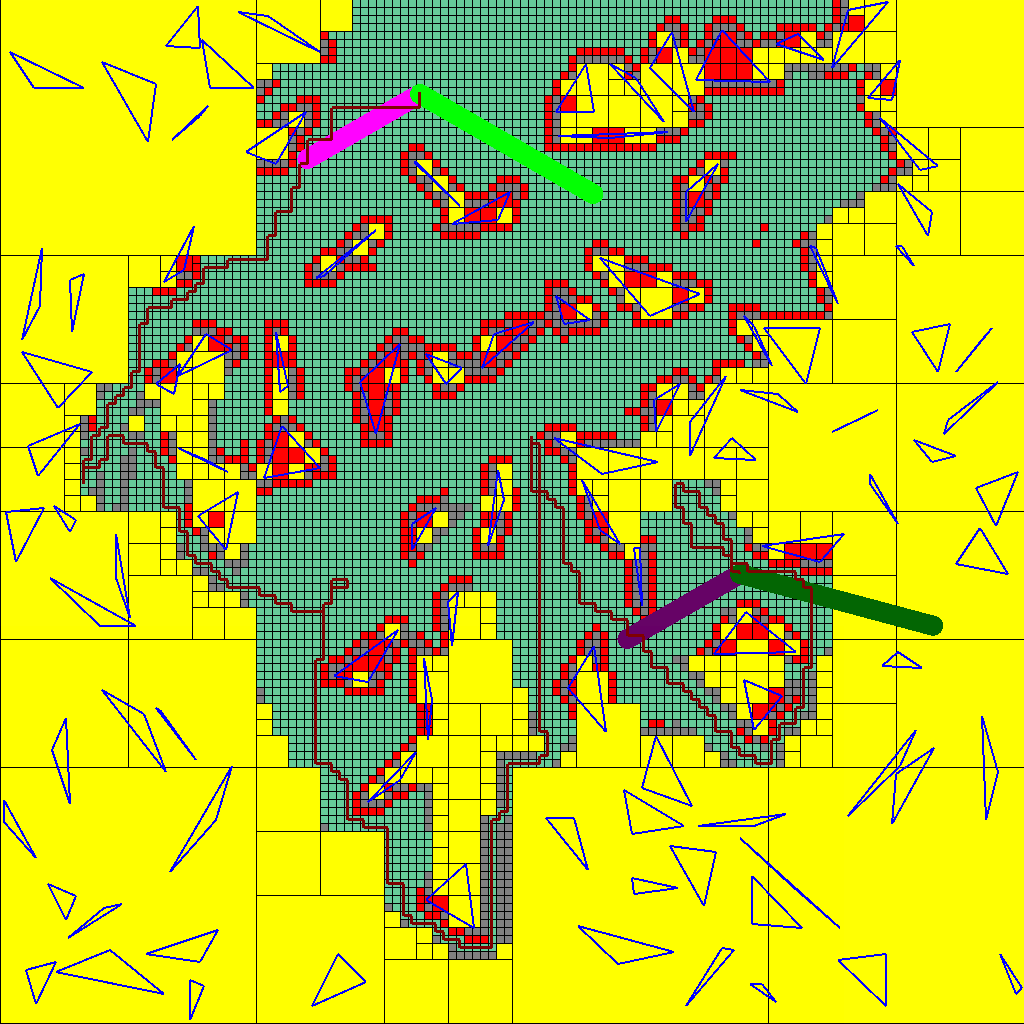}}
		\caption{100 Random Triangles Environment:
		self-crossing path found $(\kappa<0)$}
		\label{fig:100-selfcrossing}
	\end{figure}
    }%

	\dt{Literature Review.}
Our theory of resolution-exactness and SSS algorithms
apply to any robot system but we focus on
algorithmic techniques to achieve the best algorithms
for a 2-link robot.
The main competition is from
sampling approaches -- see \cite{choset-etal:bk} for a survey.
In our experiments, we compare against the well-known
PRM \cite{kslo:prm} as well as variants such as
Toggle PRM \cite{toggle-prm:12} and
lazy version \cite{lazy-toggle-prm:13}.
Another important family of sampling methods is the
RRT with variants such as RRT-connect \cite{kuffner2000rrt} and
retraction-RRT \cite{pan-zhang-manocha:retraction-rrt:10}.
The latter is
comparable to Toggle PRM's exploitation of non-free configurations,
except that retraction-RRT focuses on contact configurations.
Salzman et al \cite{salzman-solovey-halperin:tiling:15}
introduce a ``tiling technique'' for handling
non-crossing links in sampling algorithms that is quite different from
our subdivision solution \cite{luo-chiang-lien-yap:link:14}.

	\dt{Overview of Paper.}
	Section 2 describes our parametrization of the
	configuration space of $R_2$, and implicitly, its free space.
	Section 3 analyzes the forbidden angles of thick links.
	Section 4 shows our subdivision representation
	of the non-crossing configuration space.
	Section 5 describes our experimental results.
	Section 6 concludes the paper.
	Omitted proofs and additional experimental data are
	available as appendices in the full paper
		\cite{yap-luo-hsu:thicklink-arxiv:17}.
	%
	%
	%
	
	\ignore{
		In \cite{luo-chiang-lien-yap:link:14},
		we suggested and implemented a form of thick robots
		using two simple heuristics:
		(1) in computing the feature set
		of a box $B$, we replace the length $\ell$ of a link by $\ell+\tau$; 
		(2) in computing the forbidden angle range, we replace the range
		$[\alpha_1,\alpha_2]$ for a thin link by
		$[\alpha_1-\kappa_1,\alpha_2+\kappa_2]$ where
		$\kappa_i$'s 
		by adding $\tau$ to the separation computation.  This produces
		paths that are safe, but it is not resolution-exact.
		Let us clarify why:
	}
	\ignore{%
	In our implementation, the non-crossing planner typically
	suffer a negligible loss of efficiency when compared
	to the self-crossing planner.
	Nevertheless, it gives real-time performance for a variety of non-trivial
	obstacle environments such as illustrated in
	\refFig{200} (200 randomly generated triangles),
	\refFig{bugmaze}(a) (Double Bug-trap
		(cf.~[p.~181, Figure 5.13]\cite{lavalle:planning:bk}),
	\refFig{bugmaze}(b) (Maze).
      }%
	
	\ignore{
	For example, \refFig{200}(a) is an environment with 200 randomly
	generated triangles, and
	a path is found with $\vareps=4$.  If we use $\vareps=5$, then it
	returns NO-PATH as shown
	in \refFig{200}(b). This NO-PATH declaration guarantees that there is no
	path with clearance $>K\vareps$ (for some constant $K$).
	
	\begin{figure}[ht]
		\centering
		\subfloat[Path found with $\vareps=4$]{
		   \includegraphics[width=0.44\textwidth]{2-link-200b-subdiv-path.eps}}
		\subfloat[NO-PATH found with $\vareps=5$]{
		   \includegraphics[width=0.44\textwidth]{2-link-200b-subdiv-NOPATH.eps}}
		\caption{200 Random Triangles.}
		\label{fig:200}
	\end{figure}
      }%

      \ignore{	
	
	\begin{figure}[ht]
		\centering
		\subfloat[Double Bugtrap]{\includegraphics[width=0.44\textwidth]{2-link-bug2-path-noncross.eps}}
		\subfloat[Maze]{\includegraphics[width=0.44\textwidth]{2-link-maze-static-path.eps}}
		\caption{(a) Double Bugtrap, (b) Maze. }
		\label{fig:bugmaze}
	\end{figure}

      }%
	
	
	%
	

\section{Configuration Space of Non-Crossing 2-Link Robot}
	The configuration space of $R_2$
	is $\cspace \as \RR^2\times \TT$ where  
	$\TT=S^1 \times S^1$ is the torus and $S^1=SO(2)$ is the unit circle.  
	We represent $S^1$ by the interval $[0,2\pi]$ with the identification $0=2\pi$.
	Closed angular intervals of $S^1$ are denoted by
	$[s,t]$ where $s,t\in [0,2\pi]$ using the convention
		$$[s,t] \as \clauses{
			\ \set{\theta: s\le \theta\le t} & \rmif\ s\le t, \\
			\ [s,2\pi] \cup [0,t] & \rmif\ s>t.}
			$$
	In particular, $[0,2\pi]=S^1$ and $[2\pi,0]=[0,0]$.
	The standard Riemannian metric $d:S^1\times S^1\to \RR_{\ge 0}$ on
	$S^1$ is given by
	$d(\theta,\theta')=\min\set{|\theta-\theta'|, 2\pi-|\theta-\theta'|}$.
	Thus $0\le d(\theta,\theta')\le \pi$.
	
	To represent the non-crossing configuration space, we
	must be more specific about interpreting the parameters
	in a configuration $(x,y,\theta_1,\theta_2)\in \cspace$:
	there are two distinct interpretations, depending on whether
	$R_2$ is viewed as a chain robot or a spider 
\ifFullPaper{robot (see \refFig{linkrobot}(b,c)).}{ robot.}
	We choose the latter view:
	then $(x,y)$ is the \dt{footprint} of the
	joint $A_0$ at the center of the spider and
	$\theta_1,\theta_2$ are the independent angles of
	the two links.  This has some clear advantage over viewing $R_2$
	as a chain robot, but we can conceive of other advantages
	for the chain robot view. 
	That will be future research.
	%
	In the terminology of \cite{luo-chiang-lien-yap:link:14}, the
	robot $R_2$ has three named points $A_0,A_1,A_2$
\ifFullPaper{(see \refFig{linkrobot}(a))}{}
	whose \dt{footprints} at configuration
	$\gamma=(x,y,\theta_1,\theta_2)$ are given by
		{\small $$ A_0[\gamma] \as (x,y),\quad
		A_1[\gamma] \as (x,y)+ \ell_1 (\cos\theta_1,\sin\theta_1),\quad
		A_2[\gamma] \as (x,y)+ \ell_2 (\cos\theta_2,\sin\theta_2).
		$$}
	The \dt{thin footprint} of $R_2$ at $\gamma$, denoted
	$R_2[\gamma]$, is defined as the union of the line segments
	$[A_0[\gamma],A_1[\gamma]]$ and $[A_0[\gamma],A_2[\gamma]]$.
	The \dt{thick footprint} of $R_2$ is given by
	%
	%
	$\Fp_\tau(\gamma)\as D(\0,\tau)\oplus R_2[\gamma]$,
	the Minkowski sum $\oplus$ of the thin
	footprint with disc $D(\0,\tau)$ of radius $\tau$ centered at $\0$.

	The \dt{non-crossing configuration space} 
	of bandwidth $\kappa$ is defined to be
		$$\cspace(\kappa)\as \RR^2\times(\TT\setminus \Delta(\kappa))$$
	where $\Delta(\kappa)$ is the \dt{diagonal band} 
		$\Delta(\kappa)\as \set{(\theta,\theta')\in \TT:
			d(\theta,\theta')\le \kappa}\ib\TT.$
	Note three special cases:
	\bitem
      \item If $\kappa<0$ then $\Delta(\kappa)$ is the empty set.
      \item If $\kappa=0$ then $\Delta(\kappa)$ is a closed curve in $\TT$.
      \item If $\kappa\ge \pi$ then $\Delta(\kappa) = S^1$.
	\eitem
	Configurations in $\RR^2\times \Delta(0)$
	are said to be \dt{self-crossing}; all other configurations
	are \dt{non-crossing}.  Here we focus on the case $\kappa\ge 0$.
	\ignore{ 
		Of course, non-crossing constraint is important for actual robots, but
		in theoretical modeling, it is often ignored (perhaps justified
		by imagining the 2 links to lie in different planes).
		This non-crossing rotational subspace
		is topologically a cylinder with boundary.
		To our knowledge, this non-crossing configuration space
		has not been studied before.
	}
	\NOignore{
	  For our subdivision below, we will split
	  $\TT\setminus\Delta(0)$ into two connected sets:
	  $\TT_<	\as \set{(\theta,\theta')\in \TT:
	  0\le \theta<\theta'<2\pi}$ and
	  $\TT_>	\as \set{(\theta,\theta')\in \TT:
	  0\le \theta'<\theta<2\pi}$.
	}%
	For $\kappa\ge 0$,
	the diagonal band $\Delta(\kappa)$ retracts to the closed
	curve $\Delta(0)$.
	In $\RR^2$, if we omit such a set, we will get two
	connected components.  In contrast,
	that $\TT\setminus\Delta(\kappa)$ remains connected.
	CLAIM: {\em $\TT\setminus \Delta(\kappa)$ is topologically
	a cylinder with two boundary components.}
	Thus, the non-crossing constraint has
	changed the topology of the configuration space.
	To see claim, consider the standard model of $\TT$
	represented by a square with opposite sides identified as
	in \refFig{torusCylinder}(a) (we show the case $\kappa=0$).
	By rearranging the two triangles $\TT_<$ and $\TT_>$
	as in \refFig{torusCylinder}(b), our claim is
	now visually obvious.

	\begin{figure}[htb]
	    \begin{center}
		   	\scalebox{0.34}{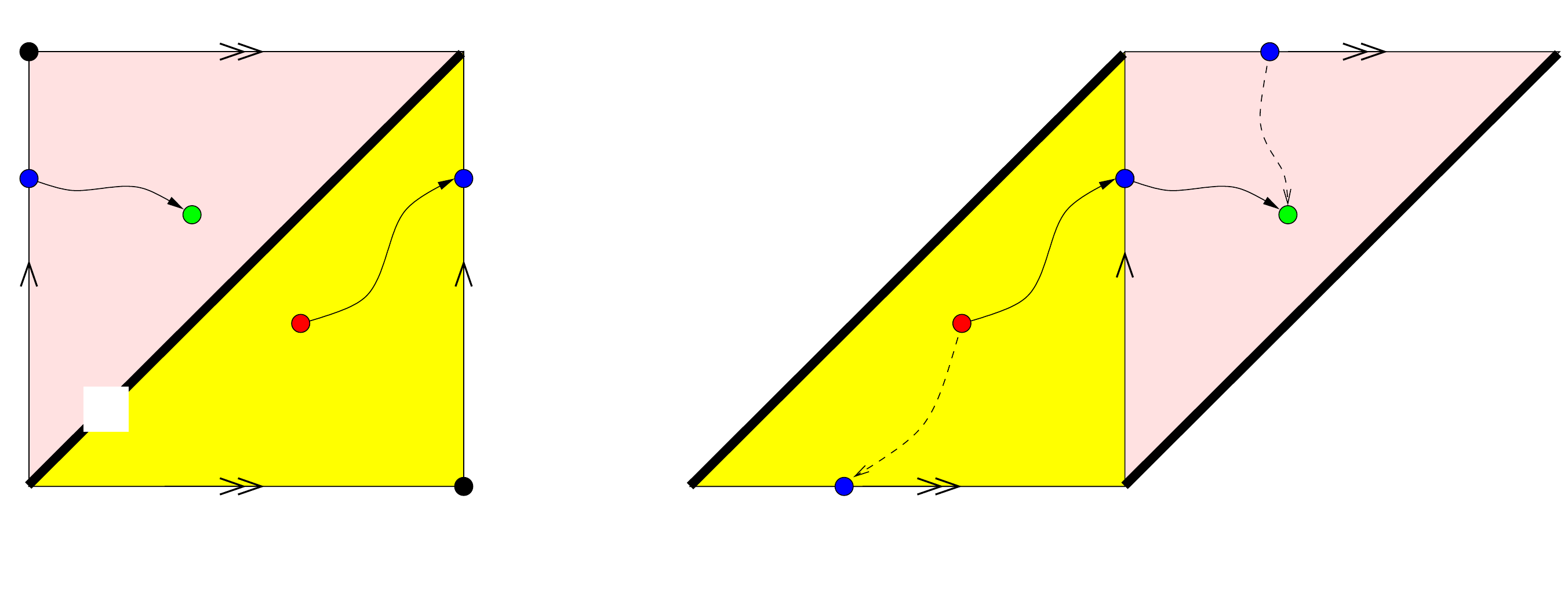}
	    	\caption{Paths in $\TT\setminus\Delta(0)$ from $\alpha\in \TT_>$ to $\beta\in \TT_<$}
	    	\label{fig:torusCylinder}
	   	\end{center}
	\end{figure}


\section{Forbidden Angle Analysis of Thick Links}
	Towards the development of a soft-predicate for thick
	links, we must first extend our analysis in 
	\cite{luo-chiang-lien-yap:link:14}
	which introduced the concept of forbidden angles for thin links.
	Let $L(\ell,\tau)$ be a single link robot
	of length $\ell>0$ and thickness $\tau\ge 0$.
	Its configuration space is $SE(2)=\RR^2 \times S^1$.
	Given a configuration $(b,\theta)\in SE(2)$, the \dt{footprint} of $L(\ell,\tau)$
	at $(b,\theta)$ is
		$$\Fp_{\ell,\tau}(b,\theta)\as L\oplus D(\0,\tau)$$
	where $\oplus$ denotes Minkowski sum, 
	$L$ is the line segment $[b, b+\ell (\cos\theta, \sin\theta)]$
	and $D(\0,\tau)$ is the disk as above.
	When $\ell,\tau$ is understood, we simply write
		``$\Fp(b,\theta)$'' instead of $\Fp_{\ell,\tau}(b,\theta)$.
	
	
	
	Let $S,T\ib \RR^2$ be closed sets.   An angle $\theta$ is \dt{forbidden}
	for $(S,T)$ if there exists $s\in S$ such that $\Fp(s,\theta)\cap T$ is non-empty.
	If $t\in \Fp(s,\theta)\cap T$, then the pair $(s,t)\in  S\times T$
	is a \dt{witness} for the forbidden-ness of $\theta$ for $(S,T)$.
	The set of forbidden angles of $(S,T)$ is called the \dt{forbidden zone} of $S,T$
	and denoted $\Forblt(S,T)$.
	Clearly, $\theta\in\Forblt(S,T)$ iff there exists
	a witness pair $(s,t)\in S\times T$.
	Moreover, we call $(s,t)$ a \dt{minimum witness} of $\theta$ if
	the Euclidean norm $\|s-t\|$ is minimum among all witnesses of $\theta$.
	If $(s,t)$ is a minimum witness, then clearly
	$s\in \partial S$ and $t\in \partial T$.
	

	\bleml{sym}
		For any sets $S,T\ib\RR^2$, we have
		$$\Forblt(S,T) = \pi + \Forblt(T,S).$$
	\eleml
	\bpf
	For any pair $(s,t)$ and any angle $\alpha$, we see that
		$$t \in \Fp(s,\alpha) \ifF\ s\in \Fp(t,\pi+\alpha).$$
	Thus, 
	there is a witness $(s,t)$ for $\alpha$ in $\Forblt(S,T)$ iff
	there is a witness $(t,s)$ for $\pi+\alpha$ in $\Forblt(T,S)$.
	The lemma follows.
	\epf

	\myPara{The Forbidden Zone of two points}
	Consider the forbidden zone $\Forblt(V,C)$
	defined by two points $V,C\in\RR^2$ with $d=\|V-C\|$.
	(The notation $V$ suggests a vertex of a translational
	box $B^t$ and $C$ suggests a corner of the obstacle set.)
	In our previous paper
	\cite{luo-chiang-lien-yap:link:14} on thin links
	(i.e., $\tau=0$), this case is not discussed for reasons
	of triviality.  When $\tau>0$, 
	the set $\Forblt(V,C)$ is more interesting.
	Clearly, $\Forblt(V,C)$ is empty iff $d>\ell+\tau$
	(and a singleton if $d=\ell+\tau$).
	Also $\Forblt(V,C)=S^1$ iff $d\le \tau$.  Henceforth, we may assume
		\beql{d}
		\tau < d < \ell+\tau.
		\eeql
	
	The forbidden zone of $V,C$ can be written in the form
		$$\Forblt(V,C) \as [\nu-\delta, \nu+\delta]$$
	for some $\nu,\delta$.  Call $\nu$ the \dt{nominal angle} and
	$\delta$ the \dt{correction angle}.  By the symmetry of the footprint,
	$\nu$ is equal to $\theta(V,C)$ (see \refFig{diamond}).
	
\ifFullPaper{
	
	    	\begin{figure}[htb]
	    	  \begin{center}
		   \scalebox{0.35}{
	    	     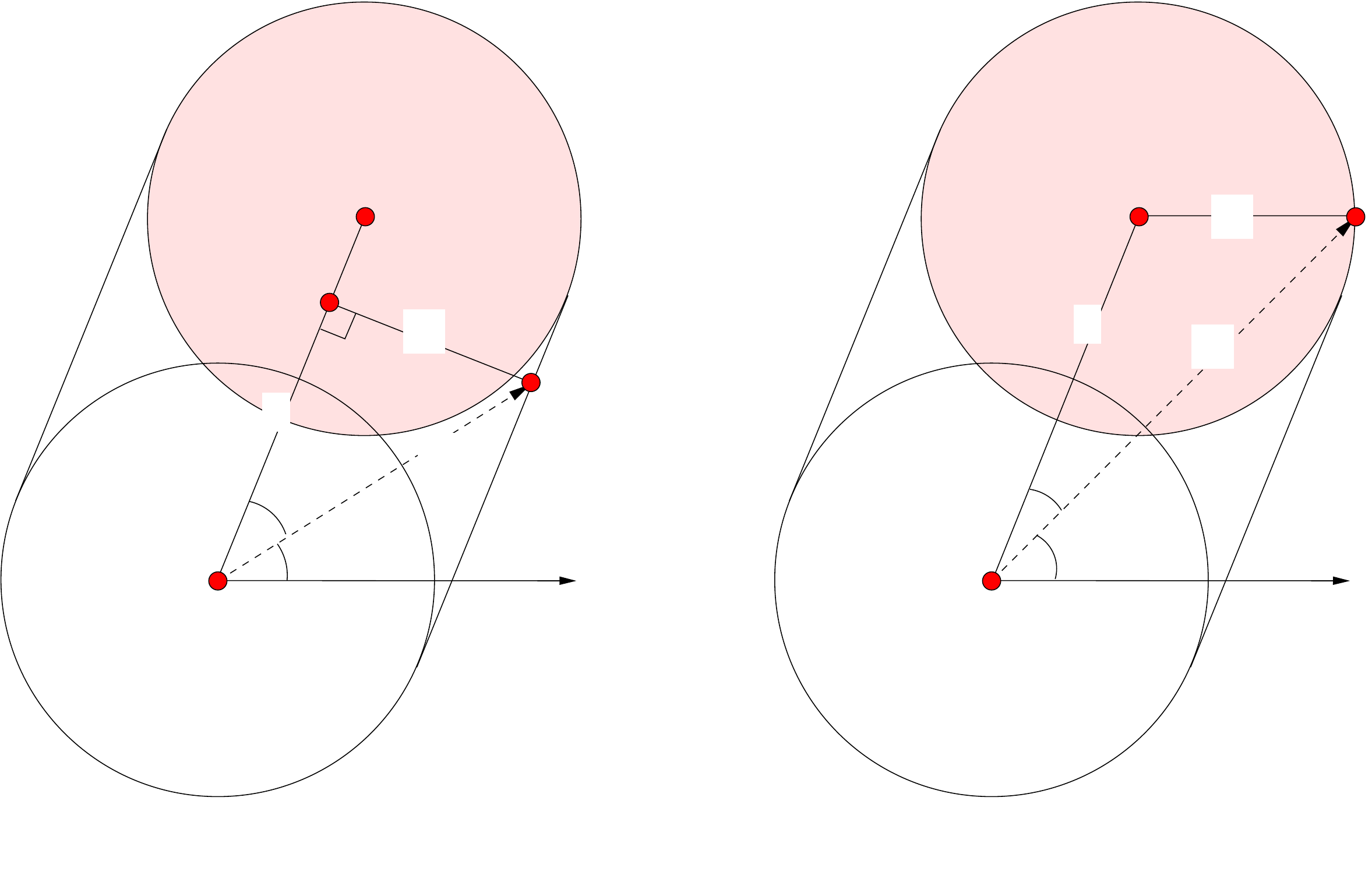}
	    	   \caption{$\Forblt(V,C)$}
	    	   \label{fig:diamond}
	    	  \end{center}
	    	\end{figure} 	
}
{
	
	    	\begin{figure}[htb]
	    	  \begin{center}
		   \scalebox{0.30}{
	    	     \input{./diamond.pdf_t}}
	    	   \caption{$\Forblt(V,C)$}
	    	   \label{fig:diamond}
	    	  \end{center}
	    	\end{figure} 	
}

	It remains to determine $\delta$. 
	Consider the configuration $(V,\theta)\in SE(2)$ of our link $L(\ell,\tau)$ where
	link origin is at $V$ and the link makes an angle $\theta$ with the
	positive $x$-axis.
	The angle $\delta$ is determined when the point $C$
	lies on the boundary of $\Fp(V,\theta)$.
	The two cases are illustrated in
	\refFig{diamond} where $\theta=\nu+\delta$ and other endpoint of the link is $U$;
	thus $\|VU\|=\ell$ and $\|VC\|=d$, and $\delta=\angle(CVU)$.
	Under the constraint \refeq{d}, there are two ranges for $d$:
	
	\benum[(a)]
	\item $d$ is short: $d^2 \le \tau^2+\ell^2$. 
	In this case, the point $C$ lies on the straight portion of the
	boundary of the footprint, as in \refFig{diamond}(a).
	From the right-angle triangle $CU'V$, we see that
	$\delta= \arcsin(\tau/d)$.
	\item $d$ is long: $d^2 > \tau^2+\ell^2$.
	In this case, the point $C$ lies on the circular portion of the
	boundary of the footprint, as in \refFig{diamond}(b).
	Consider the triangle $CUV$ with side lengths
	of $d,\ell,\tau$.   By the cosine law,
		$\tau^2 = d^2+\ell^2 -2d\ell\cos\delta$
	and thus 
		$$\delta = \arccos\paren{ \frac{\ell^2 + d^2 -\tau^2}{2d\ell}}.$$
	\eenum
	
	This proves:
	
	\bleml{point}
	Assume $\|VC\|=d$ satisfies \refeq{d}. Then
		$$\Forblt(V,C) = [\nu-\delta, \nu+\delta]$$
	where $\nu=\theta(V,C)$ and
		\beql{point}
		\delta=\delta(V,C)= \clauses{
			\arcsin(\tau/d)
				& \rmif\ d^2 \le \tau^2+\ell^2,\\
			\arccos\paren{ \frac{\ell^2 + d^2 -\tau^2}{2d\ell}}
				& \rmif\ d^2 > \tau^2+\ell^2.
		}\eeql
	\eleml
	
	\myPara{The Forbidden Zone of a Vertex and a Wall}
	Recall that the boundary of a box $B^t$ is divided into four \dt{sides},
	and two adjacent sides share a common endpoint which we call a \dt{vertex}.
	%
	We now determine $\Forblt(V,W)$ where $V$ is a vertex and $W$ a wall feature.
	Choose the coordinate axes such that $W$ lies on the $x$-axis,
	and $V=(0,-\sigma)$ lies on the negative $y$-axis, for some $\sigma>0$.
	Let the two corners of $W$ be $C, C'$
	with $C'$ lying to the left of $C$.   See \refFig{stops}.
	
\ifFullPaper{
	
	    	\begin{figure}[htb]
	    	  \begin{center}
		   \scalebox{0.28}{
	    	     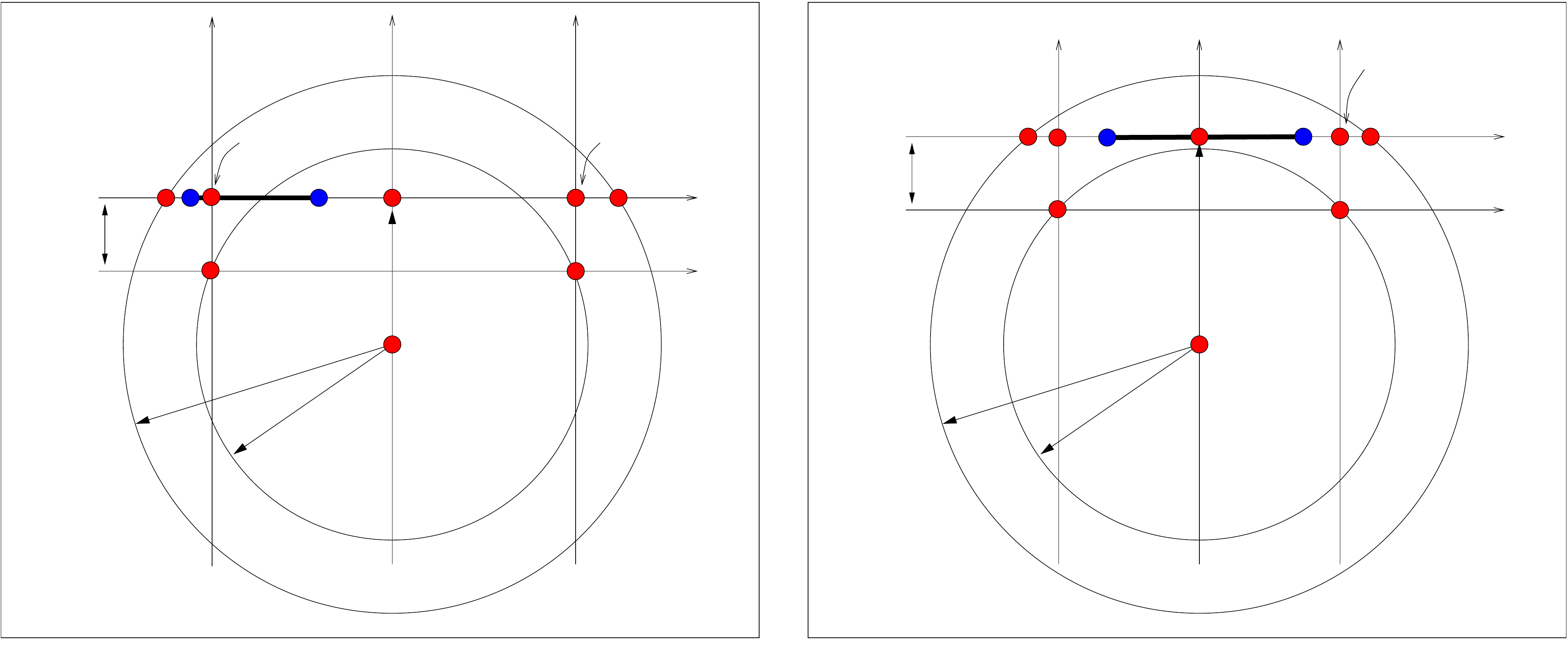}
	    	   \caption{Stop Analysis for $\Forblt(V,W)$ (assuming $\sigma>\tau$)}
	    	   \label{fig:stops}
	    	  \end{center}
	    	\end{figure} 	
}
{
	
	    	\begin{figure}[htb]
	    	  \begin{center}
		   \scalebox{0.24}{
	    	     \input{./stops.pdf_t}}
	    	   \caption{Stop Analysis for $\Forblt(V,W)$ (assuming $\sigma>\tau$)}
	    	   \label{fig:stops}
	    	  \end{center}
	    	\end{figure} 	
}
	We first show that the interesting case is when
		\beql{sigma}
		\tau <\sigma < \ell+\tau.
		\eeql
	If $\sigma\ge \ell+\tau$ then $\Forblt(V,W)$ is either
	a singleton ($\sigma=\ell+\tau$) or else is empty
	($\sigma>\ell+\tau$).
	Likewise, the following lemma shows that when $\sigma\le \tau$,
	we are to point-point case of \refLem{point}:
	
	\blem
	Assume $\sigma \le \tau$.  We have 
		$$\Forblt(V,W)=
			\clauses{ S^1 & \rmif\ D(V,\tau)\cap W\neq \es,\\
			\Forblt(V,c) & \elsE}$$
	where $c=C$ or $C'$.
	\elem
	\bpf
	Recall that we have chosen the coordinate system so that
	$W$ lies on the $x$-axes and $V=(0,-\sigma)$.
	It is easy  to see that $\Forblt(V,W)=S^1$
	iff the disc $D(V,\tau)$ intersects $W$.
	So assume otherwise.  In that case, the closest
	point in $W$ to $V$ is $c$, one of the two corners of $W$.
	The lemma is proved if we show that
		$$\Forblt(V,W)=\Forblt(V,c).$$
	It suffices to show $\Forblt(V,W)\ib \Forblt(V,c)$.
	Suppose $\theta\in\Forblt(V,W)$.  So it has a witness $(V,c')$
	for some $c'\in W$.  However, we see that the minimal
	witness for this case is $(V,c)$.  This proves that $\theta\in\Forblt(V,c)$.
	\epf
	
	In addition to \refeq{sigma}, we may also assume
	the wall lies within the annulus of radii $(\tau,\tau+\ell)$ centered
	at $V$:
		\beql{vc}
		\|VC\|,\|VC'\| \in (\tau, \ell+\tau)
		\eeql
	Using the fact that $V=(0,-\sigma)$
	and $W$ lies in the $x$-axis, we have:

	\blem
	Assume \refeq{sigma} and \refeq{vc}. \\
	Then
	$\Forblt(V,W)$ is a non-empty connected interval of $S^1$,
	$$\Forblt(V,W)=[\alpha,\beta]\ib (0,\pi).$$
	\elem
	%
	
	Our next goal is to determine the angles $\alpha,\beta$ in this lemma.
	Consider the footprints of the
	link at the extreme configurations $(V,\alpha), (V,\beta)\in SE(2)$.
	Clearly, $W$ intersects the boundary (but not interior) of these footprints,
	$\Fp(V,\alpha)$ and $\Fp(V,\beta)$.
	Except for some special configurations,
	these intersections are singleton sets.  Regardless,
	pick any $A\in W\cap \Fp(V,\alpha)$ and $B\in W\cap \Fp(V,\beta)$.
	Since $\alpha$ is an endpoint of $\Forblt(V,W)$, we
	see that $A\in (\partial W) \cap\partial (\Fp(V,\alpha))$.
	We call $A$ a \dt{left stop} for the pair $(V,W)$ because\footnote{
	  	Intuitively:
		At configuration $(V,\alpha)$, the single-link robot
		can rotate about $V$ to the right,
		but if it tries to rotate to the left,
		it is ``stopped'' by $A$. 
	} for any $\delta'>0$ small enough,
	$A \in \Fp(V,\alpha+\delta')$ while $W\cap (V,\alpha-\delta')=\es$.
	Similarly the point $B$ is called a \dt{right stop} for
	the pair $(V,W)$.
	Clearly, we can write
		$$\alpha=\theta(V,A)-\delta(V,A),\qquad
			\beta=\theta(V,B)+\delta(V,B)$$
	where $\delta(V,\cdot)$ is given by \refLem{point}.
	We have thus reduced the determination of angles $\alpha$ and $\beta$ to
	the computation of the left $A$ and right $B$ stops.
	
	We might initially guess that the left stop of $(V,W)$ is $C$,
	and right stop of $(V,W)$ is $C'$.  But the truth is a bit more subtle.
	Define the following points $X_*, X_{\max}$ on the positive $x$-axis
	using the equation:
		\beqarrays
			\|OX_*\| &=& \sqrt{(\ell+\tau)^2- \sigma^2},\\
			\|OX_{\max}\| &=&  \sqrt{\ell^2 -(\sigma-\tau)^2}.
		\eeqarrays
	These two points are illustrated in \refFig{stops}.
	Also, let $\olX_*$ and $\olX_{\max}$ be mirror reflections of
	$X_*$ and $X_{\max}$ across the $y$-axis.
	The points $X_*, \olX_*$ are the two points at
	distance $\ell+\tau$ from $V$. 
	The points $X_{\max}, \olX_{\max}$ are the left and right
	stops in we replace $W$ by the infinite line through $W$
	(i.e., the $x$-axis).

	With the natural ordering of points on the $x$-axis, we can show
	that 
		$$\olX_* < \olX_{\max} < O < X_{\max} <X_*$$
	where $O$ is the origin.   Since $\|VC\|$ and $\|VC'\|$ lie in $(\tau,\tau+\ell)$,
	it follows that
		$$\olX_* < C' < C <X_*.$$
	Two situations are shown in \refFig{stops}.
	The next lemma is essentially
	routine, once the points $X_{\max}, \olX_{\max}$ have defined:

	\bleml{stop}
	Assume \refeq{sigma} and \refeq{vc}.\\
		The left stop of $(V,W)$ is
		$$\clauses{ C' & \rmif\ X_{\max}\le C' & \qquad(L1)\\
				X_{\max} & \rmif\ C'<X_{\max}< C & \qquad(L2) \\
	      		C & \rmif\ C\le  X_{\max} & \qquad(L3)}$$		
		The right stop of $(V,W)$ is
		$$\clauses{ C & \rmif\ C\le \olX_{\max} & \qquad(R1) \\
			\olX_{\max} & \rmif\ C'< \olX_{\max}<C & \qquad(R2) \\
	      		C' & \rmif\ \olX_{\max}\le C' & \qquad(R3)}$$
	\eleml
	
	The cases (L1-3) and (R1-3) in this lemma
	suggests 9 combinations, but 3 are logically impossible:
	(L1-R1), (L1-R2), (L2-R1).  The remaining 6 possibilities for
	left and right stops are summarized in the following table:

	\bcen
	\renewcommand{\arraystretch}{1.2} 
	{\scriptsize
	\begin{tabular}{c || c|c|c|}
	  	& (R1) & (R2) & (R3)\\[1mm] \hline\hline
	  (L1) & * & * & $~(C',C')~$ \\[1mm] \hline
	  (L2) & * & $~(X_{\max},\olX_{\max})~$ & $~(X_{\max},C')~$ 
	    	\\[1mm]\hline
	  (L3) & $~(C,C)~$ & $~(C,\olX_{\max})~$ & $~(C,C')~$ \\[1mm]\hline
	\end{tabular}
	}
	\ecen
	
	Observe the extreme situations (L1-R3) or (L3-R1) where the
	the left and right stops are equal to the same corner,
	and we are reduced to the point-point analysis.
	Once we know the left and right stops for $(V,W)$, then
	we can use \refLem{point} to calculate the angles $\alpha$ and $\beta$.

	\myPara{The Forbidden Zone of a Side and a Corner}
	We now consider the forbidden zone
	$\Forblt(S,C)$ where $S$ is a side and $C$ a corner feature.
	Note that is complementary to the previous case of $\Forblt(V,W)$
	since $C$ and $V$ are points and $S$ and $W$ are line segments.
	We can exploit the principle of reflection symmetry
	of \refLem{sym}:
		$$\Forblt(S,C) = \pi + \Forblt(C,S)$$
		where $\Forblt(C,S)$ is provided by previous Lemma
		(with $C,S$ instead of $V,W$).
	
	\myPara{Cone Decomposition}
	We have now provided formulas for computing
	sets of the form $\Forblt(V,W)$ or $\Forblt(S,C)$;
	such sets are called \dt{cones}.
	We now address the problem of computing $\Forblt(B^t,W)$
	where $B^t\ib\RR^2$ is a (translational) box.
	We show that this set of forbidden angles can be written as the
	union of at most 3 cones,
	generalizes a similar result in
	\cite{luo-chiang-lien-yap:link:14}.
	\ignore{
		One technique in the case of thin links is to reduce
		$\Forbl(B^t,W)$ to the
		Vertex-Wall interaction, namely $\Forbl(V,W)$ or its dual
		$\Forbl(S,C)$.
		This remains true even when we introduce thickness, $\tau>0$.
		The other technique is to reduce 
		$\Forbl(V,W)$ to the case where
		the length $\ell$ is infinity,
		but $W$ is truncated to some $W'\ib W$, i.e.,
		$\Forbl(V,W)=\Forb_{\infty}(V,W')$.
        }%
	Towards such a cone decomposition, we first classify
	the disposition of a wall $W$ relative to a box $B^t$.
	There is a preliminary case:
	if $W$ intersects $B^t \oplus D(0,\tau)$, then we have
		$$\Forbl(B^t,W)=S^1.$$
	Call this \dt{Case (0)}.  Assuming $W$ does not intersect $B^t\oplus
	D(0,\tau)$, there are three other possibilities,
	\dt{Cases (I-III)} illustrated \refFig{wall-box}.

	    	\begin{figure}[htb]
	    	  \begin{center}
		   \scalebox{0.28}{
	    	     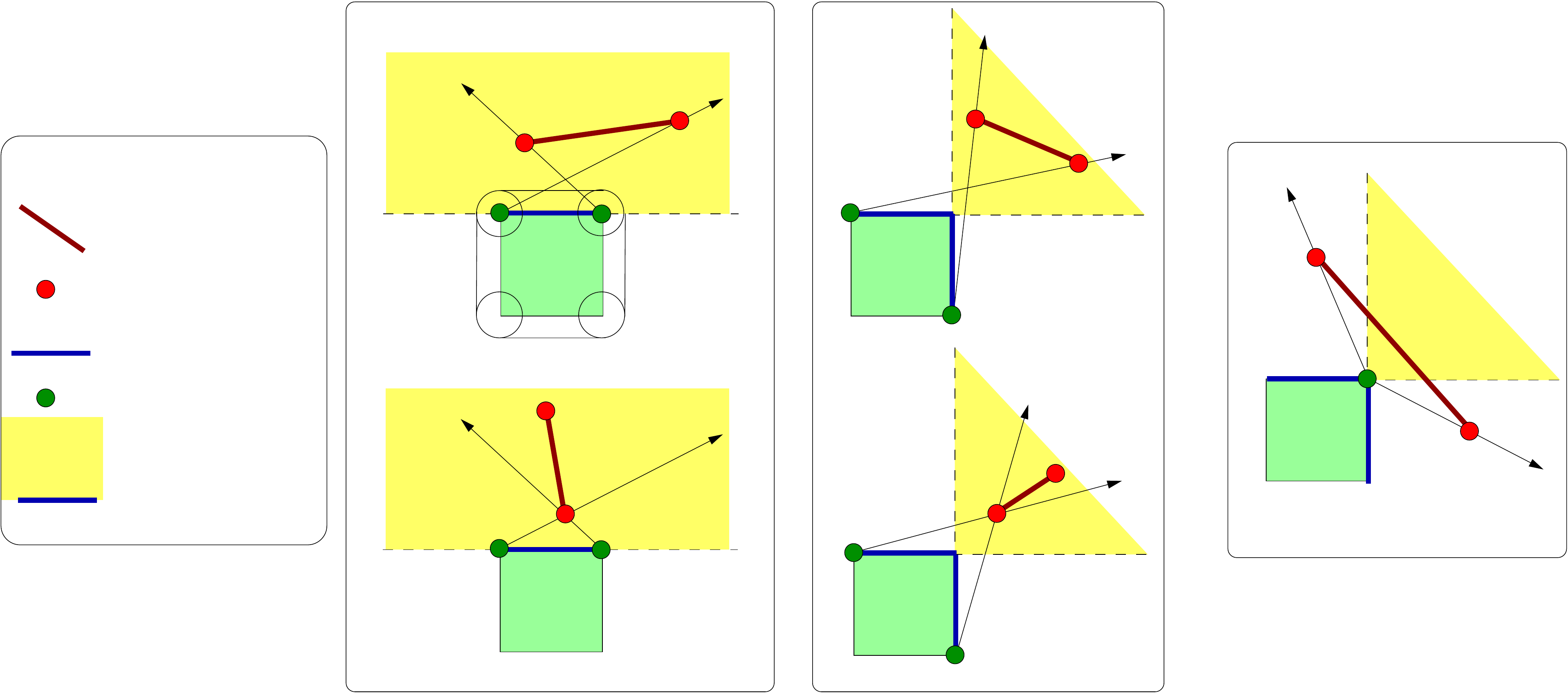}
	    	   \caption{Cases (I-III) of $\Forb_{\ell,\tau}(B^t,W)$}
	    	   \label{fig:wall-box}
	    	  \end{center}
	    	\end{figure} 	

	We first need a notation:
	if $S\ib\partial(B^t)$ is a side of the box $B^t$,
	let $H(S)$ denote the open half-space
	which is disjoint from $B^t$ and is bounded by the line through $S$.
	Then we have these three cases:

	\benum[(I)]
	\item $W\ib H(S)$ for some side $s$ of box $B^t$.
	\item $W\ib H(S)\cap H(S')$ for two adjacent sides $S,S'$ of box $B^t$.
	\item None of the above.  This implies that
	$W\ib H(S)\cup H(S')$ for two adjacent sides $S,S'$ of box $B^t$.
	\eenum

	\bthml{thick}
		$\Forblt(B^t,W)$ is the union of
		at most three thick cones.
	\ethml
	Sketch proof: 
		we try to reduce the argument to the case $\tau=0$
		which is given in
		\cite{luo-chiang-lien-yap:link:14}.
		In that case, we could write 
			$$\Forbl(B^t,W)= C_1\cup C_2\cup C_3$$
		where each $C_i$ is a thin cone or an empty set.
		In the non-empty case, the cone $C_i$
		has the form $\Forbl(S_i,T_i)$ where $S_i\ib \partial B^t, T_i\ib W$.
		The basic idea is that we now
		``transpose'' $\Forbl(S_i,T_i)$ to
		the thick version $C'_i\as \Forblt(S_i,T_i)$.
		In case $C_i$ is empty, $C'_i$ remains empty.
		Thus we would like to claim that 
			$$\Forbl(B^t,W) = C'_1\cup C'_2\cup C'_3.$$
		This is almost correct, except for one issue.
		It is possible that some $C_i$ is empty,
		and yet its transpose $C'_i$ is non empty.
		See proof in the full paper
		\cite{yap-luo-hsu:thicklink-arxiv:17}.
		In case of thin cones,
	    the $C_i$'s are non-overlapping (i.e., they may only share endpoints).
	    But for thick cone decomposition, the cones will in general overlap.

\ignore{
\section{Resolution-Exact Planning}
	The \dt{separation} of two sets $S,T\ib\RR^2$ is
	$\Sep(S,T) :=\inf\set{\|s-t\|: s\in S, t\in T}$.
	The \dt{clearance} of $\gamma\in\cspace$ relative
	to any set $\Omega\ib\RR^2$ is $\Sep(R_2[\gamma],\Omega)$, denoted
	$\Cl(\gamma,\Omega)$ or $\Cl(\gamma)$ when $\Omega$ is understood.
	A configuration $\gamma$ is \dt{$\Omega$-free} if $\Cl(\gamma,\Omega)>0$.
	Let $\cfree(\Omega)=\cfree(\Omega;R_2)$ denote the
	set of $\Omega$-free configurations of $R_2$.
	A \dt{$\Omega$-free path} (or simply ``path'') is
	a continuous function $\mu:[0,1]\to \cfree(\Omega;R_2)$;
	the \dt{clearance} of $\mu$ is $\inf\set{\Cl(\mu(t),\Omega): t\in[0,1]}$.
	The \dt{basic planning problem} for a robot $R$ is this:
	{\em given a polygonal set $\Omega\ib\RR$,
	a box $B_0\ib\cspace(R)$ and $\alpha,\beta\in B_0$,
	to find any $\Omega$-free path $\mu:[0,1]\to B_0$
	with $\mu(0)=\alpha$ and $\mu(1)=\beta$ if any such path exists;
	otherwise, return NO-PATH if no such path exists.}
	
	To avoid exact computation,
	we \cite{wang-chiang-yap:motion-planning:13,yap:sss:13} introduced
	the \dt{resolution-exact planning problem}:
	{\em given $\Omega,B_0,\alpha,\beta$ as before,
	but additionally $\vareps>0$, to find any $\Omega$-free path
	$\mu:[0,1]\to B_0$ if there exists any path with clearance $K\vareps$;
	and
	return NO-PATH if there does not exist a path with clearance
	$\vareps/K$.}
	Here, $K>1$ is any constant that depends on the
	algorithm but independent of the inputs
	$(\Omega,\alpha,\beta,B_0;\vareps)$.
	For simplicity, we do not require that
	the returned path $\mu$ have any specified clearance;
	in \cite{wang-chiang-yap:motion-planning:13}
	we require $\mu$ to have clearance $\vareps/K$.
	

\myPara{Soft-Subdivision Search}
	To construct resolution-exact planners, we use the well-known subdivision paradigm 
	\cite{brooks-perez:subdivision:83,zhu-latombe:hierarchical:91,barbehenn-hutchinson:incremental-planner:95,zhang-kim-manocha:path-non-existence:08}.
	Our subdivision framework for such planners
	is called \dt{Soft Subdivision Search} (SSS), and exploits the
	concept of soft predicates.   Appendix A reviews these concepts.
	To get a planner for any specific robot like our 2-link robot, we
	need three subroutines:
	
	\bitem
	\item Soft Predicate $\wtC$ for classifying boxes: for each box $B$,
		$\wtC(B)\in \set{\mixed,\free,\stuck}$.
		Leaf boxes that are $\mixed$ and not ``$\vareps$-small'' are
		placed in a priority queue $Q$.
	\item Search Strategy $Q.\getnext()$
		which returns the next box $B$ in $Q$ to be split.   
		There are canonical choices for $Q.\getnext()$,
		such as BFS, random choice, various A-star analogues.
	\item Split Strategy $\splittt(B)$:
		We could split $B$ into $2^d$ congruent children
		if $B$ is a $d$-dimensional box -- but this is unlikely to
		scale for $d>3$.  We may
		use global strategies that depend on state information and
		other computed parameters.
		Following \cite{luo-chiang-lien-yap:link:14}, this paper will use the T/R
		approach.
	\eitem
	
	\ignore{
	These 3 subroutines are used roughly as follows:
	there is a main while-loop.  As long as
	the queue $Q$ is non-empty, and no path has been found,
	we call $\splittt$ on $Q.\getnext()$.
	The subboxes from this split are classified using $\wtC$, and the underlying
	data structures updated.  We can detect a path as soon as the $\free$ boxes
	form a channel from $\alpha$ to $\beta$ (this is very efficiently detected
	using a Union-Find data structure).  Moreover, when the $Q$ is empty,
	we declare NO-PATH.
	}
	
	\refFig{2-link-Troom-subdivision-path}
	shows such a subdivision for our 2-link robot.  Each box $B\ib\cspace$
	is decomposed into the translation and rotational
	components: $B=B^t\times B^r$ where $B^t\ib \RR^2$ and $B^r\ib\TT$.
	Our display only shows the square $B^t$ but a user could click $B^t$
	to read off the corresponding angular ranges of $B^r$ in the panel.
	Each box $B^r$ is colored red/green/yellow/gray. 
	Red and green indicate \stuck\ and \free\ boxes.  The \mixed\
	boxes are colored yellow and gray, depending on whether its radius
	is at least $\vareps$ or not. Thus, only yellow boxes are candidates
	for splitting.
	
	
	In this paper, we will concentrate on the soft predicate $\wtC(B)$.
	The search strategy $Q.\getnext()$
	can be any of the mentioned canonical ones.
	The split strategy $\splittt(B)$ is the T/R strategy from 
	\cite{luo-chiang-lien-yap:link:14}.  The idea is that
	we split the angular range only when a box $B$ has radius $<\vareps$,
	otherwise we only split its translational subbox $B^t$.
	Moreover, the splitting of $B^r$ is not based on binary splits, but
	depends on the geometry of the obstacles.  

}%

\section{Subdivision for Thick Non-Crossing 2-Link Robot}

	A resolution-exact planner for a thin self-crossing
	2-link robot was described in
	\cite{luo-chiang-lien-yap:link:14}.
	We now extend that planner to 
	the thick non-crossing case.
	
	We will briefly review the ideas of the algorithm
	for the thin self-crossing 2-link robot.
	We begin with a box $B_0\ib\RR^2$ and it is in the subspace
	$B_0\times \TT\ib\cspace$ where our planning problem takes place.
	We are also given a polygonal obstacle set $\Omega\ib\RR^2$;
	we may decompose its boundary $\partial\Omega$
	into a disjoint union of corners (=points) and edges (=open line segments)
	which are called (boundary) \dt{features}.  
	Let $B\ib\cspace$ be a box; there is an exact classification of $B$ as
	$C(B)\in \set{\free,\stuck,\mixed}$ relative to $\Omega$.
	But we want a soft classification
	$\wtC(B)$ which is correct whenever $\wtC(B)\neq\mixed$,
	and which is equal to $C(B)$ when the width of $B$ is small enough.
	Our method of computing $\wtC(B)$ is based
	on computing a set $\phi(B)$ of features that are relevant to $B$.
	A box $B\ib \cspace$ may be
		written as a Cartesian product $B=B^t\times B^r$
		of its translational subbox $B^t\ib \RR^2$
		and rotational subbox $B^r\ib\TT$.
	In the T/R splitting method (simple version), we split $B^t$ until
	the width of $B^t$ is $\le \vareps$.  Then we do a single split
	of the rotational subbox $B^r$ into all the subboxes obtained
	by removing all the forbidden angles determined by the walls
	and corners in $\wtphi(B^t)$.  This ``rotational split'' of $B^r$
	is determined by obstacles, unlike the ``translational splits''
	of $B^t$.

\ignore{
	\myPara{Subdivision of Boxes.}
	By a \dt{box} (or $d$-box) of dimension $d\ge 1$
	we mean a set of the form $B=\prod_{i=1}^d I_i$ 
	where $d\ge 1$ and each $I_i$ is a closed interval of $\RR$ or $S^1$
	of positive length.  Such boxes are natural for doing
	subdivision in configuration spaces of the form $\RR^k \times (S^1)^{d-k}$.
	For our 2-link robots, $d=4$ and $k=2$.  
	The configuration space for a submarine or helicopter might be regarded
	as $\RR^3\times S^1$.
	
	For $i=1\dd d$, we have the notion of \dt{$i$-projection}
	and \dt{$i$-coprojection} of $d$-boxes:
	\bitem
	\item (Projection)
		$\proj[i](B) \as  \prod_{j=1, j\neq i}^d I_j$ is a $(d-1)$ dimensional box.
	\item (Co-Projection)
		$\coproj[i](B)\as I_i$ is the $i$th interval of $B$.
	\eitem
	We also define the \dt{indexed Cartesian product} $\otimes_i$
	via the identity 
		$$B = \proj[i](B)\otimes_i \coproj[i](B).$$
	Let $j=-1,0,1\dd d$.
	Two boxes $B,B'$ of dimension $d\ge 1$ are said to be \dt{$j$-adjacent}
	if $\dim(B\cap B')=j$.   Note that $B$ and $B'$ are $(-1)$-adjacent means
	they are disjoint.  When $i=d-1$, we simply say the boxes are 
	\dt{adjacent}, denoted $B :: B'$;  when $i=d$, we say
	they are \dt{overlapping}, denoted $B\circ B'$.
	The following is immediate:
	
	\blem
	Let $B, B'$ be boxes of dimension $d\ge 1$.
	\bitem
	\item If $d=1$ then $B::B'$ iff $|B\cap B'|\in\set{1,2}$.
	\item If $d>1$ then $B::B'$ iff $(\exists i=1\dd d)$ such that
		$$\proj[i](B)\circ \proj[i](B) \quad\land\quad
			\coproj[i](B)::\coproj_i(B').$$
	\eitem
	\elem
}%

\myPara{Boxes for Non-Crossing Robot.}
	Our basic idea for representing
	boxes in the non-crossing configuration space $\cspacedel$
	is to write it as a pair $(B,\XT)$  where $\XT\in \set{\LT,\GT}$, and
	$B\ib \cspace$.
	The pair $(B,\XT)$ represents the set $B\cap (\RR^2 \times \TT_\XT)$
	(with the identification $\TT_\LT=\TT_<$ and $\TT_\GT=\TT_>$).
	It is convenient to call $(B,\XT)$ an \dt{$X$-box}
	since they are no longer ``boxes'' in the usual sense.
	
	%
	%
	An angular interval $\Theta\ib S^1$ that\footnote{
		Wrapping intervals are either equal to $S^1$
			or has the form $[s,t]$ where $2\pi>s>t>0$.
	}
	contains a open neighborhood of $0=2\pi$ is said to be \dt{wrapping}.
	Also, call $B^r=\Theta_1\times \Theta_2$
	wrapping if either $\Theta_1$ or $\Theta_2$ is wrapping.  
	Given any $B^r$, we can decompose the
	set $B^r\cap (\TT\setminus\Delta(\kappa))$ into the union of
	two subsets $B^r_\LT$ and $B^r_\GT$, where
	$B^r_\XT$ denote the set $B^r \cap \TT_\XT$.
	In case $B^r$ is non-wrapping, this decomposition has the nice
	property that each subset $B^r_\XT$ is connected.
	For this reason, we prefer to work with non-wrapping boxes.
	Initially, the box $B^r=\TT$ is wrapping.  The initial split of $\TT$
	should be done in such a way that the children are all non-wrapping:
	the ``natural'' (quadtree-like) way to split $\TT$ into four congruent
	children has\footnote{
		This is not a vacuous remark -- the quadtree-like split is
		determined by the choice of a ``center'' for splitting.
		To ensure non-wrapping children, this center is necessarily
		$(0,0)$ or equivalently $(2\pi,2\pi)$.
		Furthermore, our T/R splitting method (to be introduced) does
		not follow the conventional quadtree-like subdivision at all.  
	} this property.  Thereafter, subsequent splitting
	of these non-wrapping boxes will remain non-wrapping.
	
	Of course, $B^r_\XT$ might be empty, and this is easily
	checked: say $\Theta_i=[s_i,t_i]$ ($i=1,2$).
	Then $B^r_<$ is empty iff $t_2\le s_1$.
	and $B^r_>$ is empty iff $s_2\ge t_1$.
	Moreover, these two conditions are mutually exclusive.
	
	We now modify the algorithm of \cite{luo-chiang-lien-yap:link:14} as follows:
	as long as we are just splitting boxes in the translational dimensions,
	there is no difference.  When we decide to split the rotational dimensions,
	we use the T/R splitting method of \cite{luo-chiang-lien-yap:link:14}, but
	each child is further split
	into two $X$-boxes annotated by $\LT$ or $\GT$ (they are filtered out if 
	empty).  We build the connectivity graph $G$ (see Appendix A) with these $X$-boxes
	as nodes.  This ensures that we only find non-crossing paths.  Our algorithm
	inherits resolution-exactness from the original self-crossing algorithm.

	The predicate $\isBoxEmpty(B^r, \kappa, \XT)$
	which returns true iff $(B^r_\XT)\cap (\TT\setminus\Delta(\kappa))$ is empty
	is useful in implementation.  It has a simple expression when restricted to
	non-wrapping translational box $B^r$:
	
	\blem
	\ \\ Let $B^r=[a,b]\times [a',b']$ be a non-wrapping box.
	\\(a) $\isBoxEmpty(B^r, \kappa, \LT)=\true$ iff $\kappa\ge b'-a$ or
	$2\pi-\kappa\le a'-b$.
	\\(b) $\isBoxEmpty(B^r, \kappa, \GT)=\true$ iff $\kappa\ge b-a'$ or
	$2\pi-\kappa \le a-b'$.
	\elem

\ignore{
  SHOULD DO THIS ONCE, for case $\kappa\ge 0$!!!
Next consider adjacency relationships of rotational boxes:

\blem
Let $B_i\as [s_i,t_i]\times [s'_i,t'_i]\ib\TT$ ($i=1,2$)
be non-wrapping boxes.  Then $(B_1,\XT_1) :: (B_2,\XT_2)$ iff
one of the following conditions hold:
\ \\ (a) $\XT_1\neq \XT_2$.  Wlog, $\XT_1=\LT$ and $\XT_2=\GT$.
	Then either $(t_1, s_2)=(2\pi,0)$ or $(t'_1, s'_2)=(2\pi,0)$.
\\ (b) $\XT_1=\XT_2$.  Wlog, $\XT_i=\LT$.
\elem
}%

\section{Implementation and Experiments}

We implemented our thick non-crossing 2-link 
planner in {\tt C++} and OpenGL on the Qt platform.
A preliminary heuristic version
appeared \cite{luo-chiang-lien-yap:link:14,luo:thesis}.
Our code, data and experiments are distributed\footnote{ 
{http://cs.nyu.edu/exact/core/download/.}
}
with our open source \corelib.
%
To evaluate our planner, we compare it with
several sampling algorithms
in the open source OMPL \cite{OMPL}.
Besides, based on a referee's suggestion, we also implemented
the 2-link (crossing and non-crossing)
versions of Toggle PRM and Lazy Toggle PRM (in lieu of publicly
available code).  We benefited greatly from the advice of Prof.~Denny
in our best effort implementation.
%
%
The machine we use is a MacBook Pro with 
2.5 GHz Intel Core i7 and 16GB DDR3-1600 MHz RAM.
%

Tables~1 and 2 summarize the results of two groups of experiments,
which we call \dt{Narrow Passages} and \dt{Easy Passages}.
Each row in the tables represents an experiment,
each column represents a planner.  
There are 8 planners: 3 versions of SSS, 3 versions of PRM and 2 versions of
RRT.  Only Table 1 is listed in the paper;
Table 2 (as well as other experimental results) are relegated to an appendix. 
\ifFullPaper{For visualization, we extract 4 bar charts 
from the two tables in \refFig{figure_table1_2}
}{
We extract two bar charts 
from Table~1 in \refFig{figure_table1_2}} for visualization to
show the average times and success rates
of the planners in Narrow Passages.  
Timing is in milliseconds on a Log10 scale.  E.g., the bar chart
\refFig{figure_table1_2}(a)
for Narrow Passages shows that the average time for the SSS(I) planner
in the T-Room experiment is about $2.9$.  This represents
$10^{2.9} \simeq 800$ milliseconds; indeed 
the actual value is $815.9$ as seen in Table~1.
On this bar chart, each unit represents a power of 10.
E.g., if one bar is at least $k$ units shorter
than another bar, we say the former is ``$k$ orders of magnitude'' faster
(i.e., at least $10^k$ times faster).
\dt{Conclusion:}
(1) SSS is at least an order of magnitude faster than
each sampling method, and (2) success rates of RRT-connect
and Toggle PRM are usually (but not always) best among sampling methods,
but both are inferior to SSS.

\ifFullPaper{
		\begin{figure}[ht]
			\makebox[\textwidth][c]{\includegraphics[width=1.5\textwidth]{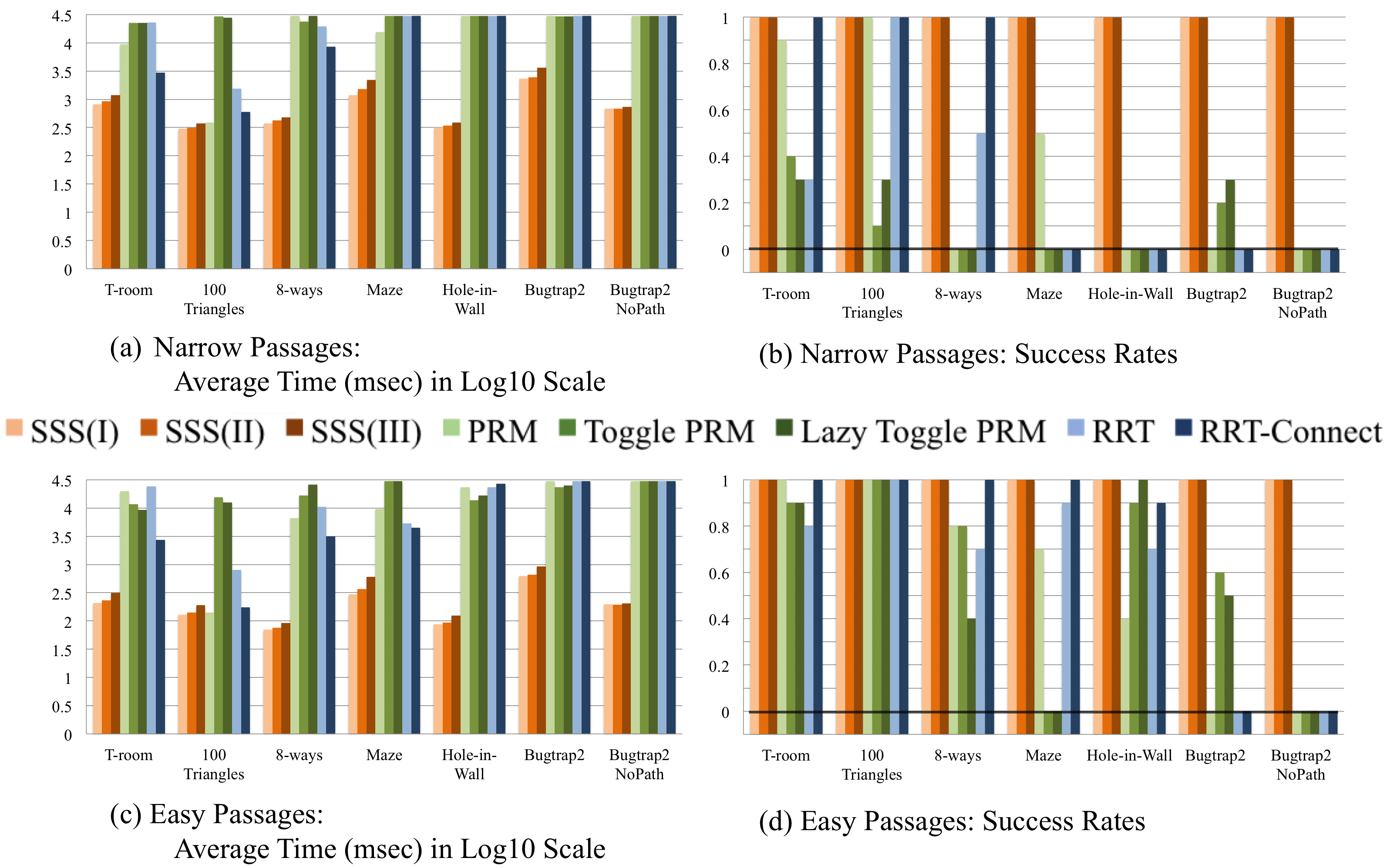}}
		\caption{ Bar Charts of Average Times and Success Rates }
		\label{fig:figure_table1_2}
		\end{figure}
}{
		\vspace*{-0.4cm}
		\begin{figure}[ht]
		\makebox[\textwidth][c]{
		    \includegraphics[width=1.2\textwidth]{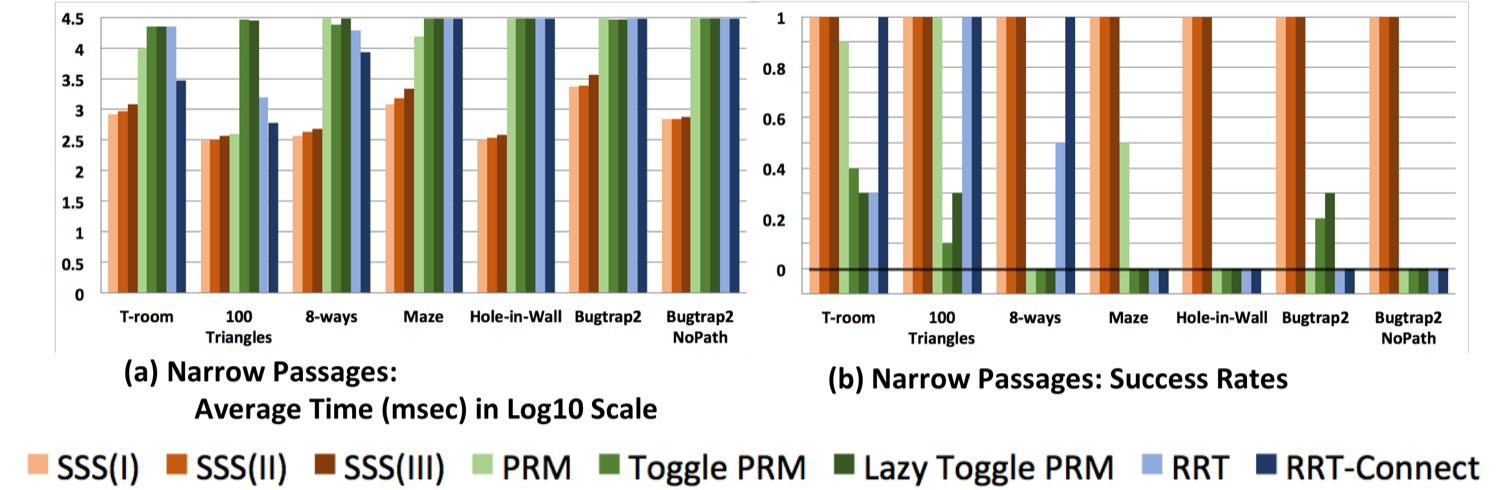}}
		\caption{ Bar Charts of Average Times and Success Rates }
		\label{fig:figure_table1_2}
		\end{figure}
		\vspace*{-0.8cm}
}

Two general remarks are in order.
First, as in our previous work, we implemented
several search strategies in SSS.  But
for simplicity, we only use the Greedy Best First (GBF)
strategy in all the SSS experiments; GBF is typically our best strategy.
Next, OMPL does not natively support articulated
robots such as $R_2$.  So in the experiments of Tables 1 and 2,
we artificially set $\ell_2=0$ for all the sampling algorithms
(so that they are effectively one-link thick robots).
This is a suboptimal experimental scenario, but it only reinforces
any exhibited superiority of our SSS methods.
In the SSS versions, we set $\ell_2=0$ for SSS(I) but
SSS(II) and SSS(III) represent (resp.)
crossing and non-crossing 2-link robots where $\ell_2$
has the values shown in the column 4 header of Table~1.
\ifFullPaper{
    As expected, SSS(I) is faster than SSS(II) which is 
    faster than SSS(III).  But even SSS(III) is faster
    than all the sampling methods by an order of magnitude
    ($> 10\times$).
}{}
\ifFullPaper{
	\begin{figure}[ht]
	  	\centering
		\subfloat[{\small Maze}]{
	    \includegraphics[width=0.24\textwidth]{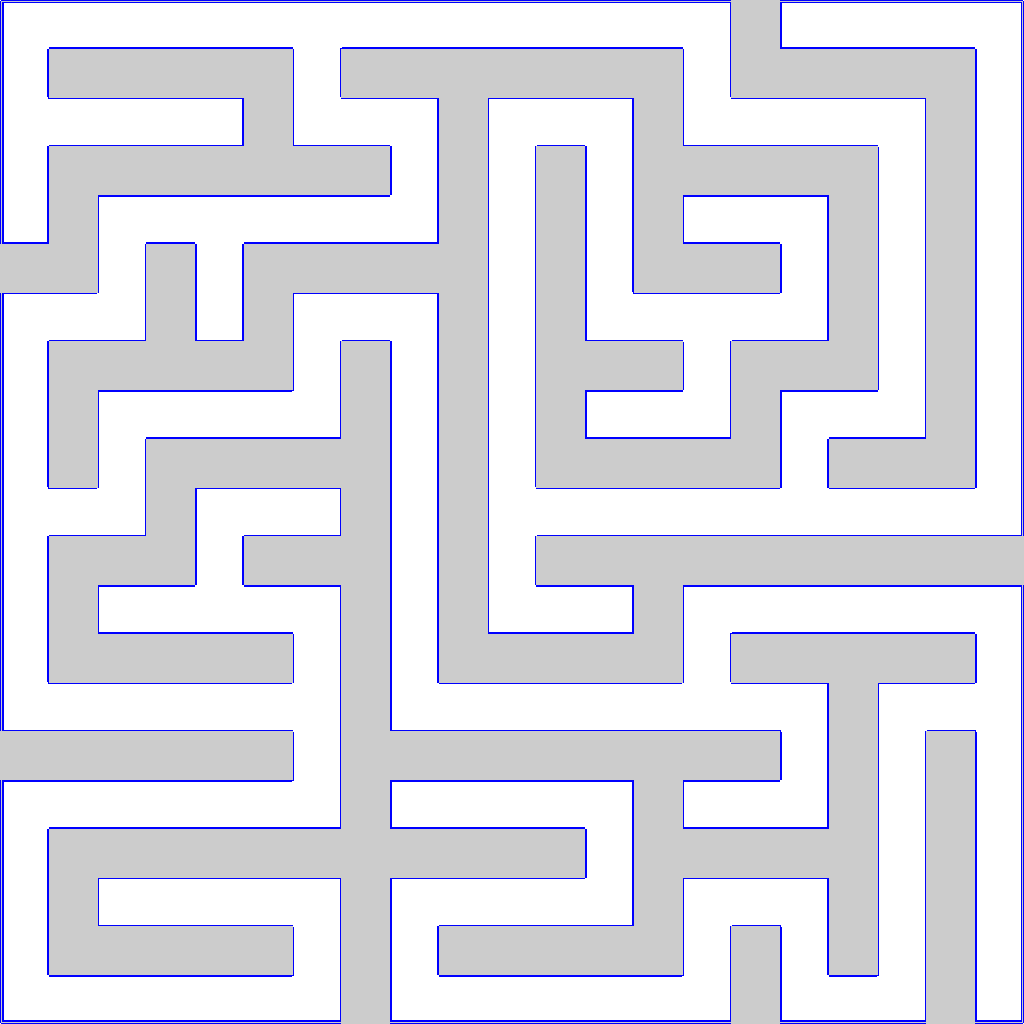}}
        \subfloat[{\small Hole-in-Wall}]{
	    \includegraphics[width=0.24\textwidth]{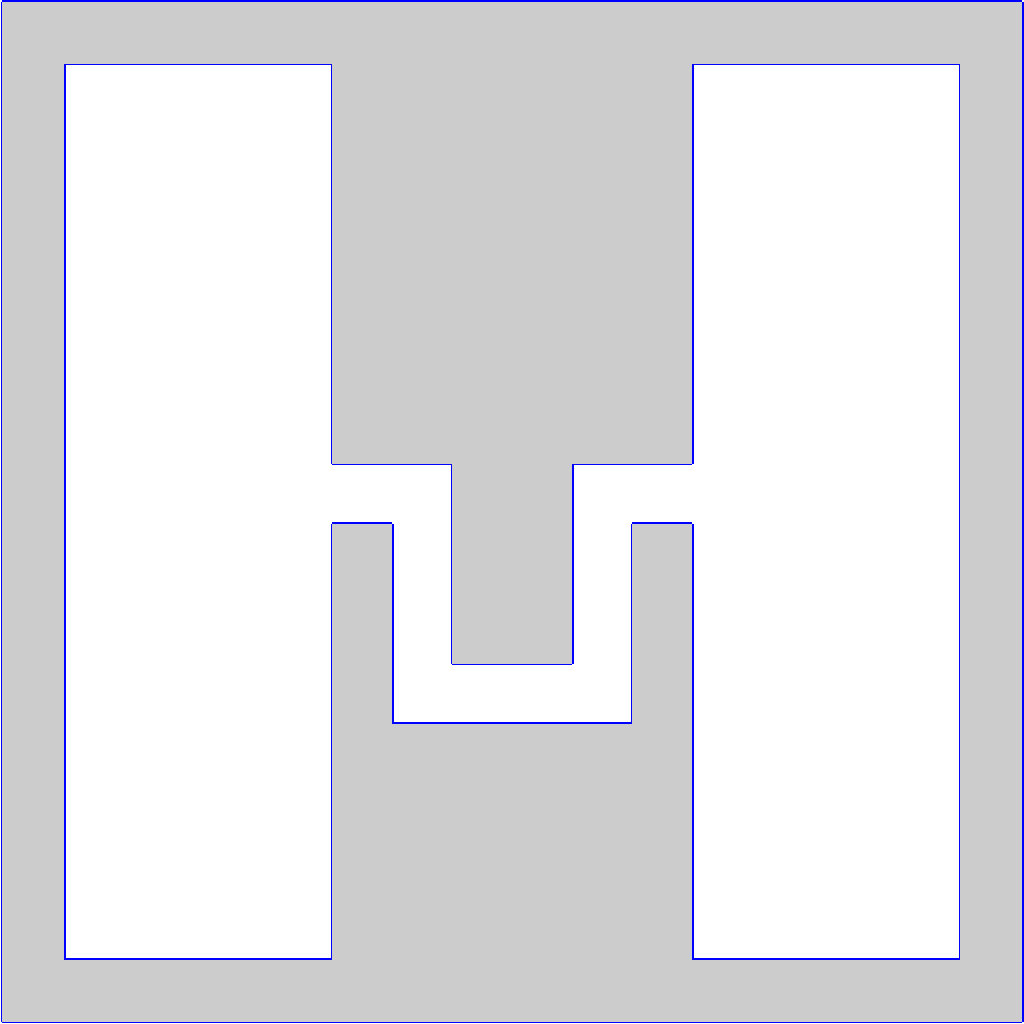}}
        \subfloat[{\small 8-Way Corridor}]{
	    \includegraphics[width=0.24\textwidth]{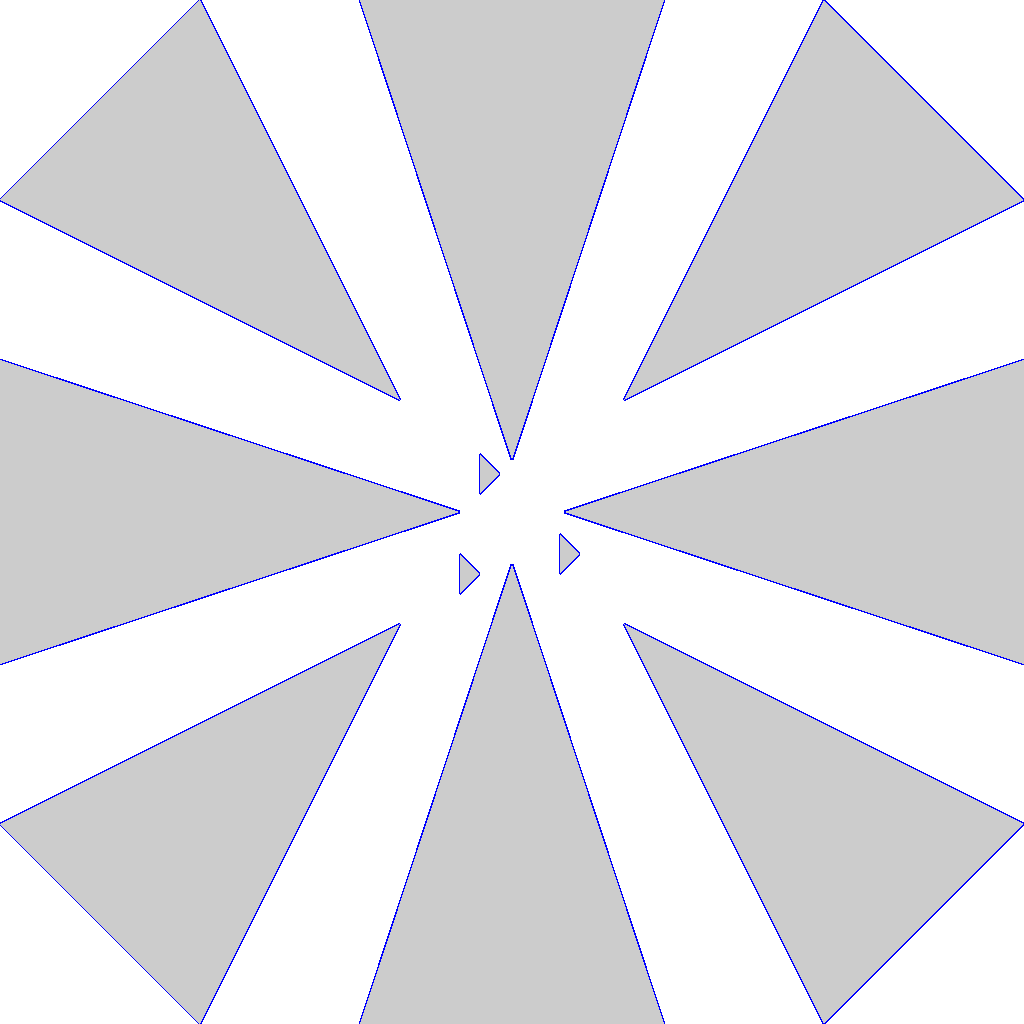}}
        \subfloat[{\small Bugtrap2}]{
	    \includegraphics[width=0.24\textwidth]{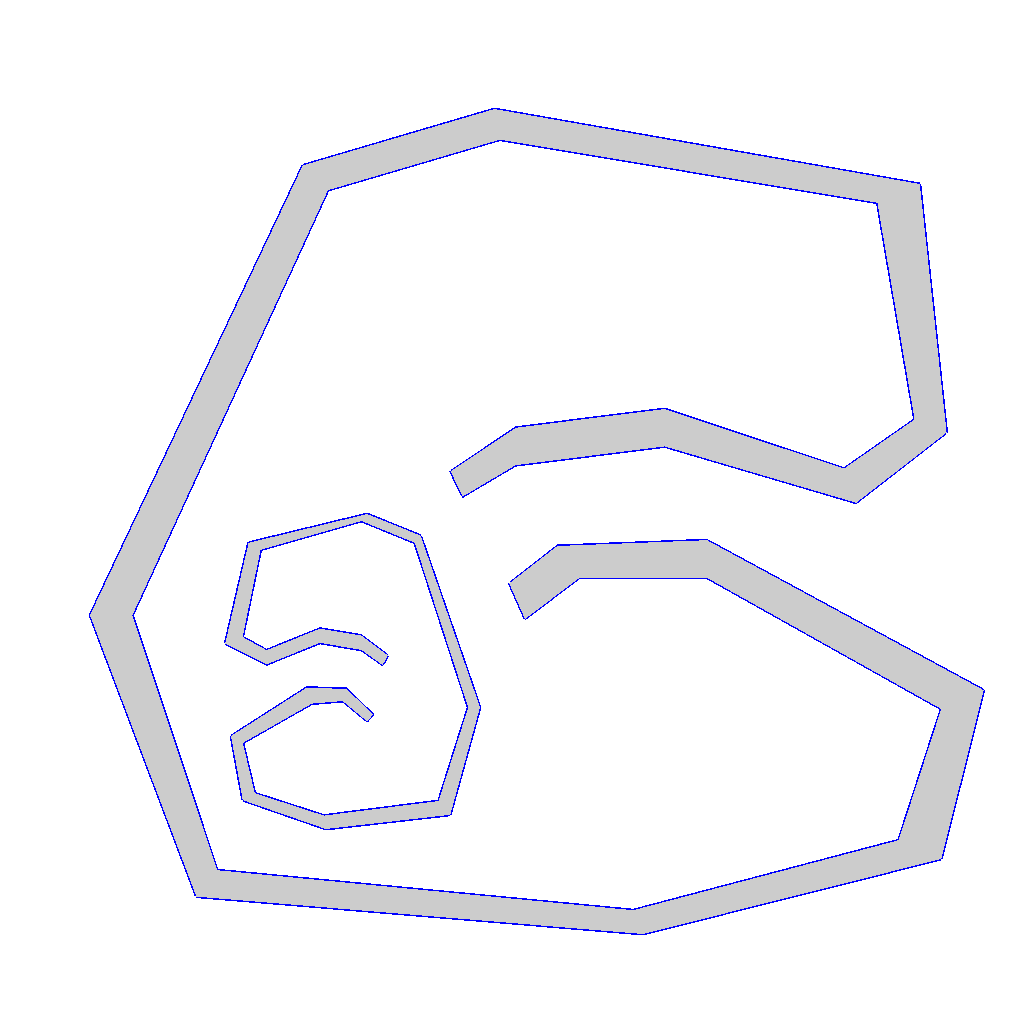}}
		\caption{More environments in our experiments}
		\label{fig:env}
	\end{figure}
} {
	\vspace*{-0.8cm}
	\begin{figure}[ht]
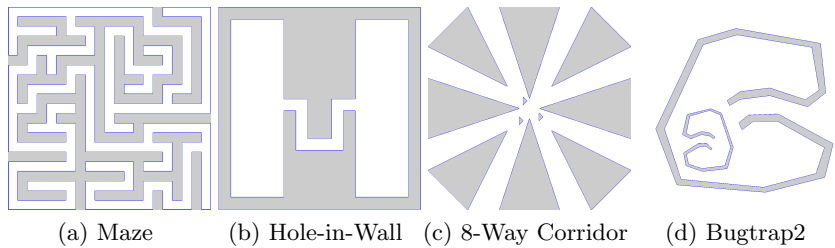

	  	\centering
		\subfloat[{\small Maze}]{
	    \includegraphics[width=0.22\textwidth]{env_maze.png}}
        \subfloat[{\small Hole-in-Wall}]{
	    \includegraphics[width=0.22\textwidth]{env_narrow.png}}
        \subfloat[{\small 8-Way Corridor}]{
	    \includegraphics[width=0.22\textwidth]{env_8-ways.png}}
        \subfloat[{\small Bugtrap2}]{
	    \includegraphics[width=0.22\textwidth]{env_bugtrap2.png}}
		\caption{More environments in our experiments}
		\label{fig:env}
	\end{figure}
	\vspace*{-0.8cm}
}

{\bf Reading the Tables:}  Each experiment (i.e., row)
corresponds to a fixed
environment, robot parameters, initial and goal configurations.
\refFig{env} depicts these environments (save for
the T-Room and 100 triangles from the introduction).
We name the experiments after the environment.
E.g., column 1 for the Maze experiment, tells us that
$\ell_1=16, \tau=10$.  The last two experiments use the
``double bugtrap'' environment, but the robot parameters for one
of them ensures NO-PATH.
For each experiment, we perform $40$ runs of the following planners:
SSS(I-III), PRM, RRT, RRT-connect (all from OMPL), 
Toggle PRM and Lazy Toggle PRM (our implementation).
Each planner produces 4 statistics:
	\bcen
	Average Time /
	Best Time /
	Standard Deviation /
	Success Rate,
	\ecen
abbreviated as \dt{Avg}/\dt{Best}/\dt{STD}/\dt{Success}, respectively.
Success Rate is the fraction of the $40$ runs for which
the planner finds a path (assuming there is one) out of $40$ runs.
But if there is no path, our SSS planner will always
discover this, so its \dt{Success} is $1$; simultaneously,
the sampling methods will time out and hence their \dt{Success} is 0.
All timing is in milliseconds (msec).
Column $2$ contains the \dt{Record Statistics}, i.e.,
the row optimum for these 4 statistics.
E.g., the Record Statistics for the T-Room experiment is 
\dt{815.9/743.6/21.9/1}.
This tells us the row optimum 
for \dt{Avg} is $815.9$ ms,
for \dt{Best} is $743.6$ ms,
for \dt{STD} is $21.9$ ms,
and for \dt{Success} is $1$.
``Optimum'' for the first three (resp., last) statistics
means minimum (resp., maximum) value.
The four optimal values may be achieved by different planners.
In the rest part of the Table, we have one column for each Planner,
showing the ratio of the planner's statistics 
relative to the Record Statistics.
The best performance is always indicated by the ratio of 1.  
E.g., for T-Room experiment,
the row maximum for \dt{Success} is 1, and it is achieved by 
all SSS planners and RRT-Connect.
The row minimum for \dt{Avg}, \dt{Best} and \dt{STD}
are achieved by SSS(I), RRT-Connect and SSS(II), resp.
We regard the achievement of row optimum for \dt{Success} and \dt{Avg}
(in that order) to be the main indicator of superiority.
Table~1 (and Table~2) show that our planner is consistently
superior to sampling planners.
    E.g., Table 1 shows that in the T-Room experiment,
    the record average time of $815.9$ milliseconds is achieved
    by SSS(I).  But SSS(III) is only 1.5 times slower,
    and the best sampling method is RRT-Connect which
    is 3.6 times slower.
    For the Maze experiment, again SSS(I) achieves the record average
    time of 1193.2 milliseconds, SSS(III) is 1.8 times slower,
    but none of the sampling methods succeeded in 40 trials.
	\begin{table*}[htbp]
	\makebox[\textwidth][c]{\includegraphics[width=1.3\textwidth]{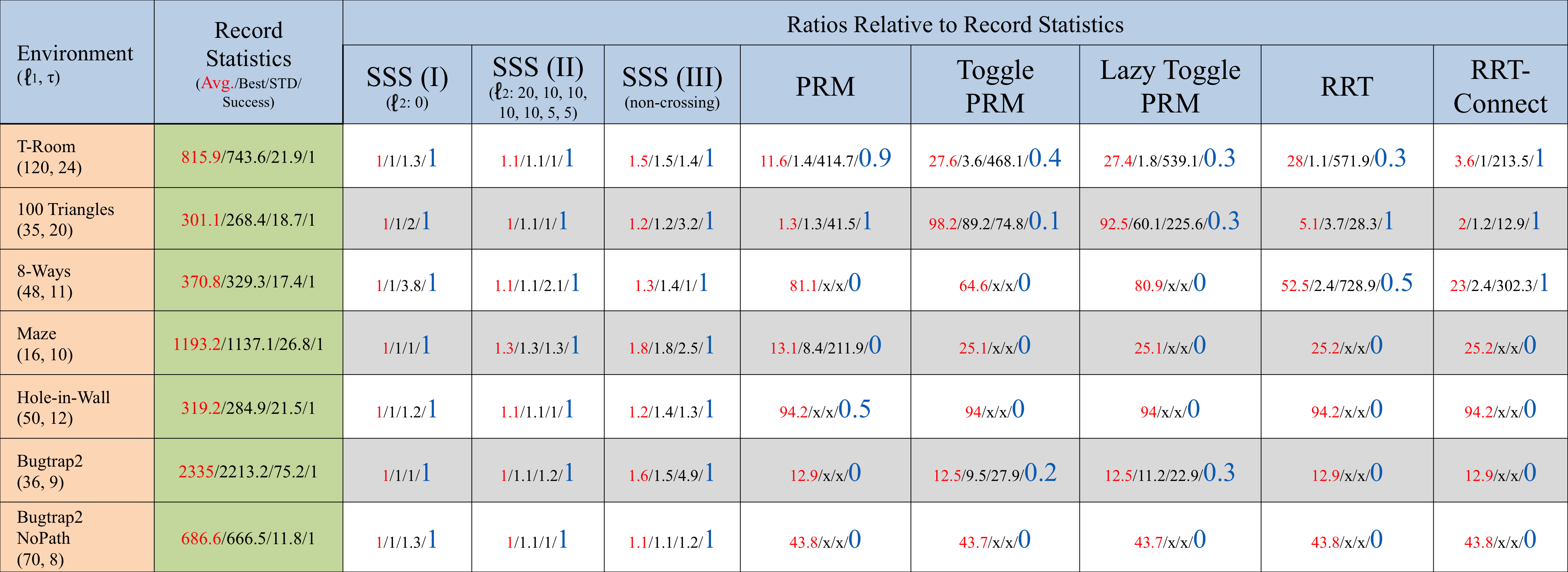}}
	\caption{ Narrow Passages }
	\end{table*}

\ifFullPaper{
    Table~1 also compares the performance of the three
    versions of our algorithm: SSS(I) (single link
    robot), SSS(II) (crossing 2-links) and SSS(III) (non-crossing 2-links).
    For both SSS(II) and SSS(III), the lengths
    $\ell_2$ of the second link in the 7 experiments are, respectively,
    $20, 10, 10, 10, 10, 5$, and $5$.  We can see that
    SSS(II) is never more than $30\%$ slower than SSS(I);
    and
    SSS(III) is never more than $80\%$ slower than SSS(I).
    Moreover, SSS(III) is always faster than all the sampling methods.
    Note that because the sampling methods are all timed out
    after 30 seconds, their average time is artificially low
    (since they are capped at 30 seconds)
    whenever they do not achieve $100\%$ success.
}{}

%

In the Appendix, we have Table~2 which is basically the same as Table~1
except that we decrease the thickness $\tau$ in order
to improve the success rates of the sampling methods.
Our planner needs an $\vareps$
parameter, which is set to $1$ in Table~1 and $2$ in Table~2
(this is reasonable in view of narrow passage demands).
Sampling methods have many more tuning parameters;
but we choose the defaults in OMPL because we saw no
systematic improvements in our partial
attempts to enhance their performance. In Toggle PRM, 
we use small $k$ for $k$-nearest neighbors in the obstacle graph and
the similar default $k$ as in OMPL in the free graph.
We set the time-out to be 30 seconds; with this cutoff,
it takes $18$ hours to produce the data of Table~1.
In the appendix, we mention some experiments to allow
the sampling algorithm up to 0.5 hour.

\ifFullPaper{
Tables~3 and 4~in Appendix B provides the detailed comparison of the 2-link  
versions of Toggle PRM and Lazy Toggle PRM against SSS,
on both Narrow and Easy Passages.  
For visualization, \refFigs{figure_table3} 
contains 4 bar charts extracted from Tables~3:
it shows the average time and success rates of 
SSS, Toggle PRM and Lazy Toggle PRM planners in Narrow Passages. 
Version (II) refers to crossing 2-link robots.
and Version (III) refers to non-crossing 2-link robots.

	\begin{figure}[h]
		\makebox[\textwidth][c]{\includegraphics[width=1.4\textwidth]{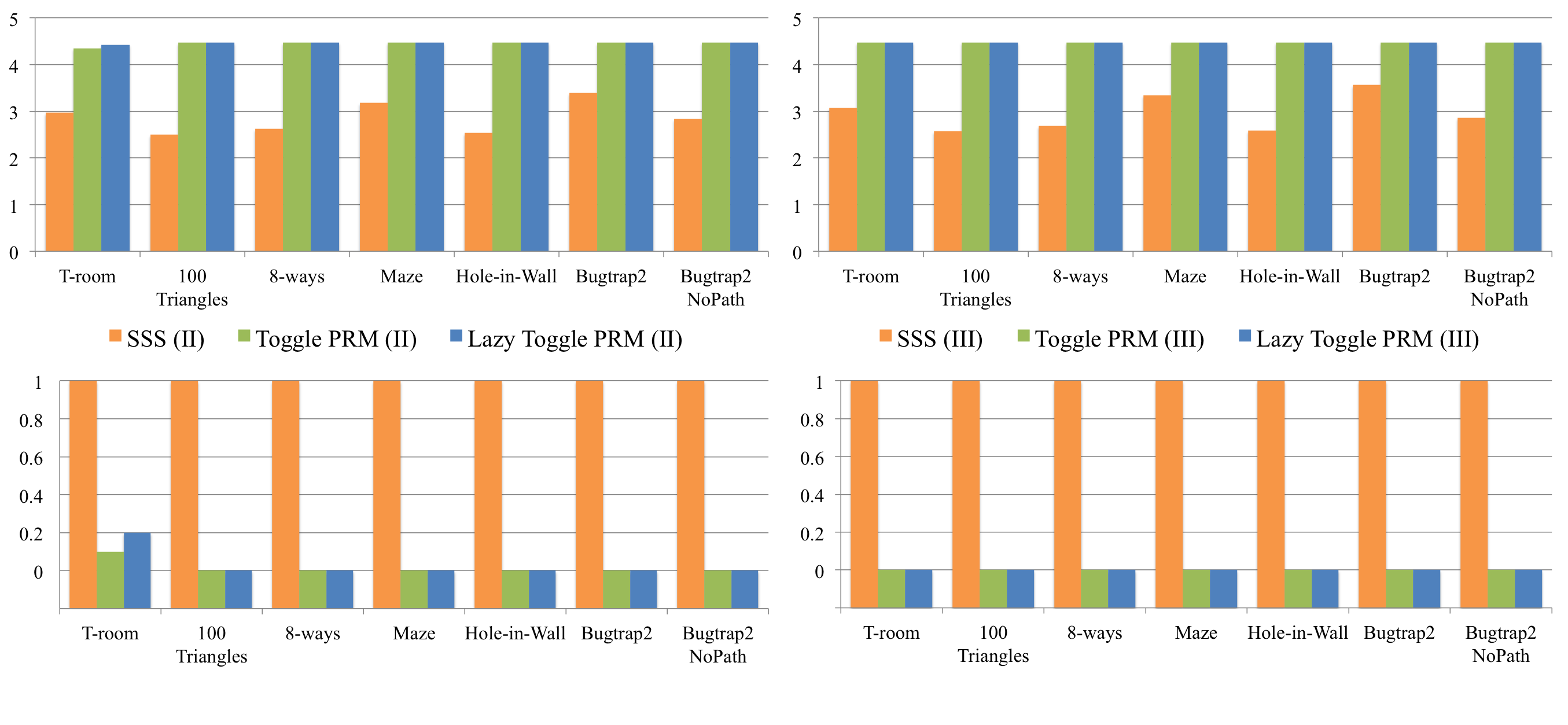}}
	\caption{Bar Charts of Average Times (Top Row) and Success Rates (Bottom Row): 
	Comparing of SSS, Toggle PRM and Lazy Toggle PRM on Narrow Passages 
	with crossing and non-crossing 2-link robots}
	\label{fig:figure_table3}
	\end{figure}

Similarly, in \refFig{figure_table4}, it shows the average times and success rates of 
SSS, Toggle PRM and Lazy Toggle PRM planners in Easy Passages. 
Conclusion: (1) as expected, from \refFig{figure_table3} and \refFig{figure_table4},
it only shows a greater lag of performance both in 
crossing and non-crossing 2-link robots setting. (2) especially, in the non-crossing case,
SSS significantly outperforms Toggle PRM and Lazy Toggle PRM.

	\begin{figure}[h]
		\makebox[\textwidth][c]{\includegraphics[width=1.4\textwidth]{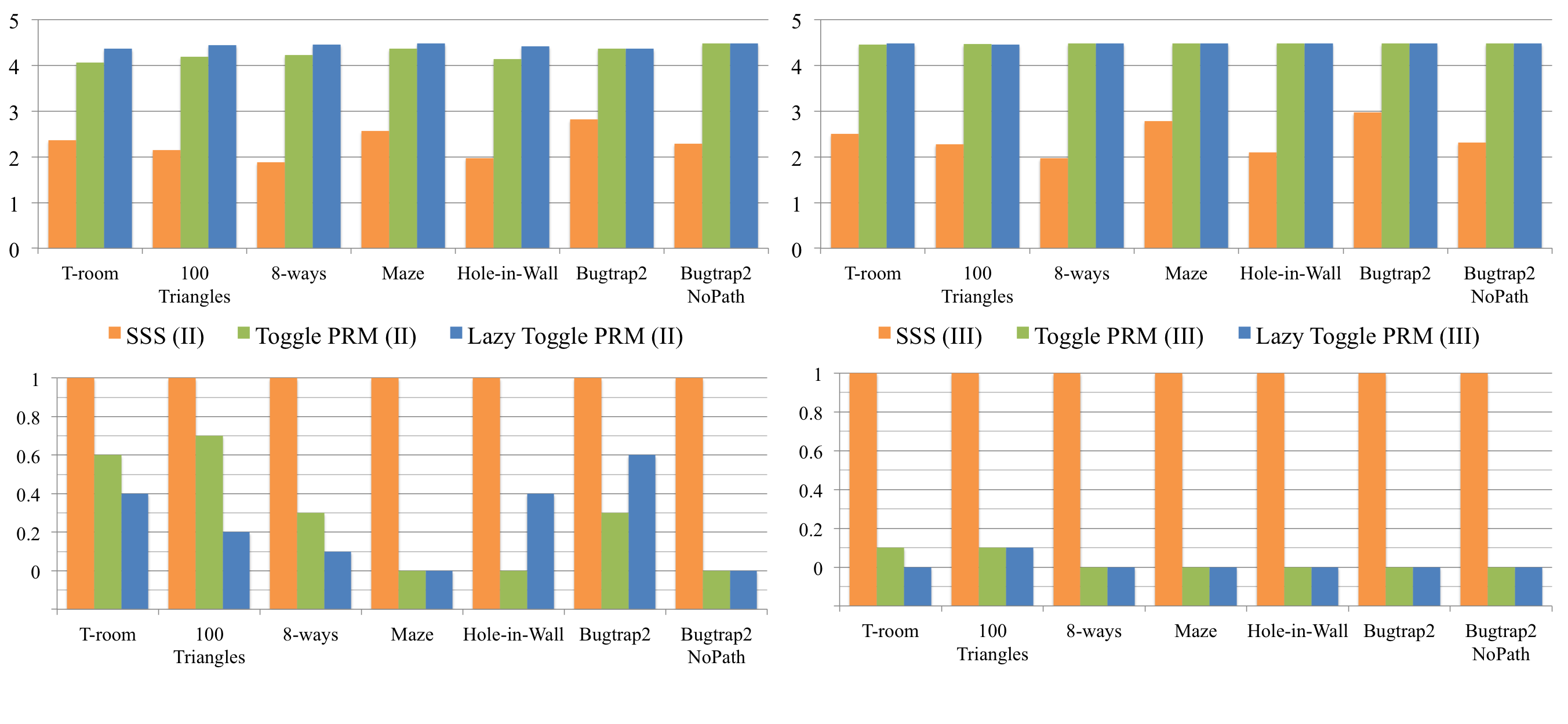}}
	\caption{Bar Charts of Average Times (Top Row) and Success Rates (Bottom Row): 
	Comparing of SSS, Toggle PRM and Lazy Toggle PRM on Easy Passages 
	with crossing and non-crossing 2-link robots}
	\label{fig:figure_table4}
	\end{figure}
}{}

\ignore{
We use various obstacle sets (most are described in
\cite{luo-chiang-lien-yap:link:14}).
Each run is a row in the Table, and has
these parameters ($\ell_1, \ell_2,S,\vareps, \kappa$) where
$\ell_i$ is the length of the $i$-th link,
$S\in \{B,D,G\}$ indicates\footnote{
	A random strategy is available, but it is never competitive.
} the search strategy
($B=\mbox{\it Breadth First Search}$ $(BFS)$,
$D=\mbox{\it Distance $+$ Size}$,
$G=\mbox{\it Greedy Best First}$ $(GBF)$).
The last parameter $\kappa\in [0,180)$ is the bandwidth of $\Delta$ in degrees.
When we run the self-crossing planner the $\kappa$ parameter is ignored.
The parameters for each run are encoded in a Makefile,
but the user may modify these parameters through the GUI interface
(see \refFig{2-link-maze-setup}).
}

\ignore{
Table~2 illustrates the ``narrow passage'' problems for
the sampling approaches: in each experiment, we choose a larger
thickness $\tau$ for which there is a path;
in other words, if we increase the thickness to $\tau +1$,
SSS will report no path.   In fact, we include a case
(the last row, Double Bug Trap-B) for which there is no path.
Again, we have three version of our algorithm with fixed $\vareps=1$.
The fair comparison would be SSS(I) against the Sampling methods,
because this runs the single link robot, but we outperform sampling
methods even with SSS(II) and SSS(III).

Instead of using default 1 second time-out, we set time-out to be 30 seconds.
We see that BIT* (resp., EST) has the best (worst)
success rate among the Sampling methods.
BIT* was able to find paths for all,
with the exception of Narrow Passage and Double Bug Trap-A.
EST could not find paths for any, with the exception of T-Room.
}%

\ignore{
	Table~I shows that the running time of the non-crossing planner is comparable
	to that of the self-crossing planner in all the examples
	(with the exception of the T-room or eg13).  Their percentage change
	is between $-44.8\%$ to $11.4\%$.
	That is because, although non-crossing planner has some
	overhead, it also filters out useless splittings earlier for the dead ends.
	The exceptional case (T-room) is explained by the fact that the
	non-crossing  planner
	must use a much longer circuitous path.
	
	Table~II shows the sensitivity of finding a path to the link length $\ell_2$,
	and to the bandwidth $\kappa$, as $\vareps$ decreases.
}%

\ignore{%
		\begin{figure}[htbp]
		\begin{center}
	        \includegraphics[scale=0.32]{2-link-maze-setup}
		\caption{ GUI Interface for Maze Example }
		\label{fig:2-link-maze-setup}
		\end{center}
		\vspace*{-2mm}
		\end{figure}

\NOignore{
	\begin{center}
	\begin{table*}[htbp]
		\centering
		\vspace*{1.5mm}
		\scriptsize
		 \parbox{0.58\textwidth}{%
		\begin{tabular}{|l|| l|| l| l || l | l |l|}\hline
			Obstacle & Configuration   &  \multicolumn{2}{c||}{Self-Crossing}
				& \multicolumn{2}{c|}{Non-Crossing} & Performance \\
			(input)  & $(\ell_1, \ell_2 , S, \epsilon, \kappa)$ & time (ms) & boxes                & time (ms) & boxes & Improvement\\
			\hline\hline
			eg2b & (88, 98, D, 2, 79)  & 1740.1      & 104663      & {\bf 1591.9} & {\bf 71123}& 8.51\%\\
			(8-way corridor)& (88, 98, D, 2, 80)  & - & -              & No Path & No Path& -\\
			    & (88, 98, D, 2, 30)  & - & -                & 1687.1 & 101287& 3.0\%\\
			    & (88, 98, D, 2, 5)  & - & -                & 1963.2 & 129394& -12.8\%\\
			\hline
			
			eg5 & (55, 50, G, 4, 95)  & {\bf 541.2}      & {\bf 22243}   & 542.3 & 27560 & -0.2\%\\
			(Double Bugtrap) & (55, 50, G, 4, 100)  & -      & -                &No Path & No Path & -\\
			    & (55, 50, G, 4, 50)  & -      & -                & 613.1       & 32157 & -13.3\%\\
			    & (55, 50, G, 4, 10)  & -      & -                & 730.3       & 42994 & -34.9\%\\
			\hline
			eg8 & (30, 25, G, 2, 7) & {\bf 31.5}     & {\bf 2215}    & 45.6       & 5214 & -44.8\%\\
			(Hsu et al.~\cite{hsu-latombe-kurniawati:foundations:06})
			     & (30, 25, G, 2, 8) & -      & -      & No Path  & No Path & -\\
			     & (30, 25, G, 2, 3) & -     & -         & 37.3       & 3514& -18.4\%\\
			\hline
			eg12 & (30, 33, D, 4, 146) & 314.4    & 19953                  & {\bf 283.3}     & {\bf 15167} & 9.9\%\\
			(Maze) & (30, 33, D, 4, 147) & -    & -                  & No Path      & No Path & -\\
			   & (30, 33, D, 4, 40) & -    & -                  & 360.9      & 22908 & -14.8\%\\
			   & (30, 33, D, 4, 10) & -    & -                  & 410.2      & 32783 & -30.5\%\\

			\hline
			eg13 & (94, 85, D, 4, 10) & {\bf 3.1}    & {\bf 616}                  & 98.9     & 12212 & -3090\%\\
			(T-Room)& (94, 85, D, 4, 11) & -    & -                  & No Path      & No Path & -\\
			   & (94, 85, D, 4, 5) & -    & -                  & 94.9      & 12068 & -2961\%\\
			\hline
			eg300 & (40, 30, G, 4, 127) &  305.7 & 8794          & {\bf 270.8}      & {\bf 7314} & 11.4\%\\
			(300 Triangles) & (40, 30, G, 4, 128) &  - & -         & No Path      & No Path & -\\
			      & (40, 30, G, 4, 40) &  - & -         & 353.6      & 11284 & -15.7\%\\
			      & (40, 30, G, 4, 10) &  - & -         & 348.4      & 12113 & -14.0\%\\
			\hline
		\end{tabular}
		}
		\qquad \qquad
	\caption{Comparison between Self-Crossing and Non-Crossing.}
	%
	\end{table*}
	\end{center}
}

\begin{table*}[htbp]
	\centering
	\parbox{0.58\textwidth}{%
	\vspace*{1.5mm}
	{\scriptsize
	\begin{tabular}{|l|| l|| l| l || l | l |l|}\hline
			Obstacle & Configuration   &  \multicolumn{2}{c||}{Self-Crossing}
				& \multicolumn{2}{c|}{Non-Crossing} & Performance \\
			(input)  & $(\ell_1, \ell_2 , S, \epsilon, \kappa)$ & time (ms) & boxes                & time (ms) & boxes & Improvement\\
			\hline\hline
			eg2a &  (85, 80, G, 8, 10)  & No Path      & No Path                & No Path & No Path & -\\ \cline{2-7}
			(8-way corridor) &(85, 80, G, {\bf 4}, 10)  & 459.0      & 33199                & {\bf 400.9} & {\bf 31390} & 12.7\%\\
			& (85, {\bf 92}, G, 4, 10)  & No Path      & No Path                & No Path & No Path & -\\ \cline{2-7}
			& (85, 92, G, {\bf 2}, 10)  & {\bf 2271.8 }     & {\bf 153425}                & 2402.3 & 192916 & -5.7\%\\
			& (85, {\bf 99}, G, 2, 10) & No Path      & No Path                & No Path & No Path & -\\ \cline{2-7}
			& (85, 99, G, {\bf 1}, 10) & {\bf 5887.4}      & {\bf 385814}                & 6190.0 & 448119 & -5.1\%\\
			& (85, {\bf 100}, G, 1, 10) & No Path      & No Path                & No Path & No Path & -\\
			\hline
				eg13 & (94, 85, D, 8, 10) & No Path      & No Path                & No Path & No Path & -\\ \cline{2-7}
				(T-Room)& (94, 85, D, {\bf 4}, 10) & {\bf 3.1}    & {\bf 616}                  & 98.9     & 12212 & -3090\%\\
				& (94, 85, D, 4, {\bf 13}) & -    & -                  & No Path      & No Path & -\\ \cline{2-7}
				& (94, 85, D, {\bf 2}, 13) & {\bf 6.2}    & {\bf 1187}                  & 417.7      & 47292 & -6637\%\\
				& (94, 85, D, 2, {\bf 14})& -    & -           & No Path      & No Path & -\\ \cline{2-7}
				& (94, 85, D, {\bf 1}, 14)& {\bf 9.8}    & {\bf 1974}           & 1553.7      & 184559 & -15754\%\\
				& (94, 85, D, 1, {\bf 15})& -    & -           & No Path      & No Path & -\\
				\hline
	\end{tabular}
	} }
	%
\caption{(a) Eg2a shows the sensitivity to length $\ell_2$ as $\vareps$ changes.
	(b) Eg13 shows the sensitivity to bandwidth $\kappa$ as $\vareps$ changes.}
	%
\end{table*}
}%

\section{Conclusion and Limitations}

We have introduced a novel and efficient planner
for thick non-crossing 2-link robots.
Our work contributes to the development of practical
and theoretically sound subdivision planners
\cite{wang-chiang-yap:motion-planning:13,yap:sss:13}.
	It is reasonable to expect a tradeoff between
	the stronger guarantees of our resolution-exact approach
	versus  
	a faster running time for sampling approaches.
	But our experiments suggest no such
	tradeoffs at all: {\em SSS is consistently superior to sampling}.
	We ought to say that
	although we have been unable to improve the sampling
	planners by tuning their parameters, it is possible
	that sampling experts might do a better job than us.
	But to actually exceed our performance, their improvement would
	have to be dramatic.  SSS has no tuning, except in the
	choice of a search strategy (Greedy Best First),
	and a value for $\vareps$.  But we do not view $\vareps$ 
	as a tuning parameter, but a value determined by the
	needs of the application.

Conventional wisdom maintains that subdivision will not scale to higher
DOF's, and our current experience have been limited to at most 4DOF.
We interpret this wisdom as telling us that
new subdivision techniques (such as the T/R splitting idea)
are needed to make higher DOF's robots perform in real-time.
This is a worthy challenge for SSS which
we plan to take up.

\ignore{
For those concerned with exactness and numerical guarantees,
Our resolution-exact theory suffers no lack here.
However, our current implementation is based on machine arithmetic,
and so the range for which such an implementation is sound would
depend on some error analysis.  Despite this caveat,
it seems fair to compare our implementation with the current software
which offer no better guarantees.
It is relatively easy to ensure arbitrary precision
using bigFloat (in \corelib) and ``lax comparison'' as described 
in \cite{wang-chiang-yap:motion-planning:13}.
}%


\bibliographystyle{abbrv}
\bibliography{test,st,yap,exact,geo,alge,math,com,rob,cad,algo,visual,gis,quantum,mesh,tnt,fluid} 

\newpage
\section*{APPENDIX A: Elements of Soft Subdivision Search}
	We review the the notion of soft predicates and how
	it is used in the SSS Framework.
	See \cite{wang-chiang-yap:motion-planning:13,yap:sss:13,luo-chiang-lien-yap:link:14}
	for more details.
	
	\myPara{Soft Predicates.}
	The concept of a ``soft predicate'' is relative to some exact
	predicate.  
	Define the exact predicate $C:\cspace\to \set{0,+1,-1}$ where 
	$C(x)=0/+1/-1$ (resp.) if configuration $x$ is semi-free/free/stuck.
	The semi-free configurations are those on the boundary of $\cfree$.
	Call $+1$ and $-1$ the \dt{definite values}, and $0$ the \dt{indefinite
	value}.
	Extend the definition to any set $B\ib\cspace$:
	for a definite value $v$, define $C(B)=v$ iff $C(x)=v$ for all $x$. 
	Otherwise, $C(B)=0$.  
	Let $\intbox(\cspace)$ denote the set of $d$-dimensional boxes in $\cspace$.
	A predicate $\wtC:\intbox(\cspace)\to\set{0,+1,-1}$ is a \dt{soft version of $C$} if
	it is conservative and convergent.  \dt{Conservative} means that
	if $\wtC(B)$ is a definite value, then $\wtC(B)=C(B)$.   \dt{Convergent} means that
	if for any sequence $(B_1,B_2,\ldots)$ of boxes, if $B_i\to p\in\cspace$ 
	as $i\to\infty$, then $\wtC(B_i)=C(p)$ for $i$ large enough. 
	To achieve resolution-exact algorithms, we must ensure
	$\wtC$ converges quickly in this sense:
	say $\wtC$ is \dt{effective} if there is a constant
	$\sigma>1$ such if $C(B)$ is definite, then $\wtC(B/\sigma)$ is definite.
	
	\myPara{The Soft Subdivision Search Framework.}
	An SSS algorithm maintains
	a subdivision tree $\TTT=\TTT(B_0)$ rooted at a given box $B_0$.
	Each tree node is a subbox of $B_0$.
	We assume a procedure $\splittt(B)$ that subdivides a given leaf box $B$
	into a bounded number of subboxes which becomes the children of $B$
	in $\TTT$.  Thus $B$ is ``expanded'' and no longer a leaf.
	For example, $\splittt(B)$ might create
	$2^d$ congruent subboxes as children.  Initially $\TTT$ has just
	the root $B_0$; we grow $\TTT$ by repeatedly expanding its leaves.
	The set of leaves of $\TTT$ at any moment constitute a subdivision of $B_0$.
	Each node $B\in \TTT$ is classified using a soft predicate $\wtC$ 
	as $\wtC(B)\in\set{\mixed,\free,\stuck/}=\set{0,+1,-1}$. 
	Only \mixed\ leaves with radius $\ge\vareps$ are candidates for expansion.
	We need to maintain three auxiliary data structures:
	\bitem
	\item
	A priority queue $Q$ which contains all candidate boxes.
	Let $Q.\getnext()$ remove the box of highest priority from $Q$.
	The tree $\TTT$ grows by splitting $Q.\getnext()$.
	\item
	A \dt{connectivity graph} $G$ whose nodes are the \free\ leaves in $\TTT$,
	and whose edges connect pairs of boxes that are adjacent,
	i.e., that share a $(d-1)$-face.
	\item
	A Union-Find data structure for connected components of $G$.
	After each $\splittt(B)$, we update $G$ and insert new
	\free\ boxes into the Union-Find data structure and perform
	unions of new pairs of adjacent \free\ boxes.
	\eitem
	
	Let $Box_{\TTT}(\alpha)$
	denote the leaf box containing $\alpha$ (similarly for $Box_{\TTT}(\alpha)$).  
	The SSS Algorithm has three WHILE-loops.  The first WHILE-loop will
	keep splitting $Box_{\TTT}(\alpha)$ until it becomes \free, or declare
	NO-PATH when $Box_{\TTT}(\alpha)$ has radius less than $\vareps$.
	The second WHILE-loop does the same for $Box_{\TTT}(\beta)$.
	The third WHILE-loop is the main one: it will keep splitting
	$Q.\getnext()$ until a path is detected or $Q$ is empty.
	If $Q$ is empty, it returns NO-PATH.  Paths are detected when
	the Union-Find data structure tells us that
	$Box_{\TTT}(\alpha)$ and $Box_{\TTT}(\beta)$ are in the same connected component.
	It is then easy to construct a path.  Thus we get:
	
	{\small
	\progb{
		{\textsc SSS Framework:}\\
		\INPUt:  Configurations $\alpha,\beta$, tolerance $\vareps>0$,
			box $B_0\in \cspace$.\\
		\> Initialize a subdivision tree $\TTT$ with root $B_0$.\\
		\> Initialize $Q, G$ and union-find data structure.\\
		1.\> While ($Box_{\TTT}(\alpha) \neq \free$)\\
		\>\>	If radius of $Box_{\TTT}(\alpha))$ is $< \vareps$, Return(NO-PATH)\\
		\>\>	Else \splittt($Box_{\TTT}(\alpha))$\\
		2.\> While ($Box_{\TTT}(\beta) \neq \free$)\\
		\>\>	If radius of $Box_{\TTT}(\beta))$ is $< \vareps$, Return(NO-PATH)\\
		\>\>	Else \splittt($Box_{\TTT}(\beta))$\\
		\> \commenT{MAIN LOOP:}\\
		3.\> While ($Find(Box_{\TTT}(\alpha))\neq Find(Box_{\TTT}(\beta))$)\\
		\>\>	If $Q_{\TTT}$ is empty, Return(NO-PATH)\\
		\>\>	$B\ass Q_{\TTT}.\getnext()$\\
		\>\>	$\splittt(B)$\\
		4.\> Generate and return a path from $\alpha$ to $\beta$ using $G$.
	}}
	
	The correctness of our algorithm does not depend on how the priority of $Q$
	is designed.  See \cite{yap:sss:13} for the correctness of this framework under
	fairly general conditions.

\section*{APPENDIX B: Detail of Experimental Results}
Table~2 follows the same statistic notations as Table~1. From Table~2, we have concluded
that Toggle PRM and RRT-Connect usually have best success rates among sampling methods.
Moreover, Toggle PRM and Lazy Toggle PRM have a chance to find the path in a short time.
But, in the double bug trap scenario, RRT-Connect cannot find a path in $40$ runs; 
in the maze scenario, both Toggle PRM and Lazy Toggle PRM cannot find a path in $40$ runs. 
With the relaxed thickness $\tau$, SSS(I) still outperforms other sampling methods.
Furthermore, even SSS(III) with non-crossing constraint is almost consistently 
superior to the sampling based planners with two exceptions: in the 100 random triangles scenario, 
PRM is about $1.36$ times faster than SSS(III) and 
RRT-Connect is approximately $1.07$ times faster than SSS(III).
But these comparisons may be
misleading: SSS(III) plans for a 2-link robot while the sampling planners
are for 1-link robots (as there were no articulated robots
native to OMPL).  RRT-Connect uses bidirectional search while
SSS(III) is unidirectional.  We are planning
to introduce bidirectional search into SSS.

\begin{center}
\begin{table*}[htbp]
	\makebox[\textwidth][c]{\includegraphics[width=1.3\textwidth]{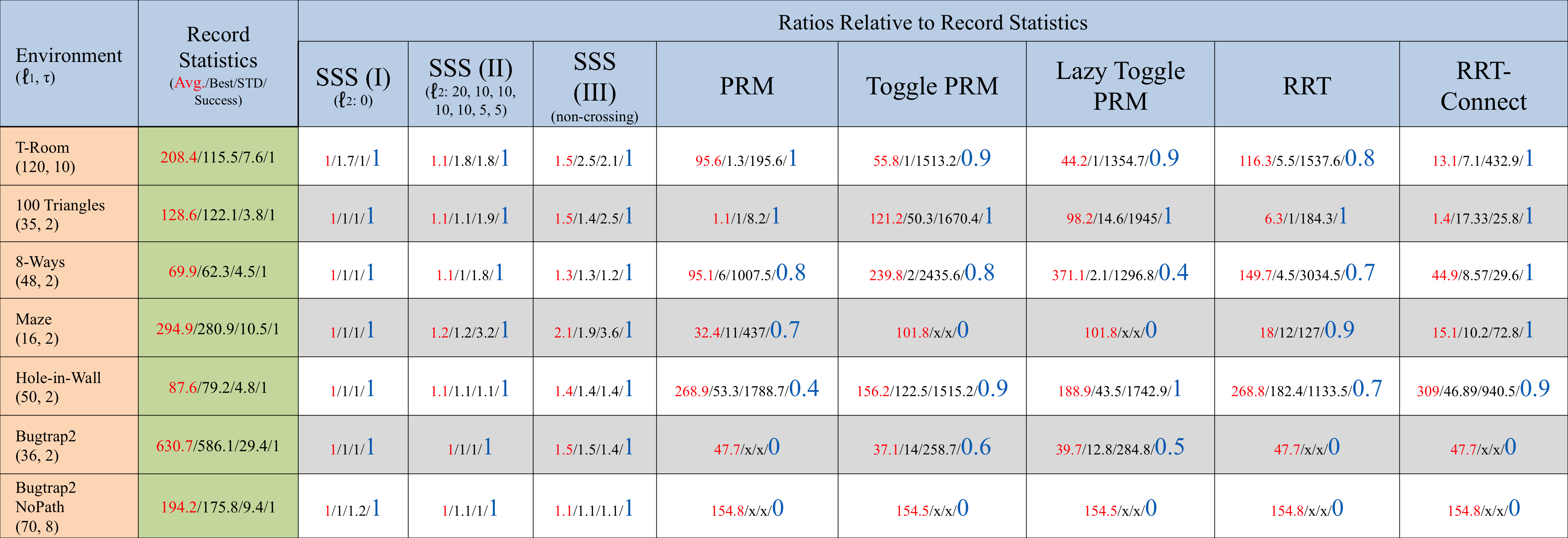}}
\caption{ Easy Passages }
\label{fig:table2}
\end{table*}
\end{center}

Before comparing SSS with Toggle PRM and Lazy Toggle PRM in Tables~3 and 4, 
we first show a sample output of Toggle PRM in the Hole-in-Wall scenario (\refFig{toggle}). 
The figure shows free graph in green and obstacle graph in red. 
Free graph means the nodes are in free configuration space and 
its edges indicate that there exists a path in free space connecting endpoints. 
Obstacle graph means that the nodes are not in free configuration space and 
its edges indicate that there exists a path in the obstacle configuration space.

\begin{figure}[ht]
	\makebox[\textwidth][c]{\includegraphics[width=1.2\textwidth]{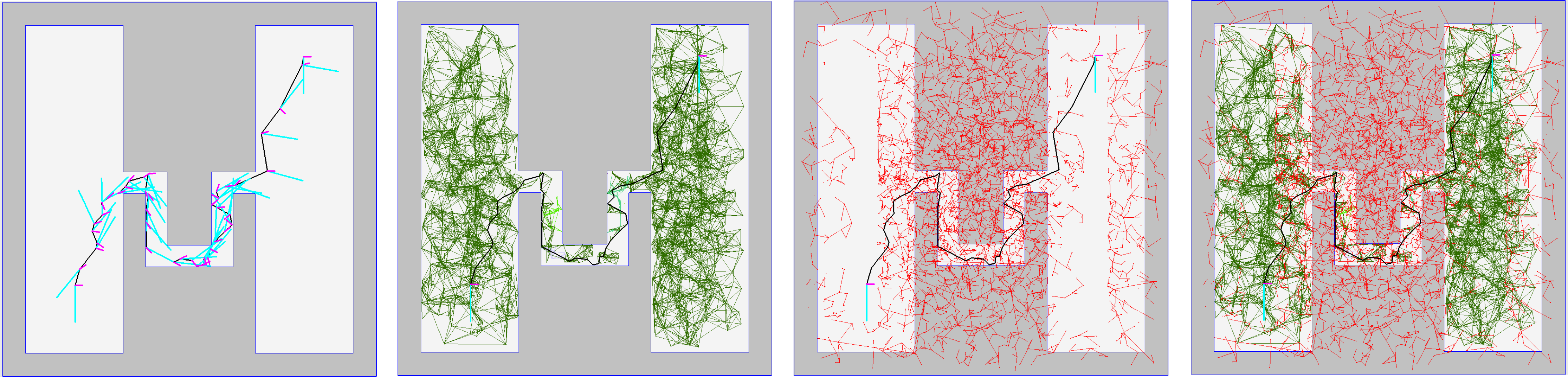}}
	\caption{ One Example Result of Toggle PRM:
	subfigures from left to right are its trace, free graph, obstacle graph, and mixed graph result. }
	\label{fig:toggle}
\end{figure}

Table~3 are experiments in Narrow Passages. While, Table~4 are in Easy Passages.
For our SSS planner, $\vareps$ is equal to 1 for Table~3 and $\vareps$ is equal to 2 for Table~4. 
In these Tables, we time out a planner in $30$ seconds. In Table~3, 
most sampling planners reach the time limitation.
On the other hand, SSS planner always produces a path or no-path within $4$ seconds. 
The worst average time of SSS is $3.646$ seconds 
when solving double bug trap environment with non-crossing 2-link robots. 
The best average time of SSS is $0.312$ seconds when solving 100 random triangles environment with crossing
2-link robots.

\begin{table*}[htbp]
	\makebox[\textwidth][c]{\includegraphics[width=1.3\textwidth]{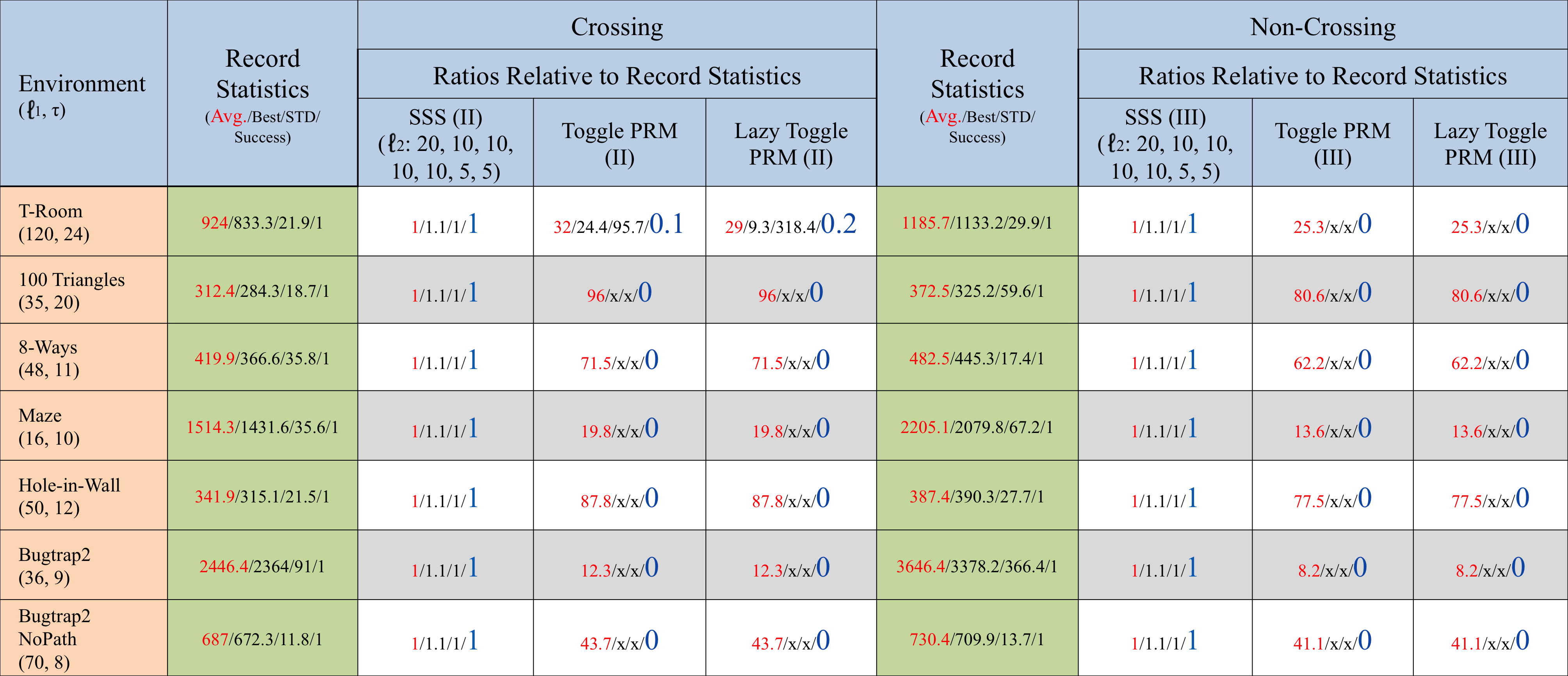}}
\caption{ Narrow Passages Result with Crossing and Non-crossing 2-link Robots }
\label{fig:table3}
\end{table*}

In Table~4, we decrease thickness $\tau$ to make it easier for Toggle PRM and Lazy Toggle PRM to find a path. 
In some situations, Toggle PRM and Lazy Toggle PRM may find the path 
almost as quickly as SSS. For example, in their best cases for the T-Room and 8-Way Corridor scenarios, 
Toggle PRM is about $1.6$ and $1.4$ times slower than SSS and Lazy Toggle PRM is approximately as fast
as SSS. But, on average, SSS outperforms Toggle PRM and Lazy Toggle PRM by at least an order 
of magnitude. In conclusion, based on Tables 3 and 4, our SSS planners is superior to Toggle
PRM and Lazy Toggle PRM with crossing and non-crossing 2-link robots.

\begin{table*}[htbp]
	\makebox[\textwidth][c]{\includegraphics[width=1.3\textwidth]{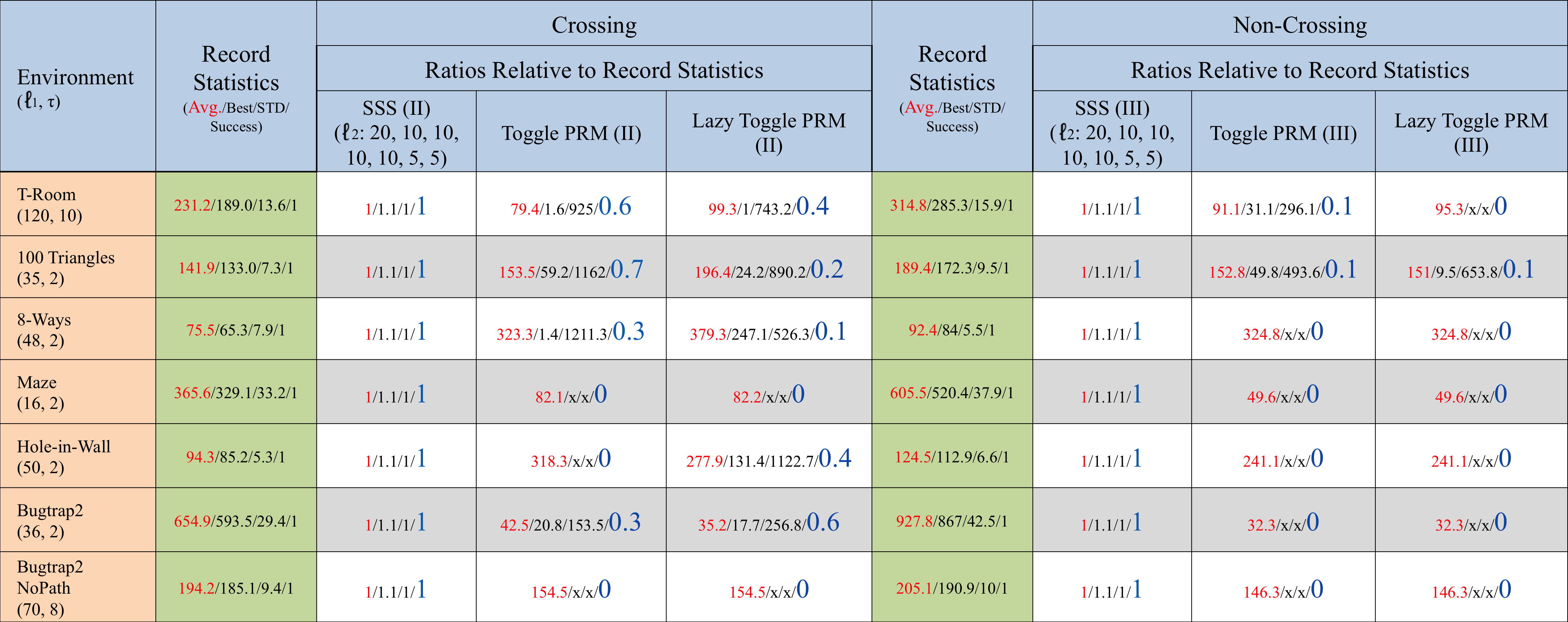}}
\caption{ Easy Passages Result with Crossing and Non-crossing 2-link Robots }
\label{fig:table4}
\end{table*}

We have conducted an exhaustive experiment using $30$ minutes timeout for Tables~3 and 4.
These results are recorded in Tables~3\textsuperscript{*} and 4\textsuperscript{*}.
However, we only do a single run of each environment.
By comparing it with our SSS planner and previous Toggle PRM results, giving 
more time does improve the success rates. For example, in the Narrow Passages
scenario (Table~3\textsuperscript{*}), by taking more time, 
Toggle PRM can find a path in the T-Room, 100 Random Triangles,
and Double Bug Trap environment in a single run with crossing 2-link robots. 
From Table~4\textsuperscript{*}, 
not only in the T-Room, 8 Way Corridors, Hole-in-Wall, and Double Bug Trap environment
with crossing 2-link robots but also in the T-Room, 100 Random Triangles, and 8 Way Corridors with 
non-crossing 2-link robots, it can make obvious improvement. In another point of view,
if sampling methods like Toggle PRM are likely to find a path, there is a trade-off between 
timeout limitation and success rates. On the other hand, in our SSS planner, there is no such trade-off.

\begin{table*}[htbp]
	\makebox[\textwidth][c]{\includegraphics[width=1.3\textwidth]{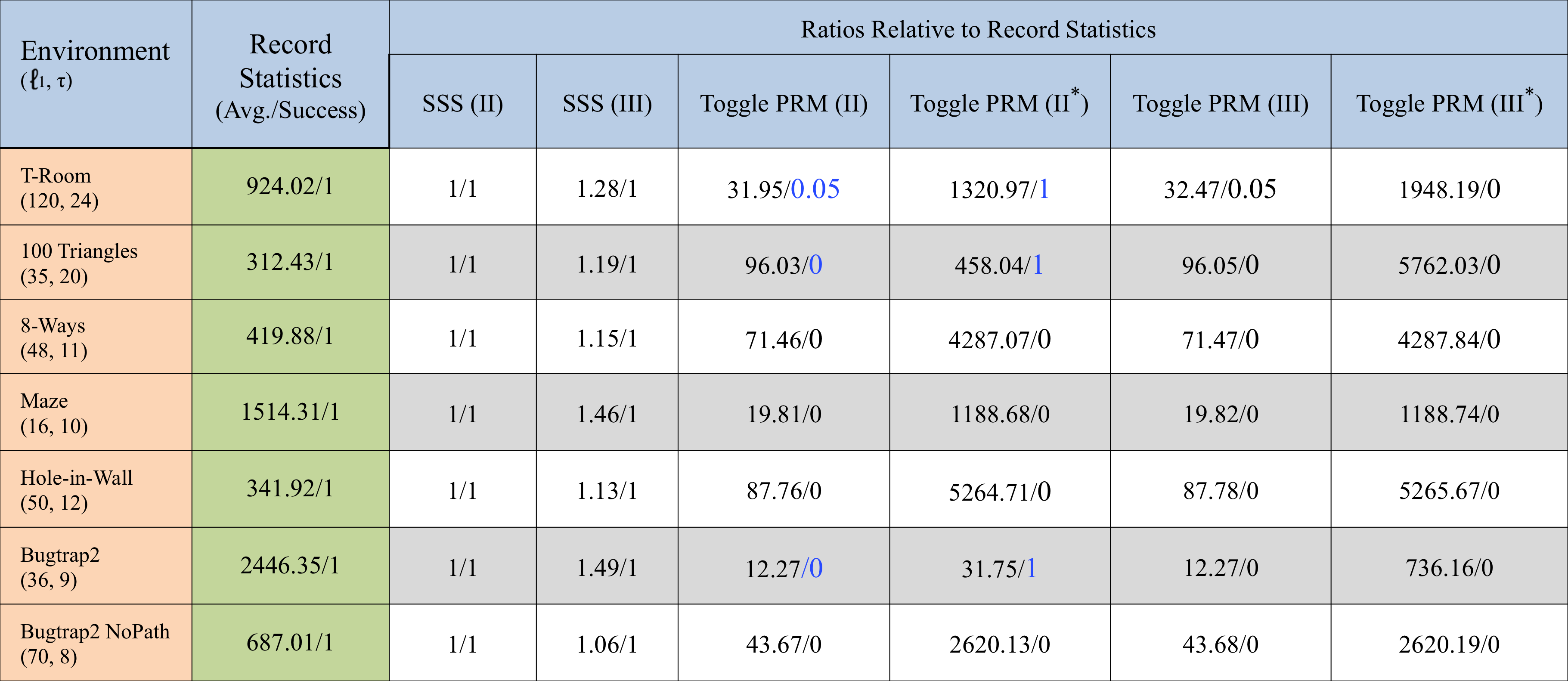}}
\tablename{ 3\textsuperscript{*}: Narrow Passages Result with Crossing and Non-crossing 2-link Robots}
\label{fig:table3star}
\end{table*}

\begin{table*}[htbp]
	\makebox[\textwidth][c]{\includegraphics[width=1.3\textwidth]{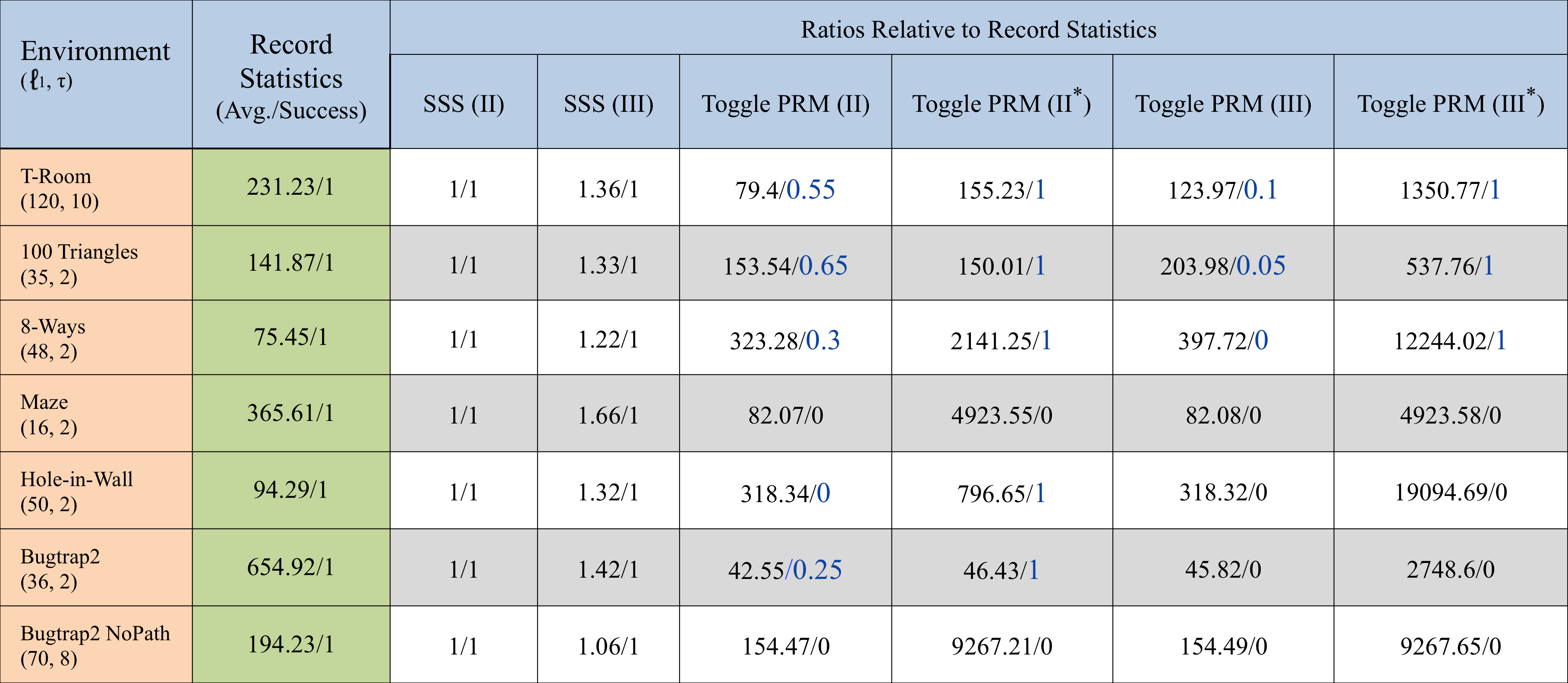}}
\tablename{ 4\textsuperscript{*}: Easy Passages Result with Crossing and Non-crossing 2-link Robots}
\label{fig:table4star}
\end{table*}

\end{document}

\cite{sucan2012the-open-motion-planning-library} 
\cite{yap:sss:13}	

%% file: lmac.tex
\ignore{
	\begin{figure}[ht]
	  	\centering
		\subfloat[Start and goal configurations]{
	   \includegraphics[width=0.36\textwidth]{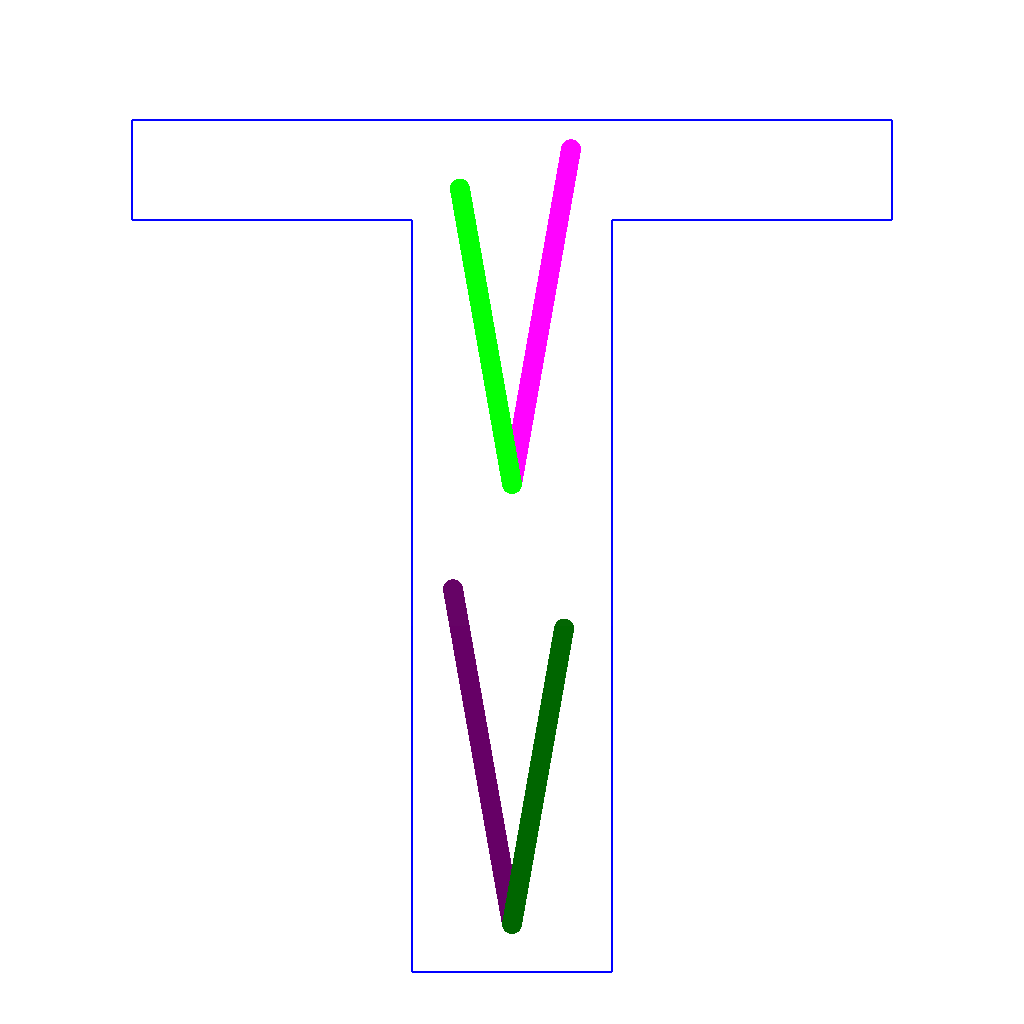}}
		\subfloat[Self-crossing path]{
	   \includegraphics[width=0.36\textwidth]{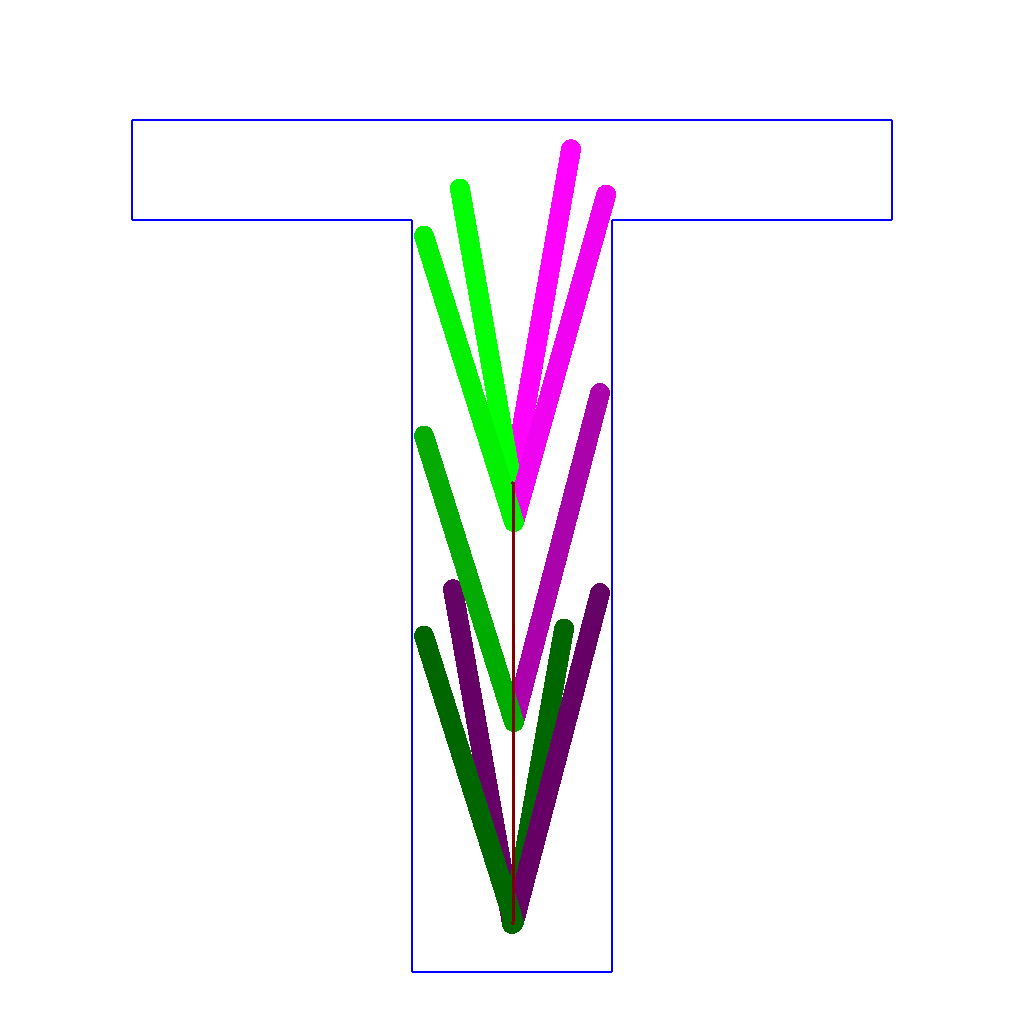}}
		\subfloat[Non-crossing path]{
	   \includegraphics[width=0.36\textwidth]{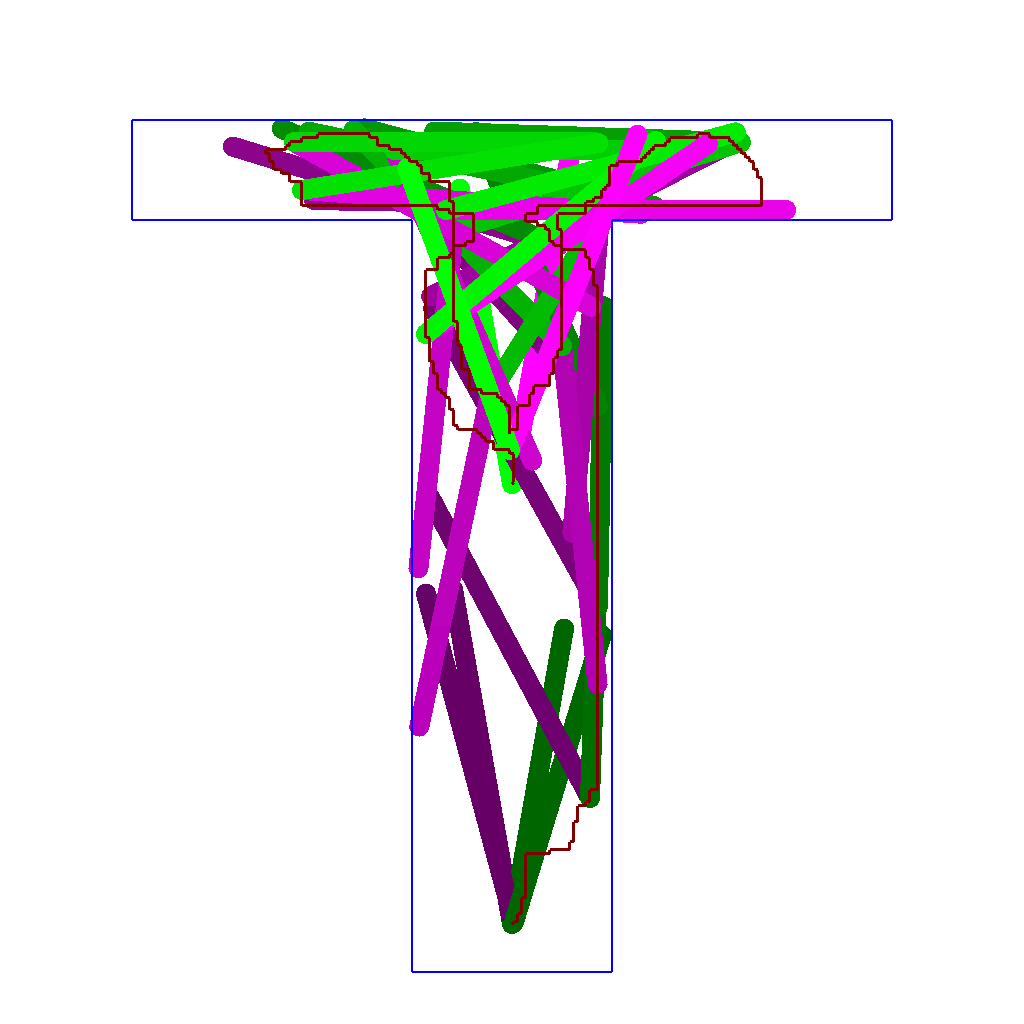}}
		\caption{T-Room Environment}
		\label{fig:Troom}
	\end{figure}
}%

	\renewcommand{\TT}{{\mathbb T}^2}		
	\newcommand{\wtC}{\wt{C}}	

	\newcommand{\Fp}{\ensuremath{Fprint}}	

	\newcommand{\free}{\ensuremath{\mathtt{FREE}}}
	\newcommand{\stuck}{\ensuremath{\mathtt{STUCK}}}
	
	\newcommand{\mixed}{\ensuremath{\mathtt{MIXED}}}
	
	\newcommand{\splittt}{\ensuremath{\mathtt Split}}
	
	\newcommand{\Sep}{\ensuremath{\mathrm Sep}}

	\newcommand{\TTT}{{\cal T}}

	\newcommand{\Cl}{\mathop{C\ell}}	
	\newcommand{\cspace}{\ensuremath{C_{space}}}
	\newcommand{\cfree}{\ensuremath{C_{free}}}

	\newcommand{\getnext}{{\tt GetNext}}

	\newcommand{\wtphi}{\wt{\phi}}

	\newcommand{\Forb}{\ensuremath{\mathrm {Forb}}}	
	\newcommand{\Forbl}{\ensuremath{\mathrm {Forb}_{\ell}}}	
	\newcommand{\Forblt}{\ensuremath{\mathrm {Forb}_{\ell,\tau}}} 
	\newcommand{\olX}{\ol{X}}	

	\newcommand{\cspacedel}{\cspace(\kappa)}
	\newcommand{\XT}{{\tt XT}}
	\newcommand{\GT}{{\tt GT}}
	\newcommand{\LT}{{\tt LT}}
	\newcommand{\isBoxEmpty}{{\tt isBoxEmpty}}

	\newcommand{\proj}[1][]{\textstyle{\mathop{\tt Proj}_{#1}}}
	\newcommand{\coproj}[1][]{\textstyle{\mathop{\tt Coproj}_{#1}}}


%% file: linkrobot.pdf_t
\begin{picture}(0,0)%
\includegraphics{linkrobot.pdf}
\end{picture}%
\setlength{\unitlength}{3947sp}%
\begingroup\makeatletter\ifx\SetFigFont\undefined%
\gdef\SetFigFont#1#2#3#4#5{%
  \reset@font\fontsize{#1}{#2pt}%
  \fontfamily{#3}\fontseries{#4}\fontshape{#5}%
  \selectfont}%
\fi\endgroup%
\begin{picture}(16031,5250)(-208,-5714)
\put(1178,-3588){\makebox(0,0)[rb]{\smash{{\SetFigFont{20}{24.0}{\familydefault}{\mddefault}{\updefault}{\color[rgb]{0,0,0}$\theta_2$}%
}}}}
\put(1764,-3708){\makebox(0,0)[lb]{\smash{{\SetFigFont{20}{24.0}{\familydefault}{\mddefault}{\updefault}{\color[rgb]{0,0,0}$\theta_1$}%
}}}}
\put(1272,-4472){\makebox(0,0)[rb]{\smash{{\SetFigFont{20}{24.0}{\familydefault}{\mddefault}{\updefault}{\color[rgb]{0,0,0}$A_0$}%
}}}}
\put(-193,-3923){\makebox(0,0)[rb]{\smash{{\SetFigFont{20}{24.0}{\familydefault}{\mddefault}{\updefault}{\color[rgb]{0,0,0}$A_2$}%
}}}}
\put(1905,-1462){\makebox(0,0)[rb]{\smash{{\SetFigFont{20}{24.0}{\familydefault}{\mddefault}{\updefault}{\color[rgb]{0,0,0}$A_1$}%
}}}}
\put(14626,-811){\makebox(0,0)[lb]{\smash{{\SetFigFont{20}{24.0}{\familydefault}{\mddefault}{\updefault}{\color[rgb]{0,0,0}$A_1$}%
}}}}
\put(4576,-2761){\makebox(0,0)[rb]{\smash{{\SetFigFont{20}{24.0}{\familydefault}{\mddefault}{\updefault}{\color[rgb]{0,0,0}$A_2$}%
}}}}
\put(4651,-3811){\makebox(0,0)[rb]{\smash{{\SetFigFont{20}{24.0}{\familydefault}{\mddefault}{\updefault}{\color[rgb]{0,0,0}$A_5$}%
}}}}
\put(4201,-4711){\makebox(0,0)[rb]{\smash{{\SetFigFont{20}{24.0}{\familydefault}{\mddefault}{\updefault}{\color[rgb]{0,0,0}$A_0$}%
}}}}
\put(13967,-4621){\makebox(0,0)[lb]{\smash{{\SetFigFont{20}{24.0}{\familydefault}{\mddefault}{\updefault}{\color[rgb]{0,0,0}$A_2$}%
}}}}
\put(14575,-5485){\makebox(0,0)[b]{\smash{{\SetFigFont{20}{24.0}{\familydefault}{\mddefault}{\updefault}{\color[rgb]{0,0,0}(d) Thick $R_2$ Robot}%
}}}}
\put(15808,-3424){\makebox(0,0)[lb]{\smash{{\SetFigFont{20}{24.0}{\familydefault}{\mddefault}{\updefault}{\color[rgb]{0,0,0}$A_0$}%
}}}}
\put(1065,-5605){\makebox(0,0)[b]{\smash{{\SetFigFont{20}{24.0}{\familydefault}{\mddefault}{\updefault}{\color[rgb]{0,0,0}(a) Thin $R_2$ Robot}%
}}}}
\put(6976,-4336){\makebox(0,0)[lb]{\smash{{\SetFigFont{20}{24.0}{\familydefault}{\mddefault}{\updefault}{\color[rgb]{0,0,0}$A_1$}%
}}}}
\put(7426,-2761){\makebox(0,0)[lb]{\smash{{\SetFigFont{20}{24.0}{\familydefault}{\mddefault}{\updefault}{\color[rgb]{0,0,0}$A_4$}%
}}}}
\put(5326,-736){\makebox(0,0)[rb]{\smash{{\SetFigFont{20}{24.0}{\familydefault}{\mddefault}{\updefault}{\color[rgb]{0,0,0}$A_3$}%
}}}}
\put(5375,-5580){\makebox(0,0)[b]{\smash{{\SetFigFont{20}{24.0}{\familydefault}{\mddefault}{\updefault}{\color[rgb]{0,0,0}(b) Chain Robot}%
}}}}
\put(10145,-5385){\makebox(0,0)[b]{\smash{{\SetFigFont{20}{24.0}{\familydefault}{\mddefault}{\updefault}{\color[rgb]{0,0,0}(c) Spider Robot}%
}}}}
\put(10366,-2896){\makebox(0,0)[rb]{\smash{{\SetFigFont{20}{24.0}{\familydefault}{\mddefault}{\updefault}{\color[rgb]{0,0,0}$A_0$}%
}}}}
\end{picture}%

%% file: torusCylinder.pdf_t
\begin{picture}(0,0)%
\includegraphics{torusCylinder.pdf}
\end{picture}%
\setlength{\unitlength}{3947sp}%
\begingroup\makeatletter\ifx\SetFigFont\undefined%
\gdef\SetFigFont#1#2#3#4#5{%
  \reset@font\fontsize{#1}{#2pt}%
  \fontfamily{#3}\fontseries{#4}\fontshape{#5}%
  \selectfont}%
\fi\endgroup%
\begin{picture}(12967,5022)(2161,-6166)
\put(11326,-6061){\makebox(0,0)[lb]{\smash{{\SetFigFont{20}{24.0}{\familydefault}{\mddefault}{\updefault}{\color[rgb]{0,0,0}(b)}%
}}}}
\put(2326,-5161){\makebox(0,0)[rb]{\smash{{\SetFigFont{20}{24.0}{\familydefault}{\mddefault}{\updefault}{\color[rgb]{0,0,0}$0$}%
}}}}
\put(5109,-4874){\makebox(0,0)[b]{\smash{{\SetFigFont{20}{24.0}{\familydefault}{\mddefault}{\updefault}{\color[rgb]{0,0,0}$\TT_>$}%
}}}}
\put(3324,-2073){\makebox(0,0)[b]{\smash{{\SetFigFont{20}{24.0}{\familydefault}{\mddefault}{\updefault}{\color[rgb]{0,0,0}$\TT_<$}%
}}}}
\put(4503,-3944){\makebox(0,0)[rb]{\smash{{\SetFigFont{20}{24.0}{\familydefault}{\mddefault}{\updefault}{\color[rgb]{0,0,0}$\alpha$}%
}}}}
\put(3051,-4633){\makebox(0,0)[b]{\smash{{\SetFigFont{20}{24.0}{\familydefault}{\mddefault}{\updefault}{\color[rgb]{0,0,0}$\Delta(0)$}%
}}}}
\put(6226,-2686){\makebox(0,0)[lb]{\smash{{\SetFigFont{20}{24.0}{\familydefault}{\mddefault}{\updefault}{\color[rgb]{0,0,0}$\gamma$}%
}}}}
\put(2176,-2686){\makebox(0,0)[rb]{\smash{{\SetFigFont{20}{24.0}{\familydefault}{\mddefault}{\updefault}{\color[rgb]{0,0,0}$\gamma$}%
}}}}
\put(2326,-3736){\makebox(0,0)[rb]{\smash{{\SetFigFont{20}{24.0}{\familydefault}{\mddefault}{\updefault}{\color[rgb]{0,0,0}$\theta_2$}%
}}}}
\put(3901,-5536){\makebox(0,0)[b]{\smash{{\SetFigFont{20}{24.0}{\familydefault}{\mddefault}{\updefault}{\color[rgb]{0,0,0}$\theta_1$}%
}}}}
\put(3883,-3031){\makebox(0,0)[lb]{\smash{{\SetFigFont{20}{24.0}{\familydefault}{\mddefault}{\updefault}{\color[rgb]{0,0,0}$\beta$}%
}}}}
\put(3976,-6061){\makebox(0,0)[lb]{\smash{{\SetFigFont{20}{24.0}{\familydefault}{\mddefault}{\updefault}{\color[rgb]{0,0,0}(a)}%
}}}}
\put(10201,-3661){\makebox(0,0)[rb]{\smash{{\SetFigFont{20}{24.0}{\familydefault}{\mddefault}{\updefault}{\color[rgb]{0,0,0}$\alpha$}%
}}}}
\put(12751,-3286){\makebox(0,0)[lb]{\smash{{\SetFigFont{20}{24.0}{\familydefault}{\mddefault}{\updefault}{\color[rgb]{0,0,0}$\beta$}%
}}}}
\put(12601,-1411){\makebox(0,0)[rb]{\smash{{\SetFigFont{20}{24.0}{\familydefault}{\mddefault}{\updefault}{\color[rgb]{0,0,0}$\gamma'$}%
}}}}
\put(8926,-5536){\makebox(0,0)[lb]{\smash{{\SetFigFont{20}{24.0}{\familydefault}{\mddefault}{\updefault}{\color[rgb]{0,0,0}$\gamma'$}%
}}}}
\put(8926,-4861){\makebox(0,0)[b]{\smash{{\SetFigFont{20}{24.0}{\familydefault}{\mddefault}{\updefault}{\color[rgb]{0,0,0}$\TT_>$}%
}}}}
\put(13951,-2086){\makebox(0,0)[b]{\smash{{\SetFigFont{20}{24.0}{\familydefault}{\mddefault}{\updefault}{\color[rgb]{0,0,0}$\TT_<$}%
}}}}
\put(2401,-5536){\makebox(0,0)[lb]{\smash{{\SetFigFont{20}{24.0}{\familydefault}{\mddefault}{\updefault}{\color[rgb]{0,0,0}$0$}%
}}}}
\put(6001,-5536){\makebox(0,0)[rb]{\smash{{\SetFigFont{20}{24.0}{\familydefault}{\mddefault}{\updefault}{\color[rgb]{0,0,0}$2\pi$}%
}}}}
\put(2326,-1786){\makebox(0,0)[rb]{\smash{{\SetFigFont{20}{24.0}{\familydefault}{\mddefault}{\updefault}{\color[rgb]{0,0,0}$2\pi$}%
}}}}
\end{picture}%

%% file: diamond.pdf_t
\begin{picture}(0,0)%
\includegraphics{diamond.pdf}
\end{picture}%
\setlength{\unitlength}{3947sp}%
\begingroup\makeatletter\ifx\SetFigFont\undefined%
\gdef\SetFigFont#1#2#3#4#5{%
  \reset@font\fontsize{#1}{#2pt}%
  \fontfamily{#3}\fontseries{#4}\fontshape{#5}%
  \selectfont}%
\fi\endgroup%
\begin{picture}(11286,7257)(-5769,-8241)
\put(-4092,-5984){\makebox(0,0)[rb]{\smash{{\SetFigFont{20}{24.0}{\familydefault}{\mddefault}{\updefault}{\color[rgb]{0,0,0}$V$}%
}}}}
\put(5502,-2748){\makebox(0,0)[lb]{\smash{{\SetFigFont{20}{24.0}{\familydefault}{\mddefault}{\updefault}{\color[rgb]{0,0,0}$C$}%
}}}}
\put(2283,-5984){\makebox(0,0)[rb]{\smash{{\SetFigFont{20}{24.0}{\familydefault}{\mddefault}{\updefault}{\color[rgb]{0,0,0}$V$}%
}}}}
\put(3518,-2840){\makebox(0,0)[rb]{\smash{{\SetFigFont{20}{24.0}{\familydefault}{\mddefault}{\updefault}{\color[rgb]{0,0,0}$U$}%
}}}}
\put(-4054,-8136){\makebox(0,0)[b]{\smash{{\SetFigFont{20}{24.0}{\familydefault}{\mddefault}{\updefault}{\color[rgb]{0,0,0}(a)}%
}}}}
\put(2406,-8126){\makebox(0,0)[b]{\smash{{\SetFigFont{20}{24.0}{\familydefault}{\mddefault}{\updefault}{\color[rgb]{0,0,0}(b)}%
}}}}
\put(-3387,-5692){\makebox(0,0)[lb]{\smash{{\SetFigFont{20}{24.0}{\familydefault}{\mddefault}{\updefault}{\color[rgb]{0,0,0}$\nu$}%
}}}}
\put(-3179,-3637){\makebox(0,0)[rb]{\smash{{\SetFigFont{20}{24.0}{\familydefault}{\mddefault}{\updefault}{\color[rgb]{0,0,0}$U'$}%
}}}}
\put(-1333,-4385){\makebox(0,0)[lb]{\smash{{\SetFigFont{20}{24.0}{\familydefault}{\mddefault}{\updefault}{\color[rgb]{0,0,0}$C$}%
}}}}
\put(2951,-5639){\makebox(0,0)[lb]{\smash{{\SetFigFont{20}{24.0}{\familydefault}{\mddefault}{\updefault}{\color[rgb]{0,0,0}$\nu$}%
}}}}
\put(-2849,-2761){\makebox(0,0)[rb]{\smash{{\SetFigFont{20}{24.0}{\familydefault}{\mddefault}{\updefault}{\color[rgb]{0,0,0}$U$}%
}}}}
\put(4311,-3943){\makebox(0,0)[rb]{\smash{{\SetFigFont{20}{24.0}{\familydefault}{\mddefault}{\updefault}{\color[rgb]{0,0,0}$d$}%
}}}}
\put(-3567,-5155){\makebox(0,0)[lb]{\smash{{\SetFigFont{20}{24.0}{\familydefault}{\mddefault}{\updefault}{\color[rgb]{0,0,0}$\delta$}%
}}}}
\put(-3430,-4494){\makebox(0,0)[rb]{\smash{{\SetFigFont{20}{24.0}{\familydefault}{\mddefault}{\updefault}{\color[rgb]{0,0,0}$\ell$}%
}}}}
\put(2852,-5032){\makebox(0,0)[lb]{\smash{{\SetFigFont{20}{24.0}{\familydefault}{\mddefault}{\updefault}{\color[rgb]{0,0,0}$\delta$}%
}}}}
\put(4283,-2854){\makebox(0,0)[lb]{\smash{{\SetFigFont{20}{24.0}{\familydefault}{\mddefault}{\updefault}{\color[rgb]{0,0,0}$\tau$}%
}}}}
\put(-2212,-4833){\makebox(0,0)[lb]{\smash{{\SetFigFont{20}{24.0}{\familydefault}{\mddefault}{\updefault}{\color[rgb]{0,0,0}$d$}%
}}}}
\put(3225,-3774){\makebox(0,0)[rb]{\smash{{\SetFigFont{20}{24.0}{\familydefault}{\mddefault}{\updefault}{\color[rgb]{0,0,0}$\ell$}%
}}}}
\put(-2383,-3817){\makebox(0,0)[lb]{\smash{{\SetFigFont{20}{24.0}{\familydefault}{\mddefault}{\updefault}{\color[rgb]{0,0,0}$\tau$}%
}}}}
\end{picture}%

%% file: stops.pdf_t
\begin{picture}(0,0)%
\includegraphics{stops.pdf}
\end{picture}%
\setlength{\unitlength}{3947sp}%
\begingroup\makeatletter\ifx\SetFigFont\undefined%
\gdef\SetFigFont#1#2#3#4#5{%
  \reset@font\fontsize{#1}{#2pt}%
  \fontfamily{#3}\fontseries{#4}\fontshape{#5}%
  \selectfont}%
\fi\endgroup%
\begin{picture}(19224,8217)(1189,-9166)
\put(14926,-2986){\makebox(0,0)[rb]{\smash{{\SetFigFont{20}{24.0}{\familydefault}{\mddefault}{\updefault}{\color[rgb]{0,0,0}$C'$}%
}}}}
\put(19726,-2836){\makebox(0,0)[lb]{\smash{{\SetFigFont{20}{24.0}{\familydefault}{\mddefault}{\updefault}{\color[rgb]{0,0,0}$x$}%
}}}}
\put(12301,-3211){\makebox(0,0)[rb]{\smash{{\SetFigFont{20}{24.0}{\familydefault}{\mddefault}{\updefault}{\color[rgb]{0,0,0}$\tau$}%
}}}}
\put(17926,-1861){\makebox(0,0)[lb]{\smash{{\SetFigFont{20}{24.0}{\familydefault}{\mddefault}{\updefault}{\color[rgb]{0,0,0}$X_{\max}$}%
}}}}
\put(18001,-2461){\makebox(0,0)[lb]{\smash{{\SetFigFont{20}{24.0}{\familydefault}{\mddefault}{\updefault}{\color[rgb]{0,0,0}$X_*$}%
}}}}
\put(16051,-5461){\makebox(0,0)[lb]{\smash{{\SetFigFont{20}{24.0}{\familydefault}{\mddefault}{\updefault}{\color[rgb]{0,0,0}$V$}%
}}}}
\put(6226,-5536){\makebox(0,0)[lb]{\smash{{\SetFigFont{20}{24.0}{\familydefault}{\mddefault}{\updefault}{\color[rgb]{0,0,0}$V$}%
}}}}
\put(2701,-6286){\makebox(0,0)[rb]{\smash{{\SetFigFont{20}{24.0}{\familydefault}{\mddefault}{\updefault}{\color[rgb]{0,0,0}$\ell+\tau$}%
}}}}
\put(3901,-6661){\makebox(0,0)[rb]{\smash{{\SetFigFont{20}{24.0}{\familydefault}{\mddefault}{\updefault}{\color[rgb]{0,0,0}$\ell$}%
}}}}
\put(5926,-3286){\makebox(0,0)[rb]{\smash{{\SetFigFont{20}{24.0}{\familydefault}{\mddefault}{\updefault}{\color[rgb]{0,0,0}$O$}%
}}}}
\put(3151,-3136){\makebox(0,0)[rb]{\smash{{\SetFigFont{20}{24.0}{\familydefault}{\mddefault}{\updefault}{\color[rgb]{0,0,0}$\olX_*$}%
}}}}
\put(9826,-3586){\makebox(0,0)[lb]{\smash{{\SetFigFont{20}{24.0}{\familydefault}{\mddefault}{\updefault}{\color[rgb]{0,0,0}$x$}%
}}}}
\put(6001,-1336){\makebox(0,0)[lb]{\smash{{\SetFigFont{20}{24.0}{\familydefault}{\mddefault}{\updefault}{\color[rgb]{0,0,0}$y$}%
}}}}
\put(8701,-3211){\makebox(0,0)[lb]{\smash{{\SetFigFont{20}{24.0}{\familydefault}{\mddefault}{\updefault}{\color[rgb]{0,0,0}$X_*$}%
}}}}
\put(2401,-3961){\makebox(0,0)[rb]{\smash{{\SetFigFont{20}{24.0}{\familydefault}{\mddefault}{\updefault}{\color[rgb]{0,0,0}$\tau$}%
}}}}
\put(8476,-2611){\makebox(0,0)[lb]{\smash{{\SetFigFont{20}{24.0}{\familydefault}{\mddefault}{\updefault}{\color[rgb]{0,0,0}$X_{\max}$}%
}}}}
\put(13731,-2476){\makebox(0,0)[rb]{\smash{{\SetFigFont{20}{24.0}{\familydefault}{\mddefault}{\updefault}{\color[rgb]{0,0,0}$\olX_*$}%
}}}}
\put(4126,-2686){\makebox(0,0)[lb]{\smash{{\SetFigFont{20}{24.0}{\familydefault}{\mddefault}{\updefault}{\color[rgb]{0,0,0}$\olX_{\max}$}%
}}}}
\put(6001,-9061){\makebox(0,0)[b]{\smash{{\SetFigFont{20}{24.0}{\familydefault}{\mddefault}{\updefault}{\color[rgb]{0,0,0}(a)}%
}}}}
\put(15901,-9061){\makebox(0,0)[b]{\smash{{\SetFigFont{20}{24.0}{\familydefault}{\mddefault}{\updefault}{\color[rgb]{0,0,0}(b)}%
}}}}
\put(5026,-3736){\makebox(0,0)[lb]{\smash{{\SetFigFont{20}{24.0}{\familydefault}{\mddefault}{\updefault}{\color[rgb]{0,0,0}$C$}%
}}}}
\put(3376,-3736){\makebox(0,0)[lb]{\smash{{\SetFigFont{20}{24.0}{\familydefault}{\mddefault}{\updefault}{\color[rgb]{0,0,0}$C'$}%
}}}}
\put(17026,-2986){\makebox(0,0)[lb]{\smash{{\SetFigFont{20}{24.0}{\familydefault}{\mddefault}{\updefault}{\color[rgb]{0,0,0}$C$}%
}}}}
\put(12601,-6286){\makebox(0,0)[rb]{\smash{{\SetFigFont{20}{24.0}{\familydefault}{\mddefault}{\updefault}{\color[rgb]{0,0,0}$\ell+\tau$}%
}}}}
\put(13801,-6661){\makebox(0,0)[rb]{\smash{{\SetFigFont{20}{24.0}{\familydefault}{\mddefault}{\updefault}{\color[rgb]{0,0,0}$\ell$}%
}}}}
\put(15901,-1336){\makebox(0,0)[lb]{\smash{{\SetFigFont{20}{24.0}{\familydefault}{\mddefault}{\updefault}{\color[rgb]{0,0,0}$y$}%
}}}}
\put(15826,-2536){\makebox(0,0)[rb]{\smash{{\SetFigFont{20}{24.0}{\familydefault}{\mddefault}{\updefault}{\color[rgb]{0,0,0}$O$}%
}}}}
\end{picture}%

%% file: wall-box.pdf_t
\begin{picture}(0,0)%
\includegraphics{wall-box.pdf}
\end{picture}%
\setlength{\unitlength}{3947sp}%
\begingroup\makeatletter\ifx\SetFigFont\undefined%
\gdef\SetFigFont#1#2#3#4#5{%
  \reset@font\fontsize{#1}{#2pt}%
  \fontfamily{#3}\fontseries{#4}\fontshape{#5}%
  \selectfont}%
\fi\endgroup%
\begin{picture}(18399,8124)(1114,-9223)
\put(11026,-3961){\makebox(0,0)[rb]{\smash{{\SetFigFont{20}{24.0}{\familydefault}{\mddefault}{\updefault}{\color[rgb]{0,0,0}$V$}%
}}}}
\put(11716,-3916){\makebox(0,0)[b]{\smash{{\SetFigFont{20}{24.0}{\familydefault}{\mddefault}{\updefault}{\color[rgb]{0,0,0}$S$}%
}}}}
\put(2401,-4636){\makebox(0,0)[lb]{\smash{{\SetFigFont{20}{24.0}{\familydefault}{\mddefault}{\updefault}{\color[rgb]{0,0,0}Corner ($C$)}%
}}}}
\put(12211,-4441){\makebox(0,0)[rb]{\smash{{\SetFigFont{20}{24.0}{\familydefault}{\mddefault}{\updefault}{\color[rgb]{0,0,0}$S'$}%
}}}}
\put(11701,-7906){\makebox(0,0)[b]{\smash{{\SetFigFont{20}{24.0}{\familydefault}{\mddefault}{\updefault}{\color[rgb]{0,0,0}$S$}%
}}}}
\put(12256,-8461){\makebox(0,0)[rb]{\smash{{\SetFigFont{20}{24.0}{\familydefault}{\mddefault}{\updefault}{\color[rgb]{0,0,0}$S'$}%
}}}}
\put(13051,-2626){\makebox(0,0)[lb]{\smash{{\SetFigFont{20}{24.0}{\familydefault}{\mddefault}{\updefault}{\color[rgb]{0,0,0}$W$}%
}}}}
\put(17116,-5896){\makebox(0,0)[rb]{\smash{{\SetFigFont{20}{24.0}{\familydefault}{\mddefault}{\updefault}{\color[rgb]{0,0,0}$V$}%
}}}}
\put(17116,-6616){\makebox(0,0)[rb]{\smash{{\SetFigFont{20}{24.0}{\familydefault}{\mddefault}{\updefault}{\color[rgb]{0,0,0}$S'$}%
}}}}
\put(8476,-3931){\makebox(0,0)[lb]{\smash{{\SetFigFont{20}{24.0}{\familydefault}{\mddefault}{\updefault}{\color[rgb]{0,0,0}$V'$}%
}}}}
\put(6661,-3946){\makebox(0,0)[rb]{\smash{{\SetFigFont{20}{24.0}{\familydefault}{\mddefault}{\updefault}{\color[rgb]{0,0,0}$V$}%
}}}}
\put(5671,-2011){\makebox(0,0)[lb]{\smash{{\SetFigFont{20}{24.0}{\familydefault}{\mddefault}{\updefault}{\color[rgb]{0,0,0}$H(S)$}%
}}}}
\put(12301,-2611){\makebox(0,0)[rb]{\smash{{\SetFigFont{20}{24.0}{\familydefault}{\mddefault}{\updefault}{\color[rgb]{0,0,0}$C'$}%
}}}}
\put(2401,-3811){\makebox(0,0)[lb]{\smash{{\SetFigFont{20}{24.0}{\familydefault}{\mddefault}{\updefault}{\color[rgb]{0,0,0}Wall ($W$)}%
}}}}
\put(13796,-3391){\makebox(0,0)[lb]{\smash{{\SetFigFont{20}{24.0}{\familydefault}{\mddefault}{\updefault}{\color[rgb]{0,0,0}$C$}%
}}}}
\put(17241,-3957){\makebox(0,0)[lb]{\smash{{\SetFigFont{20}{24.0}{\familydefault}{\mddefault}{\updefault}{\color[rgb]{0,0,0}$H(S)\cap H(S')$}%
}}}}
\put(8026,-2461){\makebox(0,0)[b]{\smash{{\SetFigFont{20}{24.0}{\familydefault}{\mddefault}{\updefault}{\color[rgb]{0,0,0}$W$}%
}}}}
\put(13276,-6736){\makebox(0,0)[rb]{\smash{{\SetFigFont{20}{24.0}{\familydefault}{\mddefault}{\updefault}{\color[rgb]{0,0,0}$W$}%
}}}}
\put(17551,-5161){\makebox(0,0)[lb]{\smash{{\SetFigFont{20}{24.0}{\familydefault}{\mddefault}{\updefault}{\color[rgb]{0,0,0}$W$}%
}}}}
\put(7126,-2911){\makebox(0,0)[rb]{\smash{{\SetFigFont{20}{24.0}{\familydefault}{\mddefault}{\updefault}{\color[rgb]{0,0,0}$C'$}%
}}}}
\put(9151,-2836){\makebox(0,0)[lb]{\smash{{\SetFigFont{20}{24.0}{\familydefault}{\mddefault}{\updefault}{\color[rgb]{0,0,0}$C$}%
}}}}
\put(7651,-6436){\makebox(0,0)[lb]{\smash{{\SetFigFont{20}{24.0}{\familydefault}{\mddefault}{\updefault}{\color[rgb]{0,0,0}$W$}%
}}}}
\put(7951,-7261){\makebox(0,0)[lb]{\smash{{\SetFigFont{20}{24.0}{\familydefault}{\mddefault}{\updefault}{\color[rgb]{0,0,0}$C$}%
}}}}
\put(12676,-7111){\makebox(0,0)[rb]{\smash{{\SetFigFont{20}{24.0}{\familydefault}{\mddefault}{\updefault}{\color[rgb]{0,0,0}$C$}%
}}}}
\put(18226,-6436){\makebox(0,0)[rb]{\smash{{\SetFigFont{20}{24.0}{\familydefault}{\mddefault}{\updefault}{\color[rgb]{0,0,0}$C$}%
}}}}
\put(16426,-4186){\makebox(0,0)[rb]{\smash{{\SetFigFont{20}{24.0}{\familydefault}{\mddefault}{\updefault}{\color[rgb]{0,0,0}$C'$}%
}}}}
\put(7561,-7861){\makebox(0,0)[b]{\smash{{\SetFigFont{20}{24.0}{\familydefault}{\mddefault}{\updefault}{\color[rgb]{0,0,0}$S$}%
}}}}
\put(7551,-3911){\makebox(0,0)[b]{\smash{{\SetFigFont{20}{24.0}{\familydefault}{\mddefault}{\updefault}{\color[rgb]{0,0,0}$S$}%
}}}}
\put(1276,-3061){\makebox(0,0)[lb]{\smash{{\SetFigFont{20}{24.0}{\familydefault}{\mddefault}{\updefault}{\color[rgb]{0,0,0}KEY:}%
}}}}
\put(2401,-5386){\makebox(0,0)[lb]{\smash{{\SetFigFont{20}{24.0}{\familydefault}{\mddefault}{\updefault}{\color[rgb]{0,0,0}Side ($S$)}%
}}}}
\put(2401,-5836){\makebox(0,0)[lb]{\smash{{\SetFigFont{20}{24.0}{\familydefault}{\mddefault}{\updefault}{\color[rgb]{0,0,0}vertex ($V$)}%
}}}}
\put(1726,-6511){\makebox(0,0)[b]{\smash{{\SetFigFont{20}{24.0}{\familydefault}{\mddefault}{\updefault}{\color[rgb]{0,0,0}$H(S)$}%
}}}}
\put(2401,-6661){\makebox(0,0)[lb]{\smash{{\SetFigFont{20}{24.0}{\familydefault}{\mddefault}{\updefault}{\color[rgb]{0,0,0}Halfspace ($H(S)$)}%
}}}}
\put(1726,-7261){\makebox(0,0)[b]{\smash{{\SetFigFont{20}{24.0}{\familydefault}{\mddefault}{\updefault}{\color[rgb]{0,0,0}$S$}%
}}}}
\put(12451,-4936){\makebox(0,0)[lb]{\smash{{\SetFigFont{20}{24.0}{\familydefault}{\mddefault}{\updefault}{\color[rgb]{0,0,0}$V'$}%
}}}}
\put(12451,-8986){\makebox(0,0)[lb]{\smash{{\SetFigFont{20}{24.0}{\familydefault}{\mddefault}{\updefault}{\color[rgb]{0,0,0}$V'$}%
}}}}
\put(5626,-5911){\makebox(0,0)[lb]{\smash{{\SetFigFont{20}{24.0}{\familydefault}{\mddefault}{\updefault}{\color[rgb]{0,0,0}$H(S)$}%
}}}}
\put(16276,-5836){\makebox(0,0)[rb]{\smash{{\SetFigFont{20}{24.0}{\familydefault}{\mddefault}{\updefault}{\color[rgb]{0,0,0}$S$}%
}}}}
\put(18976,-7411){\makebox(0,0)[b]{\smash{{\SetFigFont{20}{24.0}{\familydefault}{\mddefault}{\updefault}{\color[rgb]{0,0,0}(III)}%
}}}}
\put(14326,-8686){\makebox(0,0)[b]{\smash{{\SetFigFont{20}{24.0}{\familydefault}{\mddefault}{\updefault}{\color[rgb]{0,0,0}(IIb)}%
}}}}
\put(14251,-4636){\makebox(0,0)[b]{\smash{{\SetFigFont{20}{24.0}{\familydefault}{\mddefault}{\updefault}{\color[rgb]{0,0,0}(IIa)}%
}}}}
\put(9676,-4711){\makebox(0,0)[b]{\smash{{\SetFigFont{20}{24.0}{\familydefault}{\mddefault}{\updefault}{\color[rgb]{0,0,0}(Ia)}%
}}}}
\put(9676,-8686){\makebox(0,0)[b]{\smash{{\SetFigFont{20}{24.0}{\familydefault}{\mddefault}{\updefault}{\color[rgb]{0,0,0}(Ib)}%
}}}}
\put(8251,-7861){\makebox(0,0)[lb]{\smash{{\SetFigFont{20}{24.0}{\familydefault}{\mddefault}{\updefault}{\color[rgb]{0,0,0}$V'$}%
}}}}
\put(6901,-7861){\makebox(0,0)[rb]{\smash{{\SetFigFont{20}{24.0}{\familydefault}{\mddefault}{\updefault}{\color[rgb]{0,0,0}$V$}%
}}}}
\put(11101,-7936){\makebox(0,0)[rb]{\smash{{\SetFigFont{20}{24.0}{\familydefault}{\mddefault}{\updefault}{\color[rgb]{0,0,0}$V$}%
}}}}
\end{picture}%